\documentclass[traditabstract,longauth]{aa} 
\usepackage{graphicx}
\usepackage[varg]{txfonts}
\usepackage{lscape}
\usepackage{url}
\begin{document} 

\title{Properties of flat-spectrum radio-loud\\narrow-line Seyfert 1 galaxies}

\author{L. Foschini\inst{1},
M.~Berton\inst{2},
A.~Caccianiga\inst{1},
S.~Ciroi\inst{2},
V.~Cracco\inst{2},
B.~M. Peterson\inst{3},
E.~Angelakis\inst{4},
V.~Braito\inst{1},
L.~Fuhrmann\inst{4},
L.~Gallo\inst{5},
D.~Grupe\inst{6,7},
E.~J\"arvel\"a\inst{8,9},
S.~Kaufmann\inst{10},
S.~Komossa\inst{4},
Y.~Y.~Kovalev\inst{11,4},
A.~L\"ahteenm\"aki\inst{8,9},
M.~M.~Lisakov\inst{11},
M.~L. Lister\inst{12},
S.~Mathur\inst{3},
J.~L.~Richards\inst{12},
P.~Romano\inst{13},
A.~Sievers\inst{14},
G.~Tagliaferri\inst{1},
J.~Tammi\inst{8},
O.~Tibolla\inst{15,16,17},
M.~Tornikoski\inst{8},
S.~Vercellone\inst{13},
G.~La~Mura\inst{2},
L.~Maraschi\inst{1},
P.~Rafanelli\inst{2}
}

\institute{INAF -- Osservatorio Astronomico di Brera, via E. Bianchi 46, 23807 Merate (LC), Italy; 
\email{luigi.foschini@brera.inaf.it}
\and
Dipartimento di Fisica e Astronomia, Universit\`a di Padova, vicolo dell'Osservatorio 3, 35122 Padova, Italy;
\and
Department of Astronomy and Center for Cosmology and AstroParticle Physics, 
The Ohio State University, 140 West 18th Avenue, Columbus, OH 43210, USA;
\and
Max-Planck-Institut f\"ur Radioastronomie, Auf dem H\"ugel 69, 53121 Bonn, Germany;
\and
Department of Astronomy and Physics, Saint Mary's University, Halifax, Canada;
\and
Space Science Center, Morehead State University, 235 Martindale Dr., Morehead, KY 40351, USA;
\and
Swift Mission Operation Center, 2582 Gateway Dr., State College, PA 16801, USA;
\and
Aalto University Mets\"ahovi Radio Observatory, Mets\"ahovintie 114, FIN-02540 Kylm\"al\"a, Finland;
\and
Aalto University Department of Radio Science and Engineering, P.O.Box 13000, FI-00076 Aalto, Finland;
\and
Landessternwarte, Universit\"at Heidelberg, K\"onigstuhl, D 69117 Heidelberg, Germany;
\and
Astro Space Center of the Lebedev Physical Institute, Moscow, Russia;
\and
Department of Physics, Purdue University, West Lafayette, IN 47907, USA;
\and
INAF -- Istituto di Astrofisica Spaziale e Fisica Cosmica, 90146 Palermo, Italy;
\and
Institut de Radio Astronomie Millim\'etrique, Avenida Divina Pastora 7, Local 20, 18012 Granada, Spain;
\and
ITPA, Universit\"at W\"urzburg, Campus Hubland Nord, Emil-Fischer-Str. 31 D-97074 W\"urzburg, Germany.
\and
Mesoamerican Centre for Theoretical Physics (MCTP), Universidad Aut\'onoma de Chiapas (UNACH), 
Carretera Emiliano Zapata Km. 4, Real del Bosque (Ter\`an). 29050 Tuxtla Guti\'errez, Chiapas, M\'exico.
\and
Istituto Nazionale di Fisica Nucleare, Sezione di Trieste, I-34127 Trieste, Italy.
}

\date{Received ---; accepted ---}
\authorrunning{L. Foschini et al.}
 
\abstract{We have conducted a multiwavelength survey of 42 radio loud narrow-1ine Seyfert 1 galaxies (RLNLS1s), selected by searching among all the known sources of this type and omitting those with steep radio spectra. We analyse data from radio frequencies to X-rays, and supplement these with information available from online catalogues and the literature in order to cover the full electromagnetic spectrum. This is the largest known multiwavelength survey for this type of source. We detected 90\% of the sources in X-rays and found 17\% at $\gamma$ rays. Extreme variability at high energies was also found, down to timescales as short as hours. In some sources, dramatic spectral and flux changes suggest interplay between a relativistic jet and the accretion disk. The estimated masses of the central black holes are in the range $\sim 10^{6-8}M_{\odot}$, lower than those of blazars, while the accretion luminosities span a range from $\sim 0.01$ to $\sim 0.49$ times the Eddington limit, similar to those of quasars. The distribution of the calculated jet power spans a range from $\sim 10^{42.6}$ to $\sim 10^{45.6}$~erg~s$^{-1}$, generally lower than quasars and BL Lac objects, but partially overlapping with the latter. Once normalised by the mass of the central black holes, the jet power of the three types of active galactic nuclei are consistent with each other, indicating that the jets are similar and the observational differences are due to scaling factors. Despite the observational differences, the central engine of RLNLS1s is apparently quite similar to that of blazars. The historical difficulties in finding radio-loud narrow-line Seyfert 1 galaxies might be due to their low power and to intermittent jet activity.}


   \maketitle

\section{Introduction}
An important new discovery made with the Large Area Telescope (LAT, Atwood et al. 2009) on board the {\it Fermi Gamma-ray Space Telescope} (hereafter {\it Fermi}) is the high-energy gamma-ray emission from radio-loud narrow-line Seyfert 1 galaxies (RLNLS1s, Abdo et al. 2009a,b,c, Foschini et al. 2010). Narrow-line Seyfert 1 galaxies (NLS1s) are a well-known class of active galactic nuclei (AGNs), but they are usually considered to be radio-quiet (e.g. Ulvestad et al. 1995, Moran 2000, Boroson 2002). Thus, the first discoveries of RLNLS1s (e.g. Remillard et al. 1986, Grupe et al. 2000, Oshlack et al. 2001, Zhou et al. 2003) seemed to be exceptions, rather than the tip of an iceberg. The early surveys revealed only a handful of objects: 11 by Zhou \& Wang (2002) and Komossa et al. (2006a), and 16 by Whalen et al. (2006). 
Williams et al. (2002) analysed 150 NLS1s from the {\it Sloan Digital Sky Survey} (SDSS) Early Data Release, and only a dozen (8\%) were detected at radio frequencies and only two (1.3\%) are radio loud, i.e.  with the ratio between radio and optical flux densities greater than 10. One source is also in the present sample (J$0948+0022$, Zhou et al. 2003), while we have discarded the other (J$1722+5654$, Komossa et al. 2006b) because of its steep radio index (see Sect.~2). Most of the mildly radio-loud NLS1 galaxies of Komossa et al. (2006a) are steep-spectrum sources, and do not show indications of beaming, while three sources are more similar to  blazars. In terms of their optical emission-line properties and BH masses, the RLNLS1s are similar to the radio-quiet NLS1 (RQNLS1) population as a whole. A larger study by Zhou et al. (2006) based on SDSS Data Release 3 resulted in a sample of 2011 NLS1s, about 14\% of all the AGNs with broad emission lines. The fraction detected in the radio is $7.1$\%, similar to 
what was found by Williams et al. (2002). From this subsample, Yuan et al. (2008) culled 23 RLNLS1s with radio loudness greater than 100 and found that these sources are characterised by flat radio spectra.  Detection of flux and spectral variability and their characteristic spectral energy distributions (SEDs) suggest a blazar-like nature. 

In 2009, detection at high-energy $\gamma$ rays by Abdo et al. (2009a,c) revealed beyond any reasonable doubt the existence of powerful relativistic jets in RLNLS1s and brought this poorly known class of AGNs into the spotlight (see Foschini 2012a for a recent review). An early survey including gamma-ray detections (after 30 months of {\it Fermi} operations) was carried out by Foschini (2011a). Forty-six  RLNLS1s were found, of which seven were detected by {\it Fermi}. Of  30 RQNLS1 that served as a control sample, none were detected at $\gamma$ rays.  Additional multiwavelength (MW) data, mostly from archives, were employed in this survey; specifically, X-ray data from {\it ROSAT} were used, but yielded a detection rate of only about 60\%. 

To improve our understanding of RLNLS1s, we decided to perform a more extended and detailed study. First, we have revised the sample selection (see Sect.~2), resulting in 42 RLNLS1s. We focus here on the population that is likely beamed (i.e.  where the jet is viewed at small angles); a parallel study on the search for the parent population (i.e.  with the jet viewed at large angles) is ongoing (Berton et al., in preparation). We therefore exclude from this study RLNLS1s with steep radio spectral indices, although we keep the sources with no radio spectral index information. We requested specific observations with {\it Swift} and {\it XMM--Newton} to improve the X-ray detection rate, which is now at 90\%. Observations with these satellites were also accompanied by ultraviolet observations to study the accretion disk emission. Optical spectra were mostly taken from the SDSS archives and from the literature. For two sources, new optical spectra were obtained at the Asiago Astrophysical Observatory (Italy). New radio observations, particularly from monitoring campaigns on the $\gamma$-ray detected RLNLS1s, supplemented the archival data. More details on radio monitoring programs at Effelsberg/Pico Veleta and Mets\"ahovi will be published separately  (Angelakis et al. in preparation, L\"ahteenm\"aki et al. in preparation). Some preliminary results from the present work have already been presented by Foschini et al. (2013).

To facilitate comparison with previous work, we adopt the usual $\Lambda$CDM cosmology with a Hubble--Lema\^itre constant $H_{0}=70$~km~s$^{-1}$~Mpc$^{-1}$ and $\Omega_{\Lambda}=0.73$ (Komatsu et al. 2011). We adopt the flux density and spectral index convention $S_{\nu}\propto \nu^{-\alpha_{\rm \nu}}$.

\section{Sample selection}
The number of RLNLS1s known today is quite small compared to other classes of AGNs. We selected all the sources found in previous surveys (Zhou \& Wang 2002, Komossa et al. 2006a, Whalen et al. 2006, Yuan et al. 2008) and from individual studies (Grupe et al. 2000, Oshlack et al. 2001, Zhou et al. 2003, 2005, 2007, Gallo et al. 2006) that meet the following criteria:

\begin{itemize}
\item Optical spectrum with an H$\beta$ line width ${\rm FWHM(H}\beta)<2000$~km s$^{-1}$ (Goodrich 1989) with tolerance $+10$\%, a line-flux ratio [O\,{\sc iii}]/H$\beta<3$, and clear broad Fe\,{\sc ii} emission blends (Osterbrock \& Pogge 1985).
\item Radio loudness $RL=S_{\rm radio}/S_{\rm optical}>10$, where $S_{\rm radio}$ is the flux density at 5~GHz and $S_{\rm optical}$ is the optical flux density at $440$~nm. In cases where 5~GHz fluxes are not available, we used other frequencies --- generally 1.4~GHz --- under the hypothesis of a flat radio spectrum, (i.e.  $\alpha_{\rm r}\approx 0$).
\item Flat or inverted radio spectra ($\alpha_{\rm r}<0.5$, within the measurement errors), in order to select jets viewed at small angles. Sources with steep radio spectra (corresponding to jets viewed at large angles) are the subject of another survey (Berton et al., in preparation). Sources without spectral information and with only a radio detection at 1.4~GHz are included in our sample.
\end{itemize}

Radio loudness was recalculated on the basis of more recent data from Foschini (2011a), leading to some sources from Whalen et al. (2006) being reclassified as radio loud or radio quiet. Given the variability of the radio emission, we decided to keep all the sources which were classified as radio loud at least in one of the two samples. The resulting list of 42 sources studied in the present work is displayed in Table~\ref{tab:sample}. For each source, we searched all the data available from radio to $\gamma$ rays (see Sect.~3). It is worth noting that in this work we do not make a distinction between quasars and Seyfert galaxies, although most of the sources of the present samples are sufficiently luminous to be classified as quasars. We adopt the general acronym RLNLS1s for all the sources in the sample.

We also note that there has been some doubt about the classification of J2007$-$4434 as NLS1 because of its weak 
Fe\,{\sc ii} emission: Komossa et al. (2006a) proposed a classification as narrow-line radio galaxy, while Gallo et al. (2006) argued that since there is no quantitative criterion on the intensity of Fe\,{\sc ii}, the source can be considered to be a genuine RLNLS1. We follow the latter interpretation and include J2007$-$4434 in our sample. 

To facilitate comparison with blazars, we selected a sample of 57 flat-spectrum radio quasars (FSRQs) and 31 BL Lac objects, all detected by {\it Fermi}/LAT (Ghisellini et al. 2009, 2010, Tavecchio et al. 2010 and references therein). This sample was built by selecting all the sources in the LAT Bright AGN Sample (LBAS, Abdo et al. 2009d) with optical-to-X-ray coverage with {\it Swift} and information about masses of the central black holes and jet power. However, those works do not contain all the information we need to make a complete broad-band comparison with the present set of RLNLS1s. Therefore, we supplemented the published data in the cited works with information from online catalogues, specifically radio data at 15~GHz from the MOJAVE Project (Lister et al. 2009, 2013), ultraviolet fluxes from {\it Swift}/UVOT extracted from the Science Data Center of the Italian Space Agency (ASI-ASDC\footnote{\url{http://www.asdc.asi.it/}}), and X-ray fluxes from the {\it Swift} X-ray Point Sources catalogue (1SXPS, Evans et al. 2014).

\begin{table*}[!ht]
\caption{Sample of RLNLS1s. Columns: (1) Name of the source as used in the present work; (2) Other name often found in the literature; (3) Right Ascension (J2000); (4) Declination (J2000); (5) Redshift from SDSS or NED; (6) Galactic absorption column density [$10^{20}$~cm$^{-2}$] from Kalberla et al. (2005); (7) Full-Width Half Maximum of broad H$\beta$ emission line [km~s$^{-1}$]; (8) Peak radio flux density at 1.4~GHz from VLA/FIRST (Becker et al. 1995) or from the nearest frequency available [mJy]. The coordinates were mostly from the VLA/FIRST survey; when missing, we referred to NED.}
\centering
\begin{tabular}{cccccccl}
\hline
Name          & Alias                     & $\alpha$ & $\delta$        & $z$    & $N_{\mathrm{H}}$  & FWHM H$\beta$ & $S_{\rm 1.4\,GHz}$\\
\hline
J$0100-0200$ & FBQS~J$0100-0200$          & $01:00:32.22$ & $-02:00:46.3$ & $0.227$  &  $4.12$  & $920$     & $6.4$\\
J$0134-4258$ & PMN~J$0134-4258$           & $01:34:16.90$ & $-42:58:27.0$ & $0.237$  &  $1.69$  & $930$     & $55.0(^{*})$ \\
J$0324+3410$ & 1H~$0323+342$              & $03:24:41.16$ & $+34:10:45.8$ & $0.061$  &  $12.0$  & $1600$    & $614.3(^{**})$ \\
J$0706+3901$ & FBQS~J$0706+3901$          & $07:06:25.15$ & $+39:01:51.6$ & $0.086$  &  $8.27$  & $664$     & $5.6$\\
J$0713+3820$ & FBQS~J$0713+3820$          & $07:13:40.29$ & $+38:20:40.1$ & $0.123$  &  $6.00$  & $1487$    & $10.4$\\
J$0744+5149$ & NVSS~J$074402+514917$      & $07:44:02.24$ & $+51:49:17.5$ & $0.460$  &  $4.83$  & $1989$    & $11.9$\\
J$0804+3853$ & SDSS~J$080409.23+385348.8$ & $08:04:09.24$ & $+38:53:48.7$ & $0.211$  &  $5.26$  & $1356$    & $2.9$\\
J$0814+5609$ & SDSS~J$081432.11+560956.6$ & $08:14:32.13$ & $+56:09:56.6$ & $0.509$  &  $4.44$  & $2164$    & $69.2$\\
J$0849+5108$ & SDSS~J$084957.97+510829.0$ & $08:49:57.99$ & $+51:08:28.8$ & $0.584$  &  $2.97$ & $1811$     & $344.1$\\
J$0902+0443$ & SDSS~J$090227.16+044309.5$ & $09:02:27.15$ & $+04:43:09.4$ & $0.532$  &  $3.10$ & $2089$     & $156.6$\\
J$0937+3615$ & SDSS~J$093703.02+361537.1$ & $09:37:03.01$ & $+36:15:37.3$ & $0.179$  &  $1.22$  & $1048$    & $3.6$\\
J$0945+1915$ & SDSS~J$094529.23+191548.8$ & $09:45:29.21$ & $+19:15:48.9$ & $0.284$  &  $2.16$  & $<2000$   & $17.2$\\
J$0948+0022$ & SDSS~J$094857.31+002225.4$ & $09:48:57.29$ & $+00:22:25.6$ & $0.585$  &  $5.55$  & $1432$    & $107.5$\\
J$0953+2836$ & SDSS~J$095317.09+283601.5$ & $09:53:17.11$ & $+28:36:01.6$ & $0.658$  &  $1.25$  & $2162$    & $44.6$\\
J$1031+4234$ & SDSS~J$103123.73+423439.3$ & $10:31:23.73$ & $+42:34:39.4$ & $0.376$  &  $1.01$  & $1642$    & $16.6$\\
J$1037+0036$ & SDSS~J$103727.45+003635.6$ & $10:37:27.45$ & $+00:36:35.8$ & $0.595$  &  $5.07$  & $1357$    & $27.2$\\
J$1038+4227$ & SDSS~J$103859.58+422742.2$ & $10:38:59.59$ & $+42:27:42.0$ & $0.220$  &  $1.50$  & $1979$    & $2.8$\\
J$1047+4725$ & SDSS~J$104732.68+472532.0$ & $10:47:32.65$ & $+47:25:32.2$ & $0.798$  &  $1.31$  & $2153$    & $734.0$\\
J$1048+2222$ & SDSS~J$104816.58+222239.0$ & $10:48:16.56$ & $+22:22:40.1$ & $0.330$  &  $1.51$  & $1301$    & $1.2$\\
J$1102+2239$ & SDSS~J$110223.38+223920.7$ & $11:02:23.36$ & $+22:39:20.7$ & $0.453$  &  $1.22$  & $1972$    & $2.0$\\
J$1110+3653$ & SDSS~J$111005.03+365336.3$ & $11:10:05.03$ & $+36:53:36.1$ & $0.630$  &  $1.85$  & $1300$    & $18.6$\\
J$1138+3653$ & SDSS~J$113824.54+365327.1$ & $11:38:24.54$ & $+36:53:27.0$ & $0.356$  &  $1.82$  & $1364$    & $12.5$\\
J$1146+3236$ & SDSS~J$114654.28+323652.3$ & $11:46:54.30$ & $+32:36:52.2$ & $0.465$  &  $1.42$  & $2081$    & $14.7$\\
J$1159+2838$ & SDSS~J$115917.32+283814.5$ & $11:59:17.31$ & $+28:38:14.8$ & $0.210$  &  $1.70$  & $1415$    & $2.2$\\
J$1227+3214$ & SDSS~J$122749.14+321458.9$ & $12:27:49.15$ & $+32:14:59.0$ & $0.137$  &  $1.37$  & $951$     & $6.5$\\
J$1238+3942$ & SDSS~J$123852.12+394227.8$ & $12:38:52.15$ & $+39:42:27.6$ & $0.623$  &  $1.42$  & $910$     & $10.4$\\
J$1246+0238$ & SDSS~J$124634.65+023809.0$ & $12:46:34.68$ & $+02:38:09.0$ & $0.363$  &  $2.02$  & $1425$    & $37.0$\\
J$1333+4141$ & SDSS~J$133345.47+414127.7$ & $13:33:45.47$ & $+41:41:28.2$ & $0.225$  &  $0.74$  & $1940$    & $2.5$\\
J$1346+3121$ & SDSS~J$134634.97+312133.7$ & $13:46:35.07$ & $+31:21:33.9$ & $0.246$  &  $1.22$  & $1600$    & $1.2$\\
J$1348+2622$ & SDSS~J$134834.28+262205.9$ & $13:48:34.25$ & $+26:22:05.9$ & $0.918$  &  $1.17$  & $1840$    & $1.6$\\
J$1358+2658$ & SDSS~J$135845.38+265808.5$ & $13:58:45.40$ & $+26:58:08.3$ & $0.331$  &  $1.56$  & $1863$    & $1.8$\\
J$1421+2824$ & SDSS~J$142114.05+282452.8$ & $14:21:14.07$ & $+28:24:52.2$ & $0.538$  &  $1.28$  & $1838$    & $46.8$\\
J$1505+0326$ & SDSS~J$150506.47+032630.8$ & $15:05:06.47$ & $+03:26:30.8$ & $0.409$  &  $4.01$  & $1082$    & $365.4$\\
J$1548+3511$ & SDSS~J$154817.92+351128.0$ & $15:48:17.92$ & $+35:11:28.4$ & $0.479$  &  $2.37$  & $2035$    & $140.9$\\
J$1612+4219$ & SDSS~J$161259.83+421940.3$ & $16:12:59.83$ & $+42:19:40.0$ & $0.234$  &  $1.29$  & $819$     & $3.4$\\
J$1629+4007$ & SDSS~J$162901.30+400759.9$ & $16:29:01.31$ & $+40:07:59.6$ & $0.272$  &  $1.06$  & $1458$    & $12.0$\\
J$1633+4718$ & SDSS~J$163323.58+471858.9$ & $16:33:23.58$ & $+47:18:59.0$ & $0.116$  &  $1.77$  & $909$     & $62.6$\\
J$1634+4809$ & SDSS~J$163401.94+480940.2$ & $16:34:01.94$ & $+48:09:40.1$ & $0.495$  &  $1.66$  & $1609$    & $7.5$\\
J$1644+2619$ & SDSS~J$164442.53+261913.2$ & $16:44:42.54$ & $+26:19:13.2$ & $0.145$  &  $5.12$  & $1507$    & $87.5$\\
J$1709+2348$ & SDSS~J$170907.80+234837.6$ & $17:09:07.82$ & $+23:48:38.2$ & $0.254$  &  $4.12$  & $1827$    & $1.6$\\
J$2007-4434$ & PKS~$2004-447$             & $20:07:55.18$ & $-44:34:44.3$ & $0.240$  & $2.93$   & $1447$    & $791.0(^{***})$\\
J$2021-2235$ & IRAS~$20181-2244$          & $20:21:04.38$ & $-22:35:18.3$ & $0.185$  & $5.54$   & $460$     & $24.9(^{**})$\\
\hline
\end{tabular}
\begin{list}{}{}
\item[$^{\mathrm{*}}$] 4.85~GHz, Grupe et al. (2000).
\item[$^{\mathrm{**}}$] VLA/NVSS, Condon et al. (1998).
\item[$^{\mathrm{***}}$] 1.4~GHz, ATCA, Gallo et al. (2006).
\end{list}
\label{tab:sample}
\end{table*}

\section{Data analysis and software}
We retrieved all the publicly available observations done by {\it Swift} (Gehrels et al. 2004) and {\it XMM-Newton} (Jansen et al. 2001) on 2013 December 9. Data analysis was performed by following standard procedures as described in the documentation for each instrument. 

In the case of {\it Swift} we used {\tt HEASoft v.6.15} with the calibration data base updated on 2013 Dec 13. We analysed data of the X-Ray Telescope (XRT, Burrows et al. 2005) and the Ultraviolet Optical Telescope (UVOT, Roming et al. 2005). XRT spectral counts were rebinned to have at least 20--30 counts per bin in order to apply the $\chi^2$ test. When this was not possible, we applied the unbinned likelihood (Cash 1979). We adopted power-law and broken power-law models. The need for the latter was evaluated by using the $f-$test (cf.\ Protassov et al. 2002) with a threshold $>99$\%. The observed magnitudes (Vega System) of UVOT were dereddened according to Cardelli et al. (1989) and converted into physical units by using zero points from {\it Swift} calibration data base.  All the sources are point-like, and therefore we consider the emission from the host galaxy to be negligible; only J0324+3410 in the $V$ filter displayed some hint of host galaxy, which was properly subtracted. We did not analysed the Burst Alert Telescope (BAT, Barthelmy et al. 2005) data because the average fluxes of RLNLS1s in hard X-rays are well below the instrument sensitivity. Indeed, by looking at the two available catalogues built on BAT data, we found only one detection of J0324+3410 in both the 70-month survey of the {\it Swift}/BAT team (Baumgartner et al. 2013) and the Palermo 54-month catalogue (Cusumano et al. 2010). J0324+3410 was first detected by Foschini et al. (2009) by integrating all the available direct observations performed during the period 2006--2008 (total exposure $\sim 53$~ks). There is also another detection of J0948+0022 in the Palermo catalogue, but not confirmed by Baumgartner et al. (2013). We did not include this information in the present work. {\it Swift} results are summarised in Tables 5 and 6 of the Online Materials.

In the case of {\it XMM--Newton}, we analysed data of the European Photon Imaging Camera (EPIC) pn (Str\"uder et al. 2001) and MOS (Turner et al. 2001) detectors. We adopted the {\tt Science Analysis Software v.13.5.0} with the calibration data base updated on 2013 December 19. We excluded time periods with high-background by following the prescriptions of Guainazzi et al. (2013). The spectral modelling was done as for {\it Swift}/XRT. {\it XMM-Newton} results are summarised in Tables 5 of the Online Materials.

\begin{figure*}[t!]
\begin{center}
\includegraphics[scale=0.40]{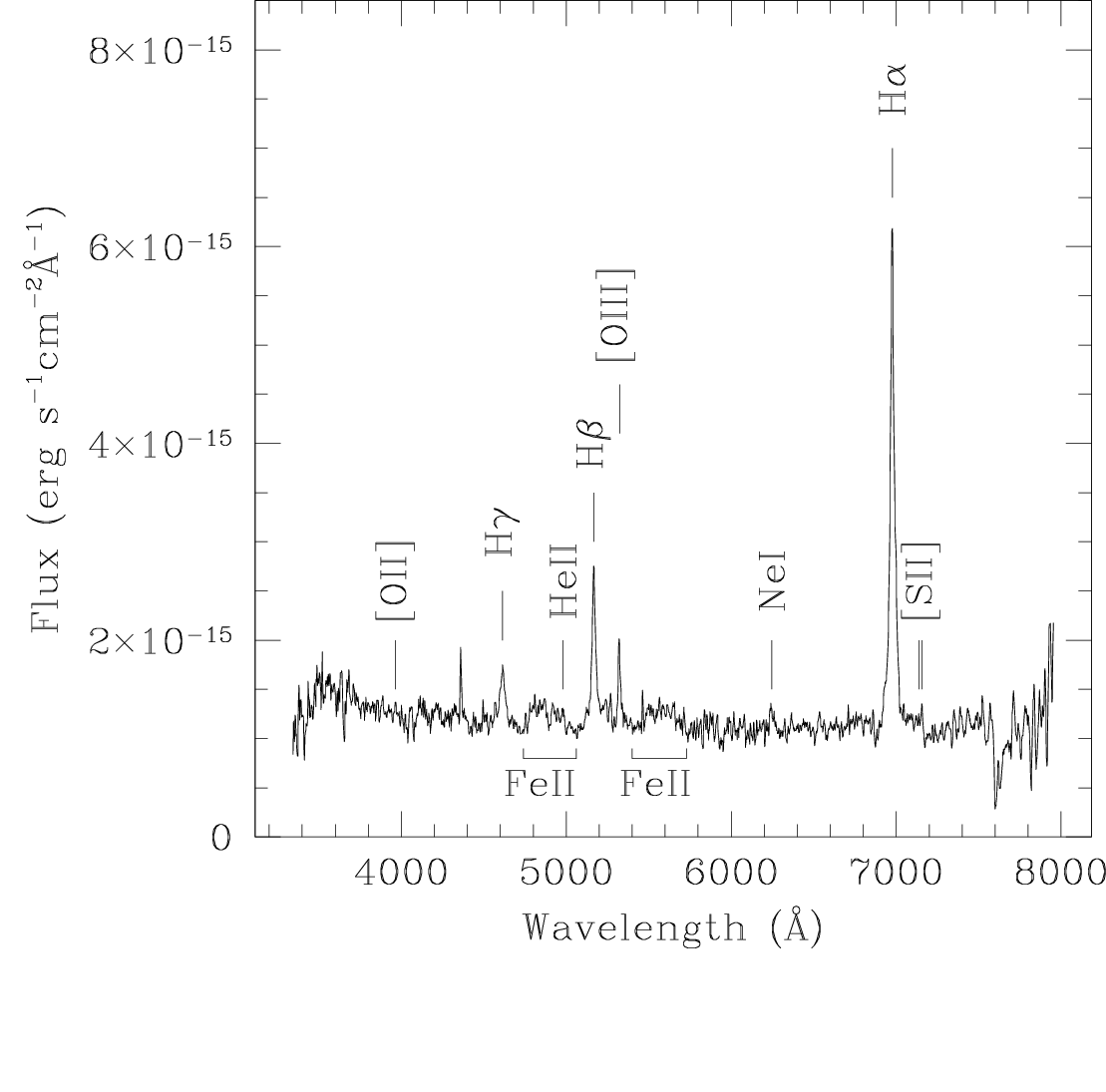}
\includegraphics[scale=0.40]{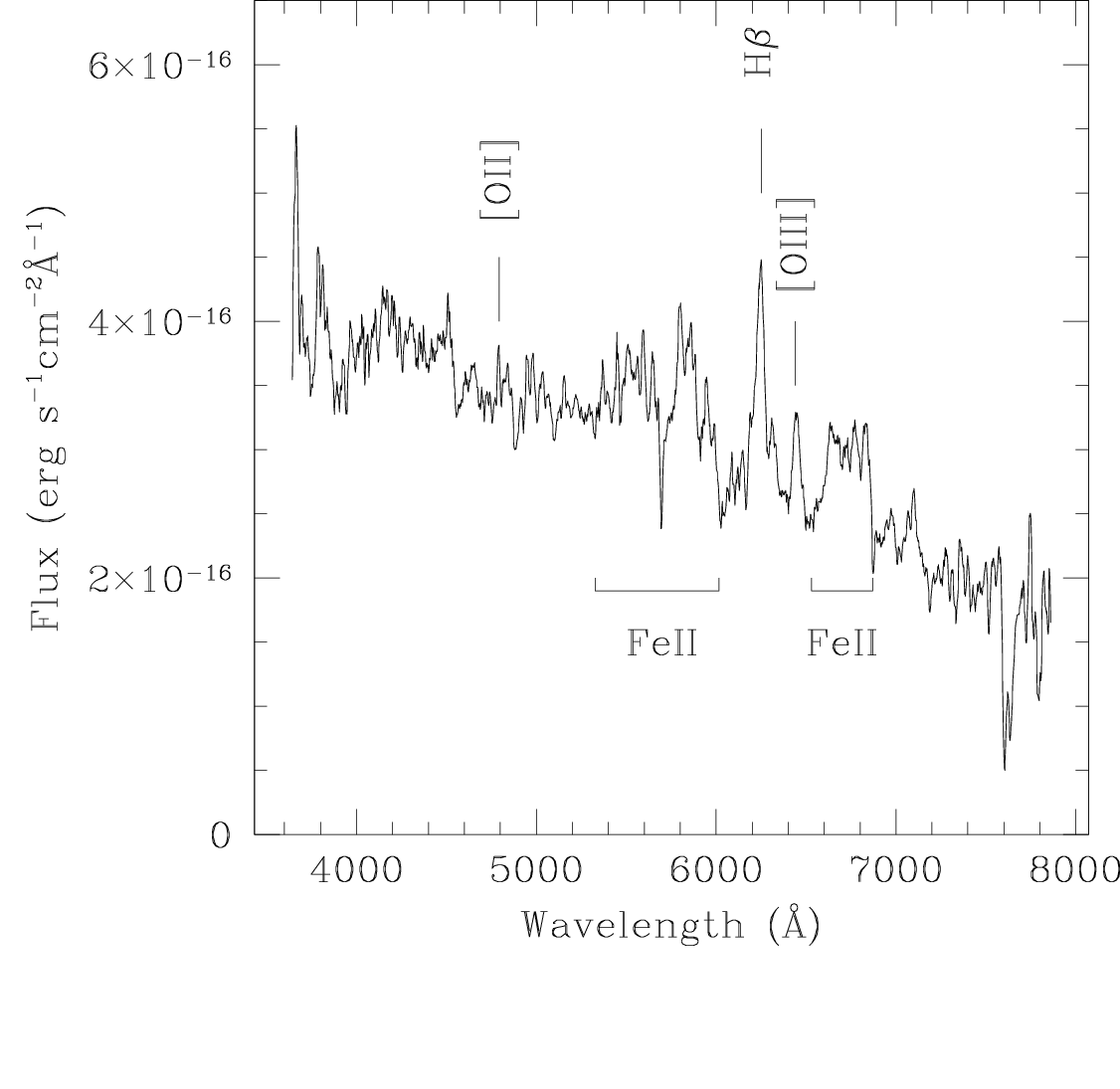}
\caption{Optical spectra of J0324+3410 ({\it left panel}) and J0945+1915
  ({\it right panel}) taken from the Asiago Astrophysical Observatory
  1.22~m telescope.}
\label{fig:asiago}
\end{center}
\end{figure*}

\subsection{Optical Data} 
Optical spectra were retrieved for 32/42 sources from SDSS DR9 
database (Ahn et al. 2012), downloaded from NED (3/42), or extracted
from figures published in the literature (2/43). Two sources,
J0324+3410 and J0945+1915, were observed with the $1.22$~m telescope
of the Asiago Astrophysical Observatory between 2013 December and 2014
January, using the Boller \& Chivens spectrograph with a 300 mm$^{-1}$
grating. The instrumental resolution was $R \approx 700$, and the
spectra covered the wavelength range between 3200 and 8000~\AA\ with a
dispersion of 2.3~\AA\,pixel$^{-1}$. The slit was oriented at $PA =
90^\circ$, with an aperture of 4.25 arcsec, corresponding to 4.7~kpc
for J0324+3410 and to 17.6~kpc for J0945+1915. The exposure time was
$3\times1200$~s for the former and $9\times1200$~s for the
latter. Data reduction was performed using the standard {\tt IRAF
v.2.14.1} tasks: the overscan was subtracted instead of the bias in
the pre-reduction steps and NeHgAr lamps were used for the wavelength
calibration. Finally the extracted spectra were combined together (see
Fig.~\ref{fig:asiago}).

We were unable to find any optical spectral data for three of the
sources in our sample.

The optical spectra were corrected for redshift and Galactic
absorption and a continuum fit was subtracted. The contribution of the
host galaxy in objects at $z > 0.1$ is typically less than 10\%
(Letawe et al. 2007). Given that the flux calibration uncertainty is
typically around 20\%, we assume that the host galaxy contribution is
negligible. Indeed the spectra, as expected, do not show any sign of
stellar absorption features, and the continuum appears to be dominated
by the AGN. For objects at $z < 0.1$ (J0324+3410 and J0706+3901), we
subtracted a template of a spiral galaxy bulge (Kinney et al. 1996) as
a test, even if no stellar features were visible. Since we did not
observe any significant change in the shape of H$\beta$, we proceeded
without any host-galaxy subtraction. We focused on the H$\beta$ region
between 4000 and 5500~\AA. To subtract Fe\,{\sc ii} multiplets, we
used a template properly created by using the online
software\footnote{\url{http://servo.aob.rs/~jelena/}} developed by
Kova\v{c}evi\'{c} et al. (2010) and Shapovalova et al. (2012).

After Fe\,{\sc ii} subtraction, we decompose the H$\beta$ line into
narrow and broad components, using the \texttt{ngaussfit} task of {\tt
IRAF}. We used three Gaussians to fit the profile, one to reproduce
the narrow component, and two others for the broad
component. Following Veron et al. (2001), we fixed the flux of the
narrow component to be 1/10 of the [O\,{\sc iii}]\,$\lambda 5007$ line
with the same velocity width.  However, given that the gas which
produces the [O\,{\sc iii}] line is often turbulent, its width can
lead to an overestimate of the H$\beta$ narrow component. For this
reason, when [O\,{\sc ii}]\,$\lambda 3727$ was clearly visible and
much narrower than [O\,{\sc iii}] lines, we used its FWHM to fix the
the width of H$\beta$ (Greene \& Ho 2005, Ho et al. 2009). In some
case, the low $S/N$ ratio required a fit with just two Gaussians, one
narrow and one broad. When necessary we also set the height of the
narrow component as a free parameter. The line centre was always left
as a free parameter.

In the case of J1348+2622, we used the Mg\,{\sc ii}\,$\lambda2798$ for
the black hole mass estimate as the H$\beta$ line it falls outside of
the spectral range. As shown by Shen et al. (2008), mass estimates
from these two lines are generally consistent.

Finally, we subtracted the narrow component and measuring both FWHM
and the line dispersion $\sigma$ only for the broad component. The
results are presented in Table~\ref{tab:masses}.

\subsection{Radio Data} Some of these sources were observed for other
programs. 37~GHz data are from the 13.7~m telescope at Mets\"ahovi
(Finland), and multiwavelength observations were done at 100-m single
dish telescope at Effelsberg (Germany, $2.64-42$~GHz) and 30-m
telescope at Pico Veleta (Spain, $86-142$~GHz). More details about the
Mets\"ahovi, and Effelsberg/Pico Veleta observations on RLNLS1s will
be published by L\"ahteenm\"aki et al. (in preparation) and Angelakis
et al. (in preparation), respectively. Some of the data have already
been published by Abdo et al. (2009a,b), Foschini et al. (2011a,
2012), Fuhrmann et al. (2011), and Angelakis et al. (2012a,b).

We also searched for publicly available observations in the VLBI
calibrated data archives. 15~GHz data are from the MOJAVE
database (Lister et al. 2009, 2013)\footnote{\url{http://www.physics.purdue.edu/MOJAVE/}}. 
VLBI results at frequencies below 15~GHz come from the VLBA and global VLBI astrometric and geodetic experiments (Beasley et al. 2002,
Fomalont et al. 2003, Petrov et al. 2005, 2006, 2008, Kovalev et al. 2007, Piner et al. 2012, Pushkarev \& Kovalev 2012). 
Calibrated visibility and image fits files are provided by
the authors in the public database\footnote{\url{http://astrogeo.org}}. 

We performed a standard CLEANing (H\"ogbom 1974) and followed the model-fitting
of the calibrated VLBI visibility data in {\tt Difmap} (Shepherd 1997). We preferred to use
circular Gaussian components unless the use of elliptical components
gave a better fit to the data. To ensure the quality of the fit, we
compared Gaussian model parameters with the results of CLEAN. The
total flux density and residual RMS appeared to be almost identical for the
two cases. All of these sources have simple radio structure, so they
are well-modelled by Gaussian components. The results are presented in
the Online Material (Table~7).

\subsection{Online catalogues and Literature} 
We supplemented these data with information from online catalogues and literature. For $\gamma$ rays, we mainly referred to Foschini (2011a), who reported the detection of 7 RLNLS1s with {\it Fermi}/LAT after 30 months of operations. When available, we reported more recent published analyses (Foschini et al. 2012, D'Ammando et al. 2013a,d, Paliya et al. 2014). No new detections have been claimed to date after Foschini (2011a). Therefore, for the non-detected sources in the present sample, we indicated the upper limit of $\sim 10^{-9}$~ph~cm$^{-2}$~s$^{-1}$ as from the {\it Fermi}/LAT performance web page\footnote{\url{http://www.slac.stanford.edu/exp/glast/groups/canda/lat_Performance.htm}}, which is the minimum detectable ($TS=25$) flux above $100$~MeV over a period of 4 years for a source with a power-law shaped spectrum with a spectral index $\alpha=1$. A summary of $\gamma$ ray characteristics found in literature is shown in the Online Material (Table~4).

For X-rays, we searched for missing sources in the {\it Chandra} X-Assist (CXA, Ptak \& Griffiths 2003)
catalogue v.4 and {\it XMM--Newton} Slew Survey Clean Sample v.1.5 (XSS,
Saxton et al. 2008). The two catalogues provide X-ray detections in
different energy bands: 0.5--8~keV for the former, and 0.2--12~keV for
the latter. The fluxes were then converted into the 0.3--10~keV band
by using {\tt WebPIMMS}\footnote{\url{http://heasarc.gsfc.nasa.gov/Tools/w3pimms.html}}
and a fixed photon index value $\Gamma =2$ ($\alpha=1$).  Some sources were not
observed by any of the above-cited satellites. In those cases, we
calculated an upper limit by using the detection limit of the {\it
ROSAT} All-Sky Survey (RASS, Voges et al. 1999, 2000).

At infrared/optical/ultraviolet wavelengths, in addition to the {\it
Swift}/UVOT data presented here, we used SDSS-III data release 9 (Ahan
et al. 2012) and 2MASS (Cutri et al. 2003). Only one source,
J2021$-$2235, remained without optical coverage from either {\it
Swift}/UVOT or SDSS, but we found $B$ and $R$ magnitudes in the US
Naval Observatory B1 catalogue (Monet et al. 2003).

We also searched the {\it WISE} all-sky catalogue (Wright et al. 2010)
for photometric data at mid-IR wavelengths (between 3.4 and
22~$\mu$m). In particular, we have used the last version of the catalogue, the {\it AllWISE} data release (November 2013). 
All the RLNLS1s of the sample are detected ($S/N>3$) in
the {\it WISE} survey at 3.4 and 4.6~$\mu$m (W1 and W2 bands,
respectively) while 41 and 37 objects are detected also at 12~$\mu$m
(W3 band) and 22~$\mu$m (W4 band) respectively.  The observed
magnitudes have been converted into monochromatic flux densities
assuming a power-law spectrum with $\alpha=2$. For the sources not
detected or detected with a $S/N<3$, we have calculated the $3\sigma$
upper limit on the flux density.

At radio frequencies, in addition to the above cited programs (see
Sect.~3.2), we have taken all available data from the
NED\footnote{\url{http://ned.ipac.caltech.edu/}} and
HEASARC\footnote{\url{http://heasarc.gsfc.nasa.gov/}} archives.

\section{Observational characteristics}
\subsection{Gamma rays} 
We found in the available literature 7/42 detections at high-energy $\gamma$ rays (17\%) sources. 
Specifically, they are:

\begin{itemize}
\item J0948+0022, the first RL-NLS1 to be detected in $\gamma$ rays
(Abdo et al. 2009a,b, Foschini et al. 2010).
\item J0324+3410, J1505+0326, and J2007$-$4434, which were detected
after the first year of {\it Fermi} operations (Abdo et al. 2009c).
\item J0849+5108, which was detected because of an outburst in 2010
(Foschini 2011a, D'Ammando et al. 2012).
\item J1102+2239, J1246+0238 (Foschini 2011a).
\end{itemize}

The spectral indices are generally steep, with a weighted average of $\alpha_{\gamma}=1.6\pm 0.3$
(median 1.7), but there is one interesting case with harder spectrum: J0849+5108 with
$\alpha_{\gamma}=1.0-1.18$ (Online Material Table~4 and 8). The average values for blazars as measured by {\it Fermi}/LAT (Ackermann et al. 2011) are $1.4\pm 0.2$ for FSRQs, $\alpha = 1.2 \pm 0.1$ for low-synchrotron peak BL Lacs, 
$\alpha = 1.1 \pm 0.1$ for intermediate-synchrotron peak BL Lacs, and 
$\alpha = 0.9\pm 0.2$ for high-synchrotron peak BL Lacs. Therefore, we conclude that the spectral
characteristics of RLNLS1s are generally similar to those of FSRQs.

Short timescale variability for factor-of-two flux changes is also reported by some authors. Specifically, Foschini (2011a) reported intraday variability for J0948+0022 and J1505+0326, while Palyia et al. (2014) found 3-hour variability of J0324+3410 during its outburst of 2013 August 28 to 2013 September 1 (see Online Material Table~9). 

\subsection{X-rays} 
\begin{figure}[t!]
\begin{center}
\includegraphics[angle=270,scale=0.35]{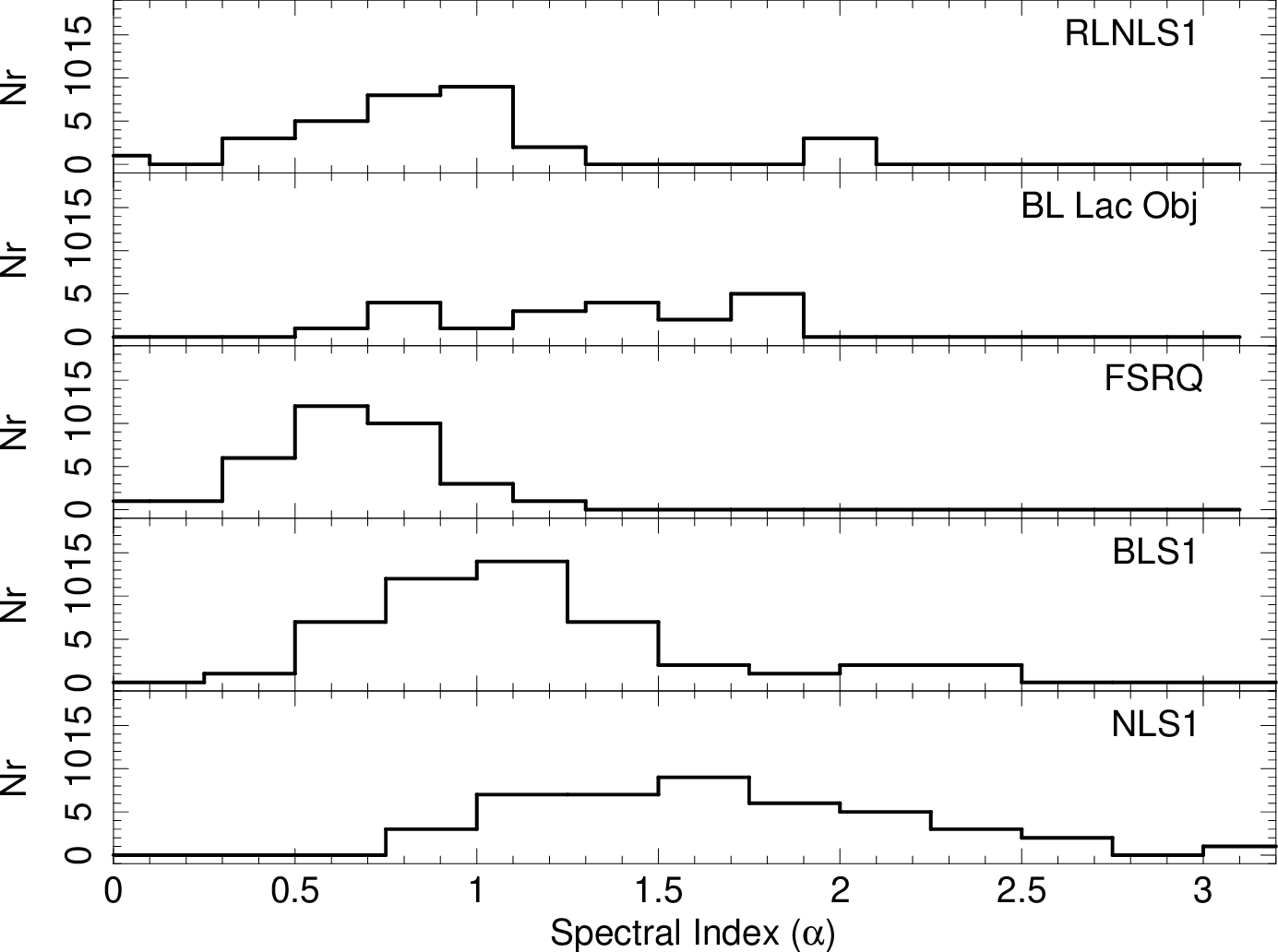}
\caption{X-ray (0.3--10~keV) spectral index distributions for the
present sample of RLNLS1s; BL Lac objects and FSRQs are from
Ghisellini et al. (2009, 2010) and Tavecchio et al. (2010); BLS1s and
RQNLS1s are from Grupe et al. (2010).}
\label{fig:xraydistrib}
\end{center}
\end{figure}

About 90\% of the sources in the present sample
(38/42) are detected in X-rays (see Table~5). The average spectral
index in the 0.3--10~keV energy range is $\alpha_{\rm X}=1.0\pm0.5$,
with a median value of $0.8$ (see Online Material Table~8), as compared with the
values of $0.58$ (FSRQs), $1.3$ (BL Lac objects), $1.1$ (BLS1s), and
$1.7$ (RQNLS1s). These values were calculated from the samples of
$\gamma-$ray blazars from Ghisellini et al. (2009, 2010) and Tavecchio
et al. (2010) and radio-quiet Seyferts from Grupe et al. (2010). The
corresponding distributions are displayed in
Fig.~\ref{fig:xraydistrib}. The average spectral indices for the
individual sources (see Online Material Table~8) are $\alpha<1$ in 23/42 cases and
$\alpha \geq 1$ in 12/42 cases. In 7/42 cases, the spectral index is
near the boundary.

We note that the X-ray spectral indices of RLNLS1s are similar to those of BLS1s, and usually harder than those of RQNLS1s. However, when compared to blazars, RLNLS1s are between the average values of FSRQs and BL Lac objects. From inspection of the SEDs (see Sect.~7), it seems that the X-ray emission of RLNLS1s could be due either to inverse-Compton (IC) radiation from a relativistic jet or from the corona of the accretion disk. This could explain why the average spectral index is softer than that of FSRQs, where the X-ray emission is dominated by the IC from the jet (see, for example, Ghisellini et al. 2010). 

In the case of four sources, there were multiple observations with sufficient exposure for individual detections (Online Material Table~5). We therefore searched for any correlation between flux and spectral slope. No significant trend was found. It is interesting to compare with radio-quiet Seyferts, where a correlation between 2--10~keV flux and the spectral index was found, indicating a steepening of the spectral shape as the flux increases (Markowitz et al. 2003). Some interesting episodes were described in the case of J0324+3410 by Foschini et al. (2009),  Foschini (2013), and Tibolla et al. (2013): the source has generally a soft spectral index, typical of NLS1s, but sometimes --- as the jet became active --- the X-ray spectrum displays a break at a 2--3~keV and a hard tail appears (see also Paliya et al. 2014). Similar behaviour has been observed in another RLNLS1, PKS~0558$-$504\footnote{This source is not in the present sample because it has a steep radio spectral index. It is included in the sample studied by Berton et al. (in preparation).}, where {\it ASCA} observed a hardening of the spectral index during an outburst, changing from $\sim 1.2$ to $\sim 0.9$ (Wang et al. 2001).  

In the case of J0324+3410, with more data (Online Material Table 5), there is no evident trend to link the change in flux with a change of the spectral slope. Although the epochs with a hard tail are concentrated in a high-flux region, there are also observations with similar fluxes that can be fit satisfactorily by a single power-law model. It is worth noting that there might be an instrumental bias: {\it Swift}/XRT has a small effective area at energies $\geq 7$~keV (Romano et al. 2005) and, therefore, the detection of the hard tail could depend on the exposure time at similar flux levels. Indeed, the exposures in the present data set ranged from 1.3 to 8.8~ks (see Online Material Table~5) and the spectral shape at shorter exposures --- having low statistics at high energies --- could be fit with just a power-law model with an index harder than usual. An observing campaign with a satellite like {\it XMM--Newton} for example, carrying X-ray instruments with a large effective area above 7~keV, could effectively monitor the spectral changes (see below the example of J0948+0022). 

Many sources of the present sample were included in previous surveys by Komossa et al. (2006a) and Yuan et al. (2008). In these studies, the X-ray characteristics were measured from {\it ROSAT} data. Komossa et al. (2006a) found spectral indices in the range 0.9--3.3, while Yuan et al. (2008) measured values between $0.37$ and $2.36$. Particularly, the spectral index of J0948+0022 was measured as $1.6\pm1.8$ by Zhou et al. (2003), $\sim 1.2$ by Komossa et al. (2006a), and $1.26\pm0.64$ by Yuan et al. (2008). The source is also present in the Williams et al. (2002) sample, who reported $\alpha_{\rm X}=1.8\pm0.5$, again on the basis of {\it ROSAT} observations. In our case, both {\it Swift} and {\it XMM--Newton} indicate a harder spectral index (average $\alpha_{\rm X}\sim 0.56$) that remains unchanged with flux variations (Online Material Table 5). This is in agreement with previous studies and MW campaigns (Abdo et al. 2009a,b, Foschini et al. 2011a, 2012). A study 
based on a long one-orbit {\it XMM--Newton} observation revealed the presence of a soft X-ray excess (D'Ammando et al. 2014, Bhattacharyya et al. 2014), which is confirmed in the present study. The break energy is between $1.72_{-0.11}^{+0.09}$~keV (D'Ammando et al. 2014) and $\sim 1.2$~keV (Bhattacharyya et al. 2014; see also our analysis in the Online Material Table~5). The low-energy spectral index is between $1.1$ and $1.3$, while the high-energy power-law has a slope 0.5--0.6. There is also a {\it Swift} observation the same day of that of {\it XMM--Newton} (2011 May 28) and we tried also to fit these data with a broken power-law model. We found $\alpha_1=\Gamma_1-1=1.8_{-0.8}^{+1.4}$, $\alpha_2=\Gamma_2-1=0.5\pm0.2$, and $E_{\rm break}=0.9\pm0.3$~keV ($\chi^2=5.4$ for 7 dof, not reported in Online Material Table~5). However, according to our threshold defined in Sect.~3, the broken power-law model is not statistically preferred over the single power-law model (95\% vs.\ a threshold of 99\%). Therefore, we conclude that the presence or absence of a soft X-ray excess is related more to an instrumental bias rather than to an effective change of the AGN. {\it ROSAT}, having a bandpass of 0.1--2.4~keV, is biased toward soft X-ray sources and therefore only captures the soft excess. {\it Swift}, with a wider energy band (0.3--10~keV) and snapshot observations, measured an average of both the soft excess and the hard tail. {\it XMM--Newton}, still operating in the 0.3--10~keV band, detected both the soft excess and the hard tail because of the longer exposure and larger effective area. Both D'Ammando et al. (2014) and Bhattacharyya et al. (2014) concluded that the excess at low energies could be due to the accretion disk/corona system, as is the usual case for many RQNLS1s (e.g.  Leighly 1999, Foschini et al. 2004, Grupe et al. 2010). 

We note that also the blazar 3C~273 displays such a soft excess and there is a correlation between the low activity of the relativistic jet and the emergence of the thermal component in the soft X-rays, which was interpreted as a signature of the jet-disk connection (Grandi \& Palumbo 2004, Foschini et al. 2006). In the present case, the instrumental bias prevents any conclusion about the X-ray component, but the optical component offers some hints that support the above hypothesis (see Sect.~7). 

The case of J2007$-$4434 is different. The X-ray spectrum as observed by {\it XMM--Newton} on 2004 April 11 shows a soft excess and a hard tail. Gallo et al. (2006) favoured the hypothesis of an accretion disk corona to generate the excess low-energy flux, while Foschini et al. (2009), on the basis of different variability characteristics (16\% and $<8$\% in the 0.2--1~keV and 2--10~keV energy bands, respectively), suggested a similarity with low-energy peaked BL Lac objects (i.e.  the low-energy component is the tail of the synchrotron emission). This seems to be confirmed by the analysis and modelling of the SED (Abdo et al. 2009c, Paliya et al. 2013b; see also the Sect.~7). In the present work, we find two more {\it XMM--Newton} archival observations performed in 2012 (May 1 and October 18): in both cases, there was no low-energy excess and the spectra were fit with a single power-law model with spectral index $\alpha \approx 0.7$. It is worth noting that the flux was about one third that of the 2004 observation, when the soft excess was detected.  

In all the other cases, {\it ROSAT} observations reported by Komossa et al. (2006a) and Yuan et al. (2008)
are generally confirmed. 

The search for variability on short timescales resulted in many significant detections of intraday variability, with flux changes greater than $3\sigma$ (Online Material Table~9). There are hour timescales for J0134$-$4258, J0324+3410, J0948+0022, J1629+4007, and J2007$-$4434. It is worth noting that the measurements reported in Table~9 (Online Material) were only made from {\it Swift}/XRT data. We also analysed {\it XMM--Newton} data and find variability on minute timescales (down to $\sim 2$ minutes with flux change at the $11\sigma$ level in the case of J0948+0022). However, we note that all {\it XMM--Newton} observations are affected by soft-proton flares: although we corrected for both the anomalous particle and photon backgrounds, we noted that minute-scale variability is detected near periods of the light curve that are excised because of high particle background. In addition, there is no confirmation of such short timescale variations in the {\it Swift} data, but it is worth noting that Itoh et al. (2013) found similar values from optical observations. These findings therefore require a much more careful dedicated analysis and confirmation with other instruments less affected by grazing-incidence particle background (i.e.  from X-ray satellites in low-Earth orbit).

The hour timescales found are much shorter than those expected in case of changing obscuration, which could be $\sim$10~hours in the most extreme case of NGC~1365 (see the review by Bianchi et al. 2012). In addition, fits of X-ray spectra do not require iron lines or obscuration in addition to the Galactic column, as expected from radio-loud AGNs, in contrast to radio-quiet AGNs (e.g.  Reeves et al. 1997). The exception seems to be  J0324+341, where Abdo et al. (2009c) reported an unresolved iron line at $E_{\rm Fe}=6.5\pm0.3$~keV with equivalent width of $147$~eV (see also Paliya et al. 2014). By integrating all the available {\it Swift} snapshots (with a total exposure time of $2.1\times 10^5$~s), we basically confirm the previous measurements: $E_{\rm Fe}=6.5\pm0.1$~keV, equivalent width $\sim 91$~eV, and $\Delta \chi^2 = 13.1$ for two additional parameters ($E_{\rm Fe}$ and normalisation; we fixed $\sigma_{\rm Fe}$ to 0.1~keV). {\it XMM--Newton} observations of J0948+0022 and J1348+2622 only show a hint of an  excess above 5~keV.

\subsection{Ultraviolet, optical, and infrared frequencies}
Spectral indices for ultraviolet and infrared frequencies were calculated by means of the two-point spectral index formula

\begin{equation}
\alpha_{\rm 12} = - \frac{\log (S_1/S_2)}{\log (\nu_1/\nu_2)},
\end{equation}

where $S_1$ and $S_2$ are the observed flux densities at frequencies $\nu_1$ and $\nu_2$, respectively. In the ultraviolet, we 
use the  {\it Swift}/UVOT observations, where $\nu_1=1.16\times 10^{15}$~Hz and $\nu_2=1.47\times 10^{15}$~Hz refer to the 
{\it uvw1} ($\lambda=2591$~\AA) and {\it uvw2} ($\lambda=2033$~\AA) bands, respectively. For infrared frequencies, we adopt the extreme filters of {\it WISE}: $\nu_1=1.36\times 10^{13}$~Hz and $\nu_2=8.82\times 10^{13}$~Hz, corresponding to $W4$ and $W1$ filters, respectively. When one of the two filters only has an upper limit, we referred to the closest other filter with a detection, either $W2$ ($\nu=2.50\times 10^{13}$~Hz) or $W3$ ($6.5\times 10^{13}$~Hz). Optical spectral indices were measured by fitting the spectra over the range $\sim 3000$--$8000$~\AA.

The average ultraviolet spectral index is $\alpha_{\rm uv}=0.7\pm1.4$ (median 0.7), and the values for the individual sources were measured from the integrated flux densities (Online Material Table~6). This spectral index can be compared with the values of $0.79$ (median 0.61) for BLS1, and $0.85$ (median 0.65) for NLS1 in the Grupe et al. (2010) sample, also based on {\it Swift}/UVOT observations. Ganguly et al. (2007) observed 14 radio-quiet low redshift ($z<0.8$) quasars with {\it Hubble Space Telescope (HST)} in the range 1570--3180~\AA~and measured an average index of $0.87$. Pian et al. (2005), also with {\it HST} over the range $\sim1570$--4780~\AA, observed 16 blazars and found $\alpha_{\rm uv}\approx 1.16$. In a previous study on a larger sample of 47 radio-loud AGNs observed in the range $1200-3000$~\AA~with {\it International Ultraviolet Explorer (IUE)}, Pian \& Treves (1993) found an average $\alpha_{\rm uv}\approx 1$, with strong emission line quasars having $\alpha_{\rm uv}\approx 1.38$, BL Lac objects with $\alpha_{\rm uv}\approx 0.97$, and radio-weak BL Lacs with $\alpha_{\rm uv}\approx 0.66$. For a  control sample of 37 objects from the Palomar--Green (PG) bright quasar survey, an  average $\alpha_{\rm uv}\approx 0.84$ was found. At the level of average values, the present sample is in agreement with the values for radio-weak blazars, PG quasars, and radio-quiet Seyferts.

The average optical spectral index of the present sample of RLNLS1s is $\alpha_{\rm o}=1.0\pm0.8$ (median $0.8$), in agreement with the previous surveys of RQNLS1 (Constantin \& Shields, 2003) and RLNLS1s (Komossa et al. 2006a, Yuan et al. 2008). A comparison with the optical slopes measured by Whalen et al. (2006) reveals similar slopes (particularly, Fig.~4 in Whalen et al. 2006), with some exceptions likely due to the source variability. For example, J1159+2838 changed from $\alpha_{\rm o}\approx -0.19$ to $0.04$, and J1358+2658 switched from $\alpha_{\rm o}\approx 0.45$ to $0.97$. Vanden Berk et al. (2001) integrated the SDSS spectra of more than 2200 quasars and found an average $\alpha_{\rm o}\approx 0.44$. They also note that by using only the low-redshift sources, the optical spectral index becomes steeper ($\alpha_{\rm o}\approx 0.65$). Our average value ($\alpha_{\rm o}\approx 1.0$) seems to be in agreement with this trend. 

\begin{figure}[t!]
\begin{center}
\includegraphics[angle=270,scale=0.35]{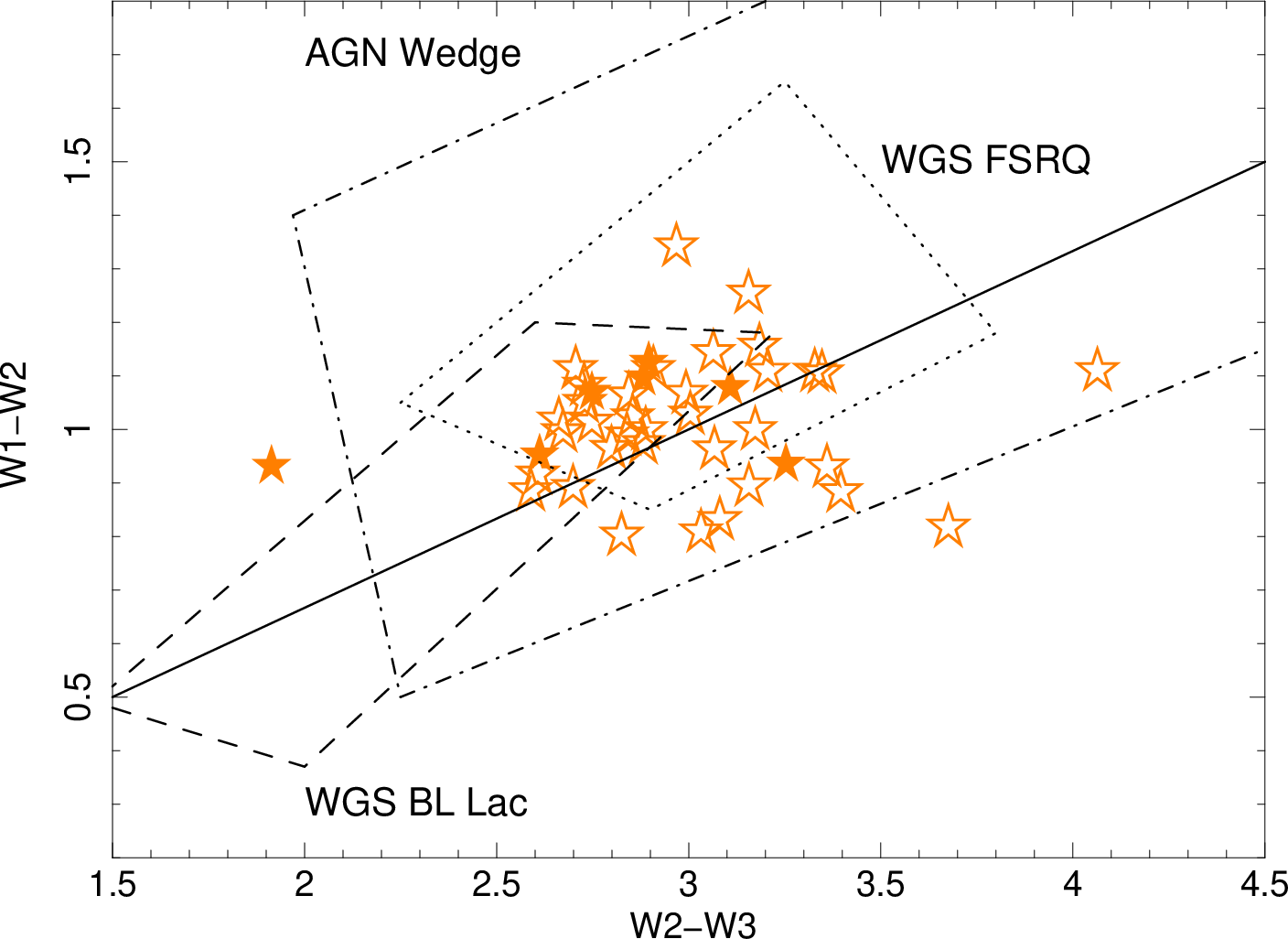}
\caption{{\it WISE} colours of the present sample of RLNLS1s (filled orange stars indicated the 
$\gamma$-ray detected RLNLS1s). Different characteristic regions are also plotted: 
the blazar {\it WISE} Gamma-ray Strip (WGS) for BL Lacs (dashed line) and FSRQs (dotted line), 
as defined by Massaro et al. (2012) and the AGN wedge (dot-dashed line) as defined by 
Mateos et al. (2012, 2013) for 
X-ray selected AGNs. The continuous line corresponded to a power-law emission 
($S_{\nu}\propto \nu^{-\alpha}$) with $\alpha$  ranging from $\sim -0.5$ to $\sim +2.5$.}
\label{fig:wise}
\end{center}
\end{figure}

The average infrared spectral index as measured from {\it WISE} is $\alpha_{\rm IR}=1.3\pm0.3$ (median 1.3), as expected in the case of synchrotron emission from a relativistic jet (see Massaro et al. 2011, D'Abrusco et al. 2012, Raiteri et al. 2014). Fig.~\ref{fig:wise} shows the distribution of {\it WISE} colours of the present sample compared with the blazar strip by Massaro et al. (2011) and the X-ray selected Type 1 and 2 AGNs by Mateos et al. (2012, 2013). While most of the RLNLS1s are in the blazar/AGN region, there are also some cases in the starburst region (cf.\ Fig.~1 in Massaro et al. 2011), suggesting the presence of intense star-formation activity (typical of NLS1s, Sani et al. 2010). Caccianiga et al. (2014) studied a steep-spectrum RLNLS1 (SDSS~J143244.91+301435.3, which is not included in the present sample because of its steep radio spectrum), and found significant star-forming activity. In that case, since the jet is likely to be viewed at a large angle, it does not overwhelm the emission from the nearby environment. The RLNLS1s of the present sample were instead selected by their flat radio spectra, to extract the beamed population, and are hence dominated by synchrotron emission. However, one source (J1505+0326) in the starburst region was detected in $\gamma$ rays, and two $\gamma$-ray RLNLS1s have $W2-W3>3$ (in the addition to the one cited earlier, there is also J0948+0022, still around the synchrotron line). Specifically, J0948+0022 and J1505+0326 were among the most $\gamma-$ray active RLNLS1s of the present sample, therefore this result could be counterintuitive (except for J0948+0022, which is still near the synchrotron line). The explanation is in the epochs of the {\it WISE} observations, performed between 2010 Jan 7 (MJD~$55203$) and Aug 6 (MJD~$55414$). Comparing with the $\gamma$-ray light curves displayed in Foschini et al. (2012), it is seen that J0948+0022 was almost undetected during the {\it WISE} observations. In the case of J1505+0326, D'Ammando et al. (2013a) detected the source by integrating over three-month time bins, but the flux in the first half of 2010 remained at low level of order $10^{-8}$~ph~cm$^{-2}$~s$^{-1}$. Therefore, it is likely that these sources could have strong star-forming activity that is sometimes overwhelmed by synchrotron emission from the jet. We also note that another RLNLS1 of the present sample, J2021$-$2235, is classified as an ultraluminous infrared galaxy (ULIRG) by Hwang et al. (2007), thus supporting the presence of intense star-forming activity. 

A search for the shortest timescale for a factor-of-two change in flux
demonstrates intraday variability for 7/42 sources (see
Table~10). There is some bias in this case, because not
all {\it Swift}/UVOT observations were performed using all six
filters. Two sources were extensively observed with almost all the six
filters (J0324+3410 and J0948+0022) and displayed intraday variability
at all wavelengths. This is in agreement of previous findings of
extreme intraday optical variability reported by Liu et al. (2010),
Itoh et al. (2013), Maune et al. (2013), and Paliya et al. (2013a). In
particular, Itoh et al. (2013) report changes on timescale of minutes
in the optical polarised flux of J0948+0022 on 2012 December 20, with
a peak degree of polarisation of 36\%.

Jiang et al. (2012) examined {\it WISE} data in a search for infrared
variability among the 23 RLNLS1 of the Yuan et al. (2008) sample. They
found three cases, also in the present sample: J0849+5108, J0948+0022,
and J1505+0326. The first two sources displayed intraday variability,
while the latter showed significant flux changes over
$\sim$6~months. The remaining 20~RLNLS1s of the Yuan et al. (2008)
sample did not show variability, most likely because of the weakness
of the sources.

A more detailed analysis of the characteristics of the optical spectra
(line, bumps, blue/red wings) will be presented elsewhere (Berton et
al. in preparation).

\subsection{Radio} 
About half of the sources (21/42) in the present
sample were only detected at $1.4$~GHz so it is not possible to
determine a radio spectral index. In the remaining half of the cases,
it was possible to estimate a spectral index between two frequencies
below $8.4$~GHz (between 1.4 and 5~GHz in 13/21 cases). About 28\% of
these sources (12/42) were detected at frequencies in the MHz range
(74--843~MHz, see Table~9). In 4/42 cases (J0324+3410, J0849+5108,
J0948+0022, and J1505+0326), spectral indices between 5--15 and 15--37
GHz are measured, because of the MW campaign of Effelsberg, 
Mets\"ahovi, and RATAN-600 (Abdo et al. 2009a,b, Foschini et al. 2011a, 2012,
Fuhrmann et al. 2011, Angelakis et al. 2012a,b). In only two cases
(J0324+3140 and J0948+0022) are there detections up to $142$~GHz at
Pico Veleta. The results are summarised in Online Material Table~8. In 7/13 cases, the
spectral indices $\alpha_{\rm r}$ were inverted. Three of these cases
were of $\gamma-$ray detected RLNLS1s. Two of these cases, J0324+3410
and J0849+5108 (two $\gamma-$ray sources), the average spectral index
was inverted at higher frequencies. The weighted mean $\alpha_{\rm r}$ was 
equal to $0.1\pm0.3$ (median 0.3). 

Comparison with blazar samples reveals similar spectral indices. Abdo
et al. (2010b) performed a linear regression of all the available data
in the 1--100~GHz frequency range of 48 blazars of the {\it Fermi} LAT
Bright AGN Sample (LBAS) and find an average value of $\alpha_{\rm
1-100\,GHz}=0.03 \pm 0.23$. They found no differences between FSRQs
and BL Lac objects. It is worth noting that the RLNLS1 J0948+0022 of
the present sample is also in the LBAS list, but it is classified
there as a low-synchrotron peak FSRQ. Abdo et al. (2010) find a radio
spectral index of $-0.645$.

Another useful comparison is with the jetted AGNs of the MOJAVE sample:
Hovatta et al. (2014) analysed the radio data of 191 AGNs (133 FSRQs,
33 BL Lac objects, 21 radio galaxies, and 4 unknown-type AGNs and
calculated the spectral index between 8.1 and 15.4 GHz. Also in this
case, there is virtually no difference between FSRQs and BL Lac
objects, $-0.22$ and $-0.19$, respectively.

Tornikoski et al. (2000) studied 47 Southern hemisphere sources,
mostly FSRQs (38), plus 6 BL Lac objects and 3 radio galaxies (see
also Ghirlanda et al. 2010 for a sample of blazars in the Southern
hemisphere) over a frequency range spanning 2.3 to 230 GHz. The
spectral indices are almost flat below 8.4 GHz, with some differences
between BL Lacs and high-polarisation quasars on one hand, and
low-polarisation quasars on the other: while the latter have a
somewhat steeper spectral index ($\alpha_{\rm 2.3-8.4\,GHz}\approx
0.05$), the former have an inverted index ($\alpha_{\rm
2.3-8.4\,GHz}\approx -0.13$). For all the sources, the spectral index
becomes steeper at higher frequencies ($\alpha_{\rm
90-230\,GHz}\approx0.79$). Nieppola et al. (2007) studied a large
sample (398) of only BL Lac objects in the Northern hemisphere and
found average values of $\alpha_{\rm 5-37\,GHz}\approx -0.25$ and
$\alpha_{\rm 37-90\,GHz} \approx 0.0$.

Our values are in agreement with these results: we find a rather flat
or inverted spectrum extending from all the available frequencies
(Online Material Table~8), as already found by Fuhrmann et al. (2011) and Angelakis et
al. (2012a,b, in preparation). However, we note that most of the radio
observations refer to the first $\gamma-$ray RLNLS1s detected, which
were immediately monitored with MW campaigns, i.e. J0324+3410,
J0849+5108, J0948+0022, and J1505+0326. All the other sources in the
present sample have been poorly observed in the radio. There are,
however, programs to increase the database on these sources at radio
frequencies. The Mets\"ahovi group is performing a multi-epoch
variability program at 22 and 37 GHz on more than 150 radio-loud AGNs,
including 38 RLNLS1s of the present sample (L\"ahteenm\"aki et al., in
preparation). Richards et al. (2014) is performing a high-spatial
resolution study on 15 RLNLS1s of the present sample with the VLBA at
5, 8, 15, and 24~GHz, including polarisation.

Analysis of archival VLBI radio maps provides some further information
(see Sect. 3.2 and Online Material Table~7). Again, the earliest $\gamma-$ray detected RLNLS1s were the
most heavily observed: J0324+3410, J0849+5108, and J0948+0022. The
source J0324+3410 (7 observations at 15~GHz) displayed a small flare
in 2011, with compactness --- defined as the core to total flux density ratio
--- of order $0.6$, which increased during the flare to
$>0.75$. Polarisation is almost stable during these observations, but
during the flare there is a marginal change in the position angle of the electric
vector (EVPA) with respect to the jet direction (from $87^{\circ}$ to
$77^{\circ}$). Also J0849+5108 showed an increase in the compactness
(from $0.75$ to $0.9$) with flux density. Between 2013 January and
July (MJD $56313-56481$), there was a swing in the EVPA (from
$168^{\circ}$ to $24^{\circ}$, with an almost stable jet direction)
that preceded a $\gamma-$ray outburst (Eggen et al. 2013),
which happened also in J0948+0022 (Foschini et al. 2011a). The latter
has been observed 17 times at 15~GHz, and many of these epochs were
already studied by Foschini et al. (2011a). The present reanalysis
basically confirms and extends the previous studies. It is worth
noting that this source is very compact (0.975) as also indicated by
previous comparison with single-dish observations (Abdo et al. 2009b,
Foschini et al. 2011a, 2012). An exceptional radio core outburst on
2013 May 5 (MJD 56417), when J0948+0022 reached a core flux density of
0.862~Jy. This followed a strong outburst at $\gamma$-rays that
occurred on 2013 January 1 (MJD 56293), as reported by D'Ammando et
al. (2013c). During the same period there was also a
swing of the EVPA, changing from $\sim 82^{\circ}$ on 2012 November 11
to $\sim 125^{\circ}$ on 2013 May 5. Moreover, Itoh et al. (2013)
reported strong variability in optical polarisation in the same epoch
(see Sect.~4.3).

The only source for which there are multifrequency observations is
J1505+0326, from 2 to 43 GHz. A detailed analysis is reported b
D'Ammando et al. (2013a). It is very compact (0.95--0.97 at 15~GHz
during flares), at level comparable to J0948+0022. A flux density excess
($\sim 0.1$~Jy) outside the core has been measured at 22~GHz in
2002. The EVPA--jet direction angle is quite unstable, changing from
$\sim 61^{\circ}$ to $-100^{\circ}$. We observed significant flux density 
increases only in the VLBI cores of all the observed sources
(Table~8), suggesting that the location of the $\gamma$-ray emission
should be very close to the central black hole.

Morphological studies of RLNLS1s have been published by Doi et
al. (2006, 2007, 2011, 2012), Gu \& Chen (2010), Giroletti et
al. (2011), and Orienti et al. (2012). The emerging characteristics
are (a) compact radio morphology, although there are kiloparsec scale
structures in some cases (Doi et al. 2012), (b) high-brightness
temperature of the core feature, indicating non-thermal processes, (c)
flat or inverted spectra (although the samples included also steep
spectrum radio sources, which are excluded from the present work), and
(d) possible links with compact steep spectrum (CSS) and gigahertz
peaked spectrum (GPS) radio sources, as also suggested by Oshlack et
al. (2001) and Gallo et al. (2006) for J2007$-$4434, Komossa et
al. (2006a), Yuan et al. (2008), and more recently by Caccianiga et
al. (2014). Doi et al. (2012) found that the detection of extended
emission is lower than expected from broad-line Seyferts and they
suggest it could be due to the lower kinetic power of jets in low-mass
AGNs, rather than the young age of the source. Interestingly, also the
radio core of RQNLS1s and Seyferts display non-thermal
characteristics, suggesting some link with jets (Giroletti \& Panessa
2009, Doi et al. 2013).

Angelakis et al. (in preparation) have studied the variability of four
RLNLS1s detected at $\gamma$ rays (J0324+3410, J0849+5108, J0948+0022,
J1505+0326) at different radio frequencies. Brightness temperature
measurements indicate minimum Doppler factors from $1.3$ for
J0324+3410 to $4.2$ in the case of J1505+0326.

The search for short variability resulted in only one case of
variability on timescales of days: we measure an upper
limit of $2.6$~days for J0948+0022 at 37~GHz. In the other cases, we
find variations on timescales of about one month, but this is likely
due to the one-month sampling rate of Effelsberg
observations. Mets\"ahovi (37~GHz) observed at a more intense sampling
rate during some MW campaigns (e.g.  Abdo et al. 2009b, Foschini et
al. 2011a, 2012). Nieppola et al. (2007) reported variability on
timescales of hours for some BL Lac objects, for example,
$\sim$8~hours for S5~$0716+71$, $\sim $1~hour for AO~0235+164, and
$\sim$6~hours for OJ~287. In the case of RLNLS1s, clearly higher
sampling rate MW campaigns are required.

\begin{table*}[ht!]
\caption{Mass and Accretion luminosity estimated from optical
data. Columns: (1) Name of the source; (2) Line dispersion
$\sigma_{\rm line}$ of the broad component of H$\beta$ [km~s$^{-1}$];
(3) FWHM of the broad component of H$\beta$ [km~s$^{-1}$]; (4)
H$\beta$ luminosity [$10^{42}$~erg~s$^{-1}$]; (5) Black hole mass
[$10^7\, M_\odot$]; (6) Disk luminosity [$10^{44}$~erg~s$^{-1}$]; (7) Disk luminosity 
[Eddington units]; (8) method adopted: A, from Asiago spectra; L,
spectra from literature; M, derived from optical magnitudes; N,
spectra downloaded from NED; S, from SDSS spectra.}  
\centering
\begin{tabular}{lccccccc} 
\hline Source & $\sigma_{\rm line}$ & FWHM & $L_{\rm H\beta}$ & M & L$_{\rm disk}$ & L$_{\rm disk}$/L$_{\rm Edd}$ & Method\\ 
\hline
J$0100-0200$ & 982 & 920 & $-$ & 4.0 & $25.2$ & 0.49 & M\\
J$0134-4258$ & 1632 & 1241 & $2.77$ & 7.1 & $8.61$ & 0.09 & L (Grupe et al. 2000)\\ 
J$0324+3410$ & 1791 & 1868 & $0.53$ & 3.6 & $1.50$ & 0.03 & A\\ 
J$0706+3901$ & 1839 & 1402 & $0.16$ & 2.0 & $0.43$ & 0.02 & N\\ 
J$0713+3820$ & 2041 & 1901 & $9.28$ & 21.2 & $31.0$ & 0.11 & N\\
J$0744+5149$ & 2122 & 1989 & $-$ & 26.5 & $49.2$ & 0.14 & M\\
J$0804+3853$ & 1588 & 1523 & $2.74$ & 6.7 & $8.51$ & 0.10 & S\\
J$0814+5609$ & 2759 & 2777 & $6.10$ & 31.0 & $19.9$ & 0.05 & S\\
J$0849+5108$ & 1330 & 2490 & $1.32$ & 3.2 & $3.94$ & 0.09 & S\\
J$0902+0443$ & 1491 & 1781 & $2.02$ & 5.0 & $6.17$ & 0.09 & S\\
J$0937+3615$ & 1343 & 2192 & $0.59$ & 2.1 & $1.68$ & 0.06 & S\\
J$0945+1915$ & 1730 & 2818 & $2.52$ & 7.6 & $7.80$ & 0.08 & A\\
J$0948+0022$ & 1548 & 1639 & $3.73$ & 7.5 & $11.8$ & 0.12 & S\\
J$0953+2836$ & 2407 & 2749 & $3.14$ & 16.6 & $9.84$ & 0.05 & S\\
J$1031+4234$ & 3153 & 1822 & $2.08$ & 22.9 & $6.37$ & 0.02 & S\\
J$1037+0036$ & 985 & 1776 & $1.75$ & 2.0 & $5.31$ & 0.20 & S\\
J$1038+4227$ & 1615 & 1917 & $2.62$ & 6.8 & $8.12$ & 0.09 & S\\
J$1047+4725$ & 1474 & 2237 & $5.80$ & 8.6 & $18.9$ & 0.17 & S\\
J$1048+2222$ & 1742 & 718 & $1.39$ & 5.7 & $4.16$ & 0.06 & S\\
J$1102+2239$ & 1940 & 2181 & $3.56$ & 11.5 & $11.2$ & 0.08 & S\\
J$1110+3653$ & 1230 & 2081 & $1.07$ & 2.4 & $3.14$ & 0.10 & S\\
J$1138+3653$ & 1231 & 1542 & $1.29$ & 2.7 & $3.85$ & 0.11 & S\\
J$1146+3236$ & 1737 & 1977 & $3.79$ & 9.5 & $12.0$ & 0.10 & S\\
J$1159+2838$ & 1907 & 2728 & $0.052$ & 1.2 & $0.13$ & 0.01 & N\\
J$1227+3214$ & 694 & 1567 & $0.51$ & 0.52 & $1.42$ & 0.21 & S\\
J$1238+3942$ & 940 & 1229 & $1.16$ & 1.5 & $3.42$ & 0.18 & S\\
J$1246+0238$ & 1667 & 1756 & $1.80$ & 5.9 & $5.45$ & 0.07 & S\\
J$1333+4141$ & 1589 & 2942 & $1.73$ & 5.3 & $5.25$ & 0.08 & S\\
J$1346+3121$ & 1074 & 1503 & $0.64$ & 1.4 & $1.83$ & 0.10 & S\\
J$1348+2622$ & 2192 & 3361 & $4.09$ & 5.3 & $13.0$ & 0.19 & S (based on Mg\,{\sc ii}$\,\lambda2798$.)\\
J$1358+2658$ & 1471 & 1805 & $3.20$ & 6.3 & $10.1$ & 0.12 & S\\
J$1421+2824$ & 1589 & 1724 & $7.18$ & 11.2 & $23.7$ & 0.16 & S\\
J$1505+0326$ & 1409 & 1337 & $0.41$ & 1.9 & $1.12$ & 0.05 & S\\
J$1548+3511$ & 1557 & 2217 & $4.37$ & 8.3 & $14.0$ & 0.13 & S\\
J$1612+4219$ & 777 & 1200 & $0.87$ & 0.88 & $2.53$ & 0.22 & S\\
J$1629+4007$ & 1246 & 1410 & $2.00$ & 3.5 & $6.10$ & 0.13 & S\\
J$1633+4718$ & 945 & 931 & $0.36$ & 0.79 & $0.98$ & 0.10 & S\\
J$1634+4809$ & 1856 & 1763 & $1.95$ & 7.7 & $5.94$ & 0.06 & S\\
J$1644+2619$ & 1129 & 1486 & $0.51$ & 1.4 & $1.42$ & 0.08 & S\\
J$1709+2348$ & 2377 & 1256 & $0.95$ & 2.4 & $2.79$ & 0.09 & S\\
J$2007-4434$ & 1869 & 2844 & $1.62$ & 7.0 & $4.8$ & 0.052 & L (Drinkwater et al. 1997)\\ 
J$2021-2235$ & 491 & 460 & $-$ & 3.75 & $2.91$ & 0.60 & M\\ 
\hline
\end{tabular}
\label{tab:masses}
\end{table*}

\section{Estimates of masses and accretion luminosities} 
The masses of the central black holes are given by
\begin{equation} M = f \left( \frac{R_{\rm BLR}\sigma_{\rm line}^2}{G}
\right),
\label{eq:mass}
\end{equation} where $R_{\rm BLR}$ is the size of the broad-line
region (BLR) measured by reverberation or estimated from scaling
relations, $\sigma_{\rm line}$ is the line dispersion (or second
moment of the line profile), $G$ is the gravitational constant, and
$f$ is a dimensionless scale factor of order unity (Peterson et
al. 2004). We used the line dispersion, because it is less affected by inclination, Eddington ratio, and line profile (Peterson et al. 2004, Collin et al. 2006). We estimate the BLR radius by using the relationship
between the luminosity of the H$\beta$ line and the radius of the BLR
($R_{\rm BLR}$) from the relationship of Greene et al. (2010),

\begin{equation} 
\log \left[ \frac{R_{\rm BLR}}{\rm 10\,light\,days}\right] = 0.85 + 0.53\log \left[ \frac{L({\rm H\beta})}{10^{43}\,\rm erg\,s^{-1}}\right].
\label{eq:sizeblr}
\end{equation} 

Following Collin et al. (2006), we adopt $f=3.85$.

\begin{figure}[t!]
\begin{center}
\includegraphics[angle=270,scale=0.35]{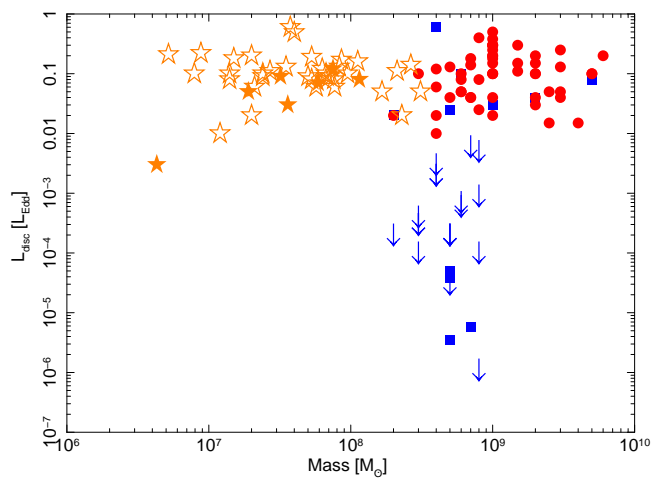}
\caption{Accretion disk luminosity [Eddington units] vs. mass of the
central black hole [$M_{\odot}$]. The orange stars are the RLNLS1s of
the present sample (see Table~\ref{tab:masses}) and filled orange
stars indicate those detected at $\gamma$ rays; the red circles are
the FSRQs, and the blue squares are the BL Lac objects (blue arrows
indicates upper limits in the accretion luminosity) from Ghisellini et
al. (2010). We noted some BL Lacs with strong accretion disk, in the region occupied by 
FSRQs: these are the so-called intruders (Ghisellini et al. 2011, Giommi et al. 2012).} 
\label{fig:massaccretion}
\end{center}
\end{figure}

The size of the BLR gives also the luminosity of the accretion disk
($R_{\rm BLR}\propto L_{\rm disk}^{1/2}$, e.g. Ghisellini \& Tavecchio
2009; see also Bentz et al. 2013), which in turn has been normalised
to the Eddington value
\begin{equation} L_{\rm Edd}= 1.3\times 10^{38}\left(
\frac{M}{M_{\odot}}\right) \, \rm [erg~s^{-1}].
\end{equation} By using $L({\rm H\beta})$ --- instead of the continuum
at $5100\,$\AA\ (or another wavelength), which is more conventional
and generally more accurate --- to estimate the size of the BLR and
the accretion disk luminosity, we avoid the problem of contamination
of the flux by either the jet or the host galaxy.

The results are displayed in Table~\ref{tab:masses}. In three cases
(3/42), no optical spectra were found. Therefore, we estimated the
line dispersion from the value of FWHM found in literature by using
the ratio FWHM/$\sigma_{\rm line}=1.07$, which is the average over
the known values of the present sample (39/42). This value is
consistent with what expected from NLS1s (cf.\ Peterson 2011). From
the available optical magnitudes near $5100\,$\AA $\,$ we estimated
the disk luminosity and the size of the BLR and then used
eq.~(\ref{eq:mass}) to estimate the mass. We note that these sources
have Eddington ratios slightly greater than the others of the sample:
this can be understood because with the photometry it is not
possible to disentangle the contribution of the disk from that of the
jet. We note that our values are in agreement with the results available in the vast majority of literature on RLNLS1s 
(e.g. Komossa et al. 2006, Whalen et al. 2006, Yuan et al. 2006) and on NLS1s in general (e.g. Peterson 2011).

A comparison of these data with the corresponding data for the blazar
sample is shown in Fig.~\ref{fig:massaccretion}. It is evident that
RLNLS1s occupy a unique parameter space among AGNs with relativistic
jets that corresponds to lower masses and high Eddington rates\footnote{This part of text has been changed after November 4, 2019. The following sentence has been removed: ``It is worth noting one outlier, J2007$-$4434, has a low Eddington rate
($0.003L_{\rm Edd}$): this was also one of the RLNLS1s whose nature is
questionable on account of its weak Fe\,{\sc ii} emission (Gallo et
al. 2006, Komossa et al. 2006)''. In the previous versions, we referred to Oshlack et al. (2001) for the optical spectrum of J$2007-4434$. However, when following up the 2019 October $\gamma-$ray outburst of this source (Berton et al. 2019), we found an error in the y-axis of the optical spectrum displayed in Fig. 2 of Oshlack et al. (2001). The same spectrum was originally recorded by Drinkwater et al. (1997), but the original 1997 figure has a flux greater than that of Oshlack by a factor $\sim 200$. After adopting the correct flux value, J$2007-4434$ is now fully consistent with the typical values of the NLS1s distributions. This change of flux affects Table~\ref{tab:masses}, Fig.~\ref{fig:massaccretion}, and Fig.~\ref{fig:seds6}, left panel.}.  There is apparently an unoccupied
area of parameter space corresponding to low black hole masses and
low Eddington ratios. It is not possible to say whether this is real
or a selection effect.  Possible candidates to occupy this region are
low-luminosity AGNs (e.g.  M81, Alberdi et al. 2013), although there is
debate about the nature of their radio emission (see Paragi et
al. 2013). Moreover, no low-luminosity AGN has yet been detected in
high-energy $\gamma$ rays.

\begin{figure}[t!]
\begin{center}
\includegraphics[angle=270,scale=0.35]{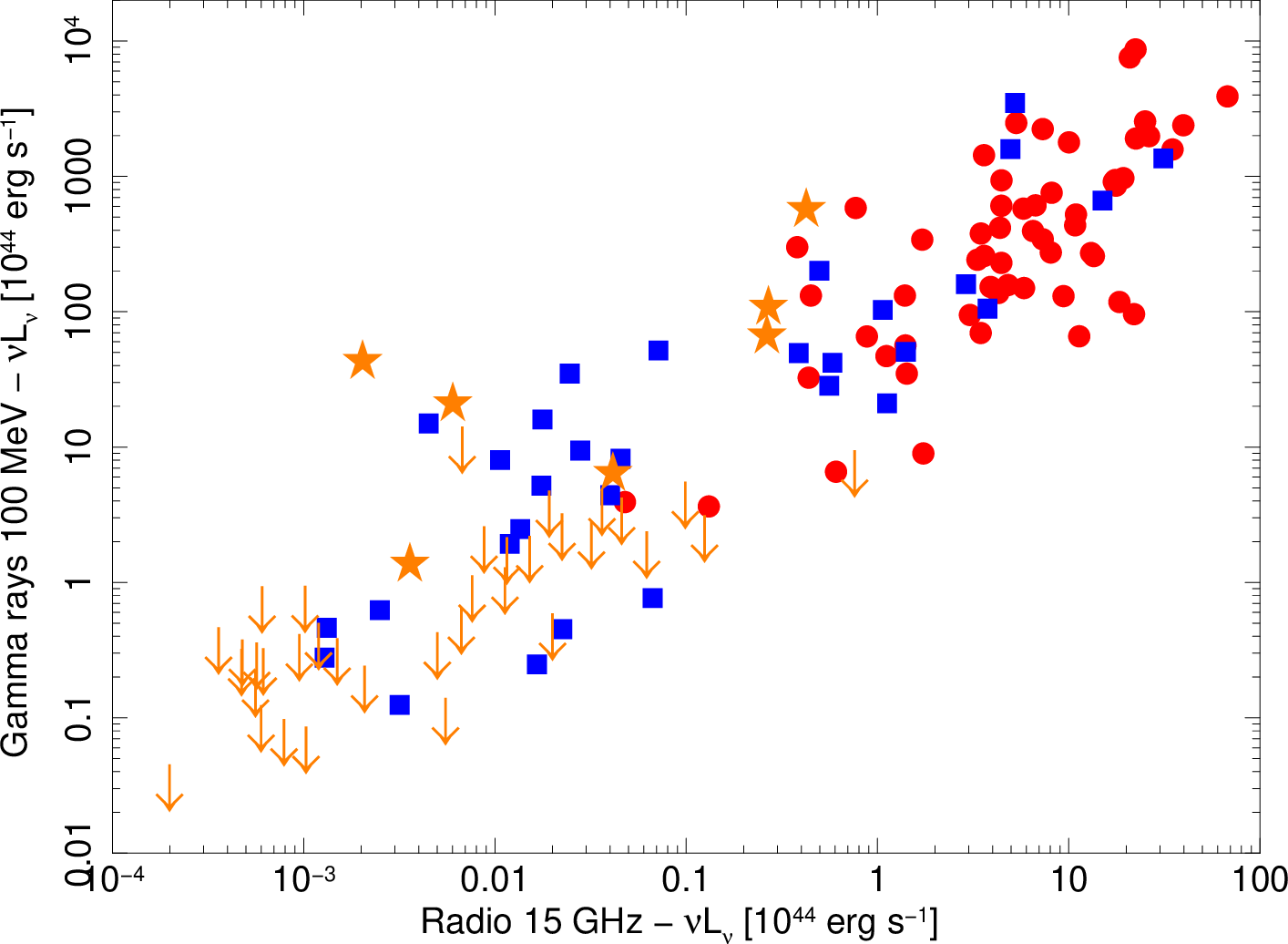}\\
\includegraphics[angle=270,scale=0.35]{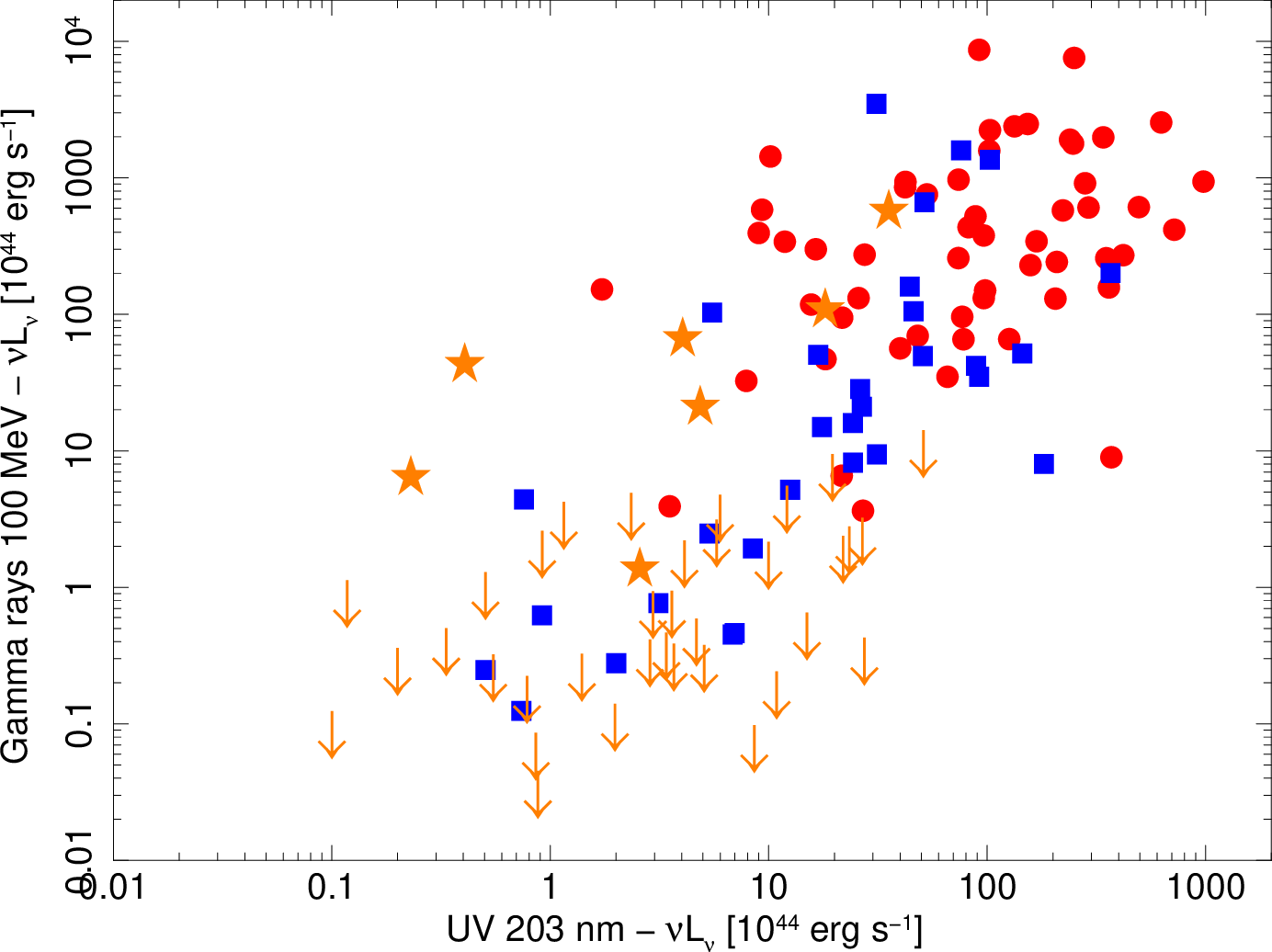}\\
\includegraphics[angle=270,scale=0.35]{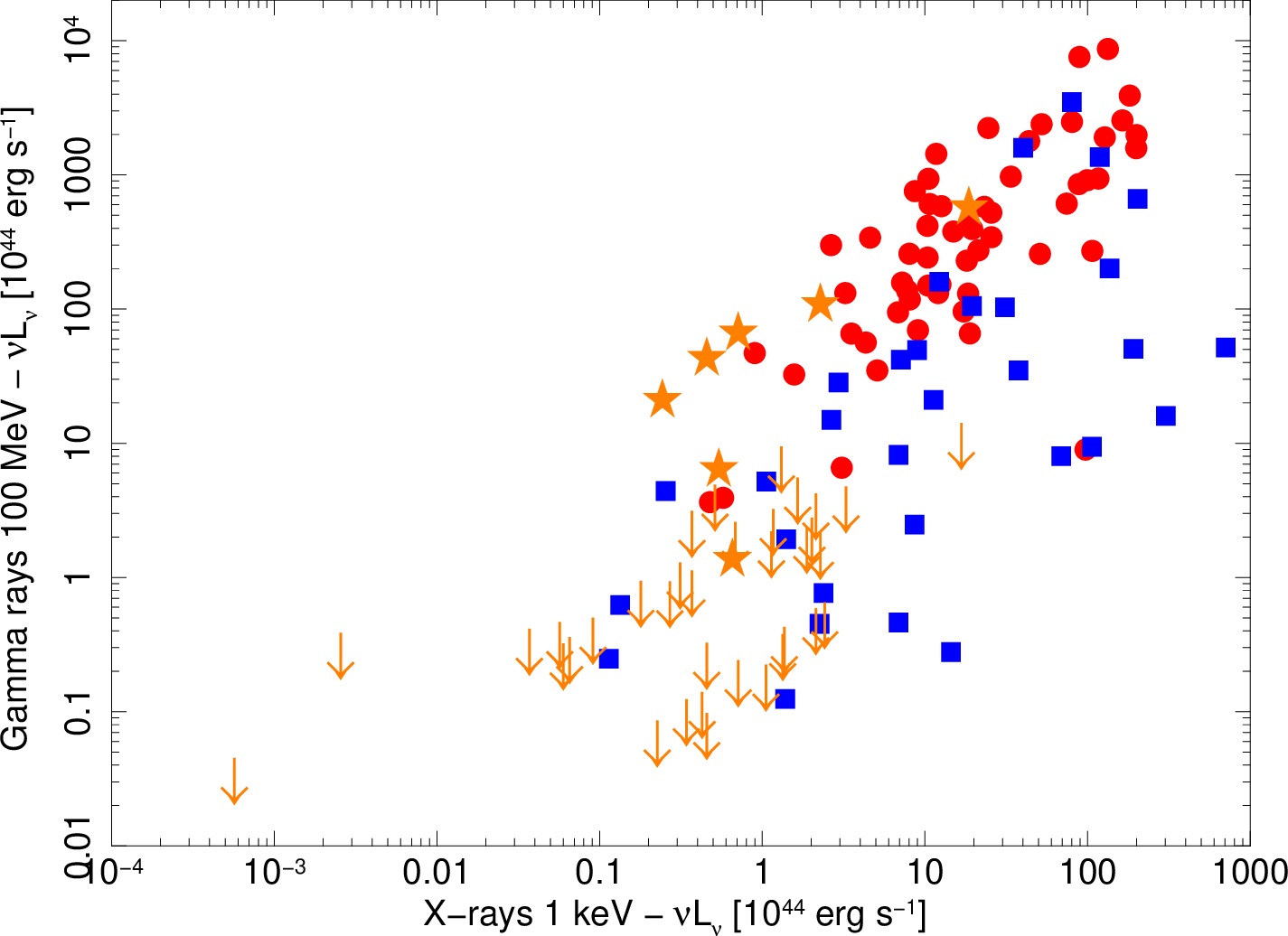}

\caption{Gamma-ray luminosity at 100~MeV compared with radio
luminosity at 15~GHz ({\it top panel}), ultraviolet luminosity at 203~nm ({\it mid panel}), X-ray luminosity at 1~keV ({\it bottom panel}). The orange stars are the RLNLS1s of the present sample detected in $\gamma$ rays, while upper limits are reported for the others (grey arrows); the red circles are the FSRQs and the blue squares are the BL Lac objects.}
\label{fig:monochromo}
\end{center}
\end{figure}

\section{Monochromatic luminosities}
Another comparison with blazars can be done via $\nu L_{\nu}-\nu L_{\nu}$ plots in Fig.~\ref{fig:monochromo}. Starting from the available data, we normalised the fluxes to four reference frequencies or wavelengths or energies: 15~GHz for radio observations, 203~nm for ultraviolet wavelengths, 1~keV for X-rays, and 100~MeV for $\gamma-$rays. While for most blazars, radio observations at 15~GHz were available from the MOJAVE project, the same was not true for RLNLS1s. For half of the RLNLS1s (21/42), there were radio data at 1.4~GHz either from NVSS or FIRST surveys. In some cases, there were also data at 5, 8.4, 17, or 20~GHz (the two latter frequencies are used in the Southern hemisphere). We extrapolated the 15~GHz flux by using the average radio spectral indices in Online Material Table~8. 

The situation is slightly better at ultraviolet wavelengths, because of the availability of {\it Swift}/UVOT observations, many of them specifically requested for this survey. For those sources with incomplete data, we used the bluest photometric data available and corrected by using the average UV spectral index in Online Material Table~8. We adopted the average spectral indices also to normalise the integrated fluxes or upper limits in the 0.3--10~keV and 0.1--100~GeV bands. 

The monochromatic fluxes were then $K$-corrected by
\begin{equation}
S_{\rm \nu,rest}=S_{\rm \nu}(1+z)^{\alpha_{\nu}-1},
\end{equation}
where $S_{\rm \nu,rest}$ is the rest-frame monochromatic flux at the frequency $\nu$, $S_{\rm \nu}$ is the observed monochromatic flux at frequency $\nu$, $z$ is the redshift, and $\alpha_{\nu}$ is the spectral index at the frequency $\nu$. The corrected monochromatic fluxes were then converted into luminosities. The results are displayed in Fig.~\ref{fig:monochromo}.

\begin{figure*}[t!]
\begin{center}
\includegraphics[angle=270,scale=0.35]{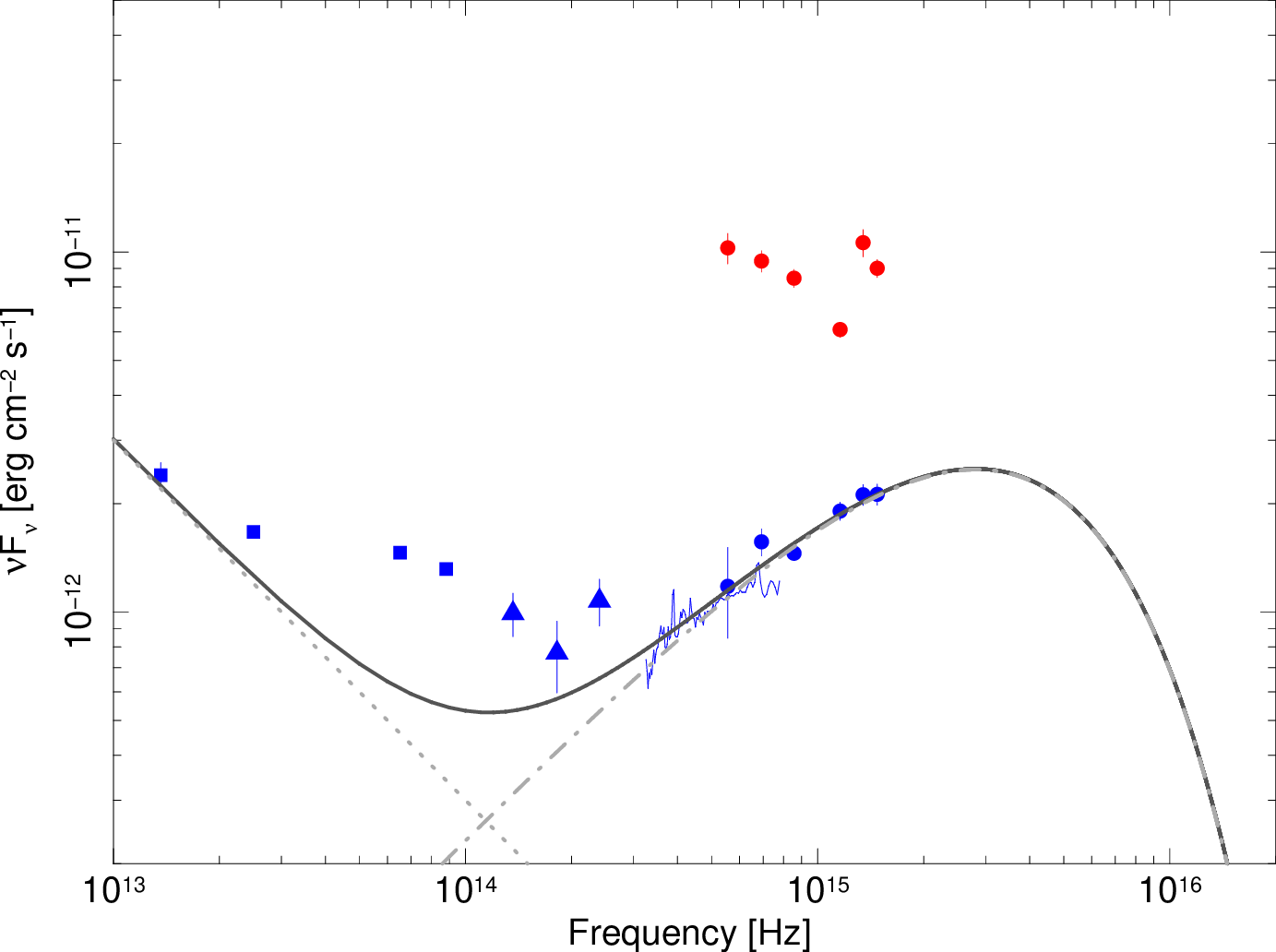}
\includegraphics[angle=270,scale=0.35]{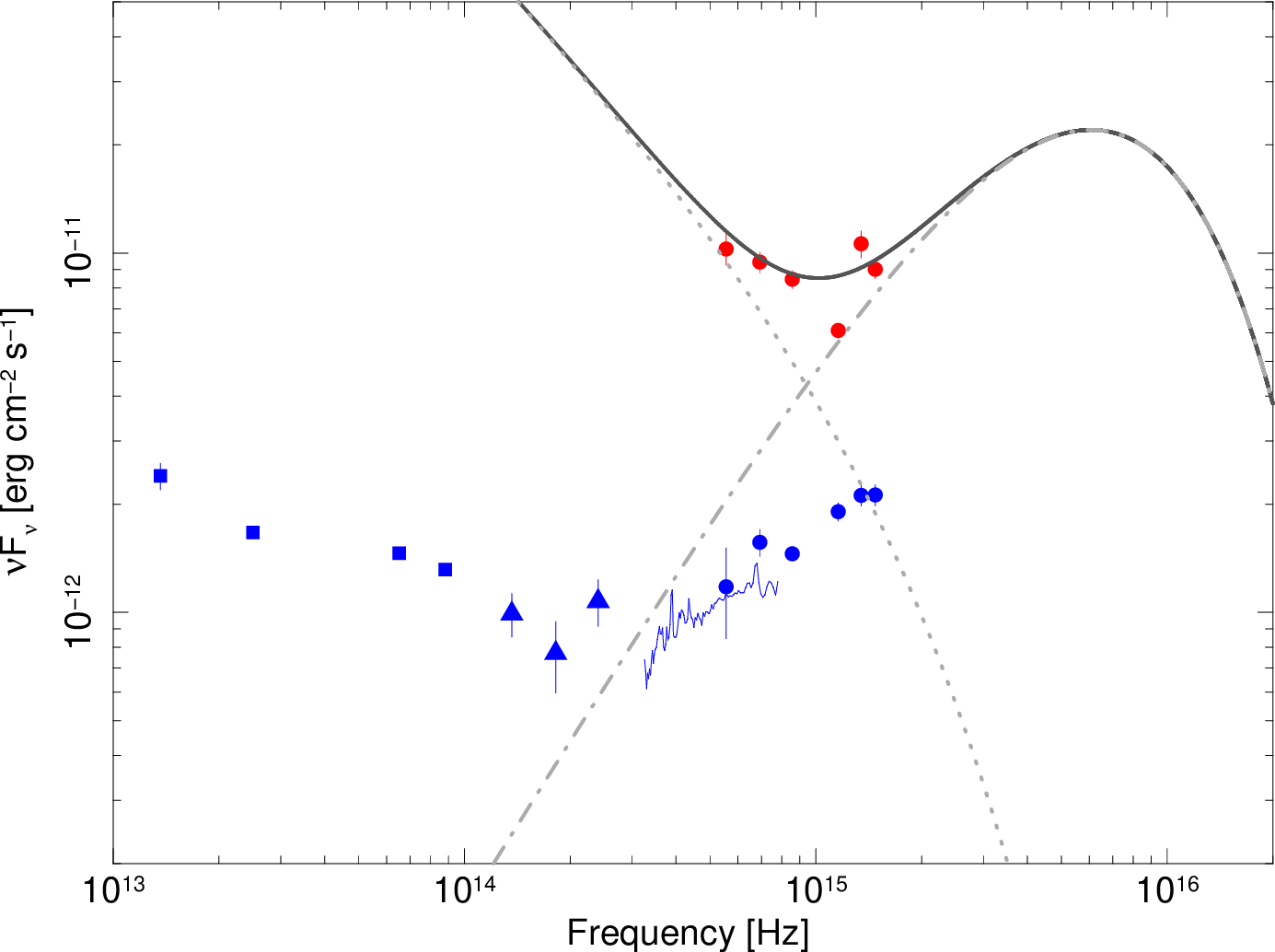}
\caption{Zoom of the SED of J0948+0022 in the infrared-to-ultraviolet range. Data are from: WISE (filled squares), 2MASS (filled triangles), SDSS (continuous line), {\it Swift}/UVOT (filled circles). ({\it left panel}) Blue refer to lowest observed activity state (LS, 2009 May 15); ({\it right panel}) red to highest activity state (HS, 2012 December 30). The grey dot-dashed line represents a model of standard accretion disk as expected in the case of J0948+0022 ($M=7.5\times 10^7 M_{\odot}$); the grey dotted line represents the synchrotron emission; the continuous grey line is the sum of the two models.}
\label{fig:zoom0948sed}
\end{center}
\end{figure*}

Fig.~\ref{fig:monochromo} shows that RLNLS1s are the low-luminosity tail of FSRQs, as already noted by Foschini et al. (2013). While at radio and ultraviolet frequencies RLNLS1s share the same region of BL Lac objects, the two populations diverge from each other at 1~keV, where BL Lac objects move to greater X-ray luminosities, indicating a different origin of the emission (synchrotron for BL Lacs, disk corona or inverse-Compton for RLNLS1s). 

We noted one possible outlier in the radio--$\gamma$ panel: J1102+2239 was detected at $\gamma$-ray flux above what is expected from the trend of the other sources. There could be several explanations: the $\gamma$-ray activity could be limited to a small time interval, the radio measurements, extrapolated from 1.4~GHz measurements from FIRST and NVSS, were likely done during periods of low activity of the sources, or it could even be an indication of some artefact in the $\gamma$-ray detection. Further studies could solve the conundrum.  

There are also two sources with very low X-ray fluxes in the X-ray/$\gamma$-ray panel, J0100$-$0200 and J0706+3901. In both cases, the X-ray flux was measured by {\it Chandra} in 2003 and there were no simultaneous data at other wavelengths. 

We stress the difference between RLNLS1s and BL Lac objects. Fig.~\ref{fig:monochromo} shows the observed luminosities at different frequencies: RLNLS1s and BL Lacs occupy similar regions and generally overlap at radio and UV frequencies. However, while BL Lac objects have low power and masses comparable to those of FSRQs, RLNLS1s have low power and lower masses (see Fig.~\ref{fig:massaccretion}). Indeed, when normalised for the mass of the central black hole, the jet power of RLNLS1s and FSRQs are of the same order of magnitude, as shown by Foschini (2014) and references therein. It is worth stressing that the normalisation is not linear, but it is necessary to divide the jet power by $M^{1.4}$, according to the theory developed by Heinz \& Sunyaev (2003) and confirmed by Foschini (2011b, 2012b,c, 2014). 

\section{Spectral energy distribution}
Fig.~\ref{fig:seds1}-\ref{fig:seds6} display the observed SEDs of all RLNLS1 in the present sample, assembled from data extracted from observations at different epochs and archives, as discussed in Sect.~3. The most complete SEDs are mostly those of the RLNLS1s significantly detected at gamma rays, i.e.  J0324+3410, J0849+5108, J0948+0022, J1505+0326, J2007$-$4434. The modelling of these SEDs already has been presented and discussed in other papers (e.g.  Abdo et al. 2009b,c, Foschini et al. 2011a, 2012, D'Ammando et al. 2012, 2013a,b, Paliya et al. 2014). 

There are also a few more cases (e.g,. J0814+5609, J1047+4725, J1548+3511, J1629+4007) with fairly good sampling because of previous specific interest. For example, J1629+4007 was long observed because it was thought to be an example of a high-frequency peaked FSRQ (Padovani et al. 2002, Falcone et al. 2004). It is evident from the SED (Fig.~15) that the strong X-ray emission is not due to synchrotron radiation, but rather to the disk corona (see also Maraschi et al. 2008). In other cases, it seems simply that the source fell into the field of view of other targets. We noted a strong change in radio flux density at 5~GHz in J1047+4725: early observations performed in 1987 with the Green Bank 91~m telescope ($\sim3.\!'5$ angular resolution) found a flux density of $\sim 0.4$~Jy (Becker et al. 1991, Gregory \& Condon 1991), while an observation with the VLA at 8.4~GHz in 1990 ($0\farcs2$ angular resolution) measured a flux density of $\sim 0.3$~Jy (Patnaik et al. 1992). A VLBA observation at 5~GHz with milliarcsecond resolution in 2006 resulted in a flux density of $\sim 33$~mJy (Helmboldt et al. 2007). One explanation for this could be the presence of significant extended emission, which is integrated in the low angular resolution of Green Bank 91~m and VLA telescopes, while is resolved in the milliarcsecond images of VLBA. 

Other sources displayed extreme variability, specifically at optical wavelengths: for example, the SDSS optical spectrum of J0849+5108 (observed in 2000) is about one order of magnitude lower than the optical observations made with {\it Swift} after the detection at $\gamma$-rays by {\it Fermi} (2011--2013). A similar case is J1159+2838, while the optical spectrum of J0953+2836 is about two orders of magnitude brighter than in the {\it Swift} observations. Spectral changes at optical frequencies, due to the jet activity, are also observed. Just as an example, we focus on the infrared-to-ultraviolet band of J$0948+0022$, which is the best sampled source, being the first to be detected at $\gamma$-rays. Fig.~\ref{fig:zoom0948sed} displays the two extreme states from the available data: the lowest activity state (LS, 2009 May 15, see also Abdo et al. 2009b, Foschini et al. 2012) and the highest state (HS, 2012 December 30). In both cases, we model the synchrotron emission (dotted grey line) with a power-law model with an exponential cutoff. The disk emission (dashed grey line) is the standard Shakura--Sunyaev model as expected from a black hole of $M=7.5\times 10^7 M_{\odot}$ (see Table~\ref{tab:masses}). Previous modelling (Abdo et al. 2009b, Foschini et al. 2012) supposed constant disk luminosity equal to $L_{\rm disk}\sim 0.4L_{\rm Edd}$, as measured by fitting the optical/UV emission with a standard Shakura--Sunyaev disk. In the present work, we obtained from the emission lines a value of $L_{\rm disk}\sim 0.12L_{\rm Edd}$. The difference could be due to a contamination of the jet emission in the optical/UV photometry fit, which is removed by using the emission lines. However, since the optical spectrum of SDSS was observed in 2000 and the MW campaign used for the SED modelling were obtained in 2008--2011, it is also possible that the Eddington ratio really changed. 

Although the lowest flux points were not simultaneous, since many of them are smoothly connected, it is reasonable to assume that they refer to a common state of low jet activity. Therefore, we modelled them together as low activity state (LS): in this case, the expected peak of the emission from the accretion disk at $12$\% of the Eddington luminosity is at $\sim 3.0\times 10^{15}$~Hz. For the high state (HS), we have just one {\it Swift}/UVOT observation and we adopted a standard disk at $40$\% of the Eddington value, peaking at $\sim 4.1\times 10^{15}$~Hz. The cutoff of the synchrotron emission was set to $8\times 10^{14}$~Hz in the LS and increased to $1.5\times 10^{15}$~Hz in the HS. We underscore that this is just another option, in addition --- and not in contrast --- to the previous model. As the expected peak of the disk emission in the far UV, outside the range of our observations, we cannot clearly distinguish between different possibilities. We also note that at infrared wavelengths, there is an excess that is likely attributable to the dusty torus and/or the host galaxy, as suggested also by the {\it WISE} colours (see Sect.~4.3). 

Changes in the activity of a relativistic jet, as for J0948+0022, could explain the differences in the spectral slopes of J0849+5108, and J1159+2838. All these sources have an optical spectrum with a slope different from that derived from the optical/UV photometry. Moreover,  some sources show optical/UV slopes decreasing with increasing frequencies (e.g.  J0804+3853, J0937+3615, J1031+4234, J1038+4227, J1102+2239, J1138+365, J1227+3214), while there were other cases with the opposite trend (e.g.  J0134$-$4258, J0324+3410, J0814+5609, J1348+2622, J1548+3511, J1629+400). These also have a flat X-ray slope, with some evidence of a soft-excess. An inspection of their corresponding central black hole masses and Eddington ratios did not reveal any trend. On the basis of the J0948+0022 behaviour, we favour the interpretation of the same central engine observed in a different combination of jet--disk states.  

\begin{figure*}[ht!]
\begin{center}
\includegraphics[angle=270,scale=0.35]{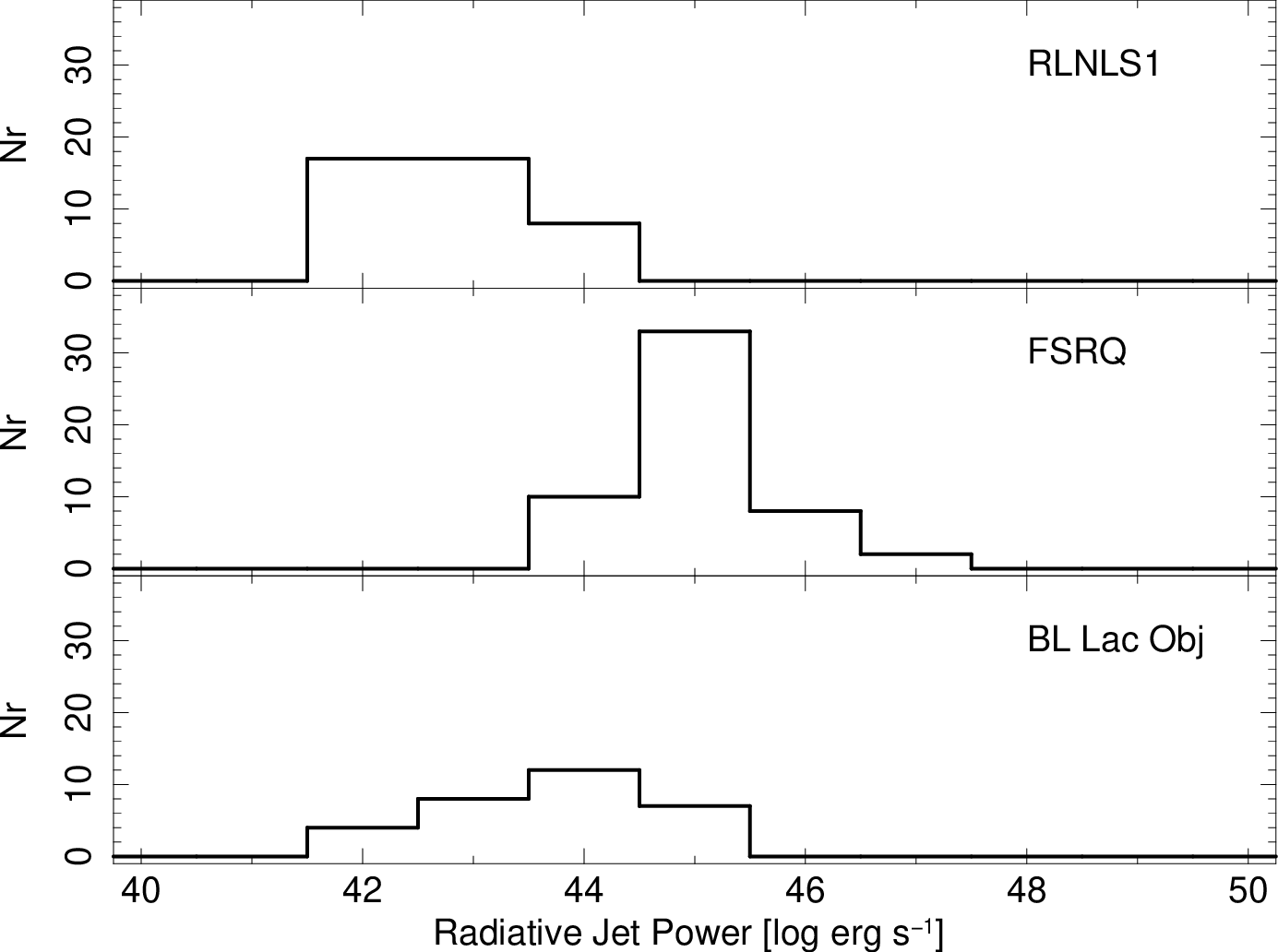}
\includegraphics[angle=270,scale=0.35]{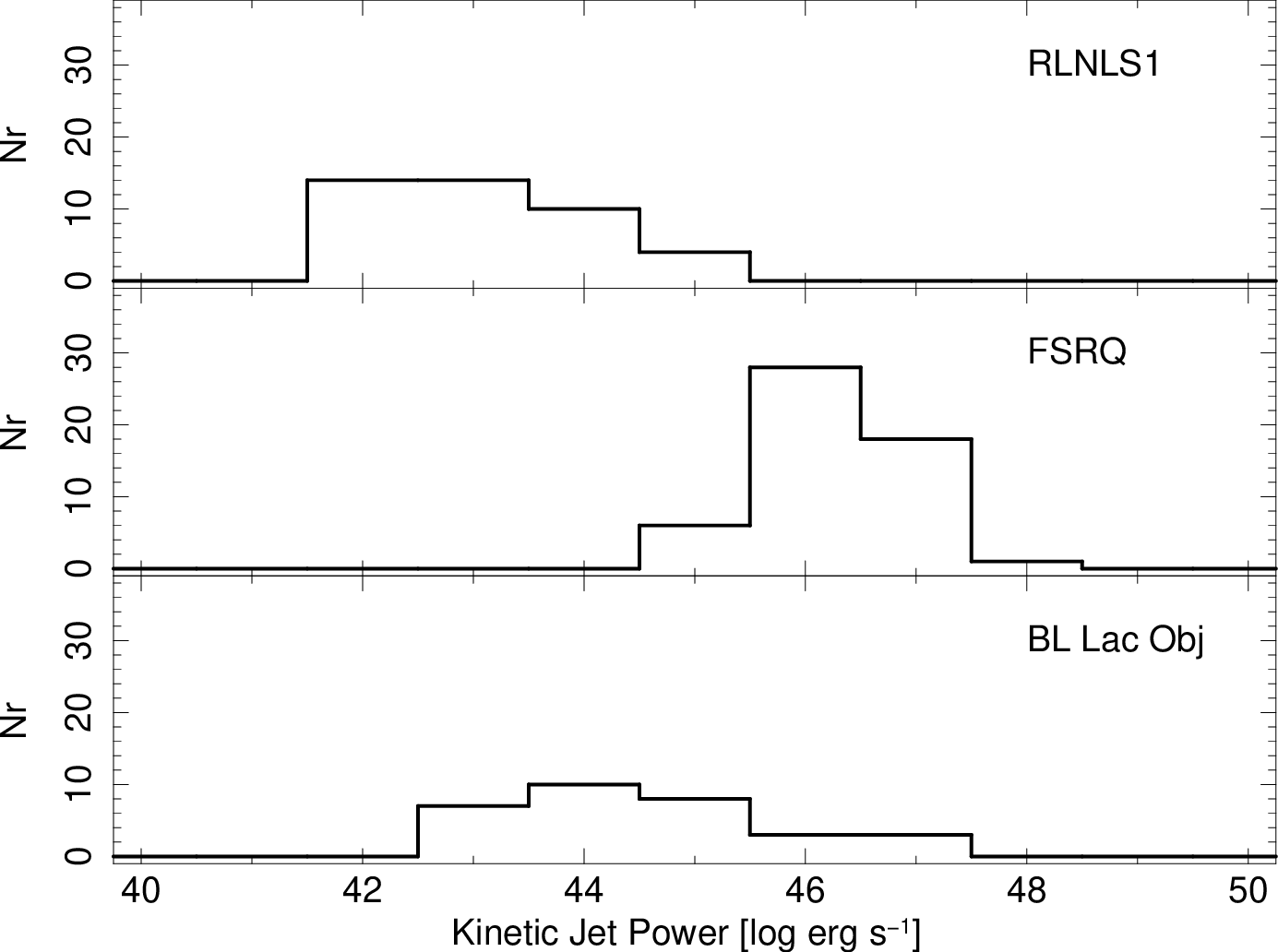}
\caption{Jet power distribution: ({\it left panel}) radiative, ({\it right panel}) kinetic. RLNLS1sdata are from the present work (Table~\ref{tab:jetpow}), while values for FSRQs and BL Lac objects are from Ghisellini et al. (2010).}
\label{fig:powdistr}
\end{center}
\end{figure*}

\begin{table*}[t!]
\caption{Estimated jet power from radio core measurements at 15~GHz according to the relationships in Foschini (2014). Columns display the logarithm of the radiative, kinetic (protons, electrons, magnetic field), and total jet powers [erg~s$^{-1}$].}
\centering
\begin{tabular}{cccc}
\hline
Source & $\log P_{\rm rad}$ & $\log P_{\rm kin}$ & $\log P_{\rm tot}$ \\
\hline
           J$0100-0200$   &     42.88     &   43.06    &    43.28\\
           J$0134-4258$   &     43.28     &   43.53    &    43.72\\
           J$0324+3410$   &     43.17     &   43.40    &    43.60\\
           J$0706+3901$   &     42.22     &   42.27    &    42.55\\
           J$0713+3820$   &     42.67     &   42.81    &    43.05\\
           J$0744+5149$   &     43.55     &   43.86    &    44.03\\
           J$0804+3853$   &     42.59     &   42.71    &    42.96\\
           J$0814+5609$   &     43.88     &   44.26    &    44.41\\
           J$0849+5108$   &     44.58     &   45.09    &    45.21\\
           J$0902+0443$   &     44.33     &   44.79    &    44.92\\
           J$0937+3615$   &     42.56     &   42.67    &    42.92\\
           J$0945+1915$   &     43.37     &   43.64    &    43.83\\
           J$0948+0022$   &     44.72     &   45.27    &    45.38\\
           J$0953+2836$   &     44.25     &   44.70    &    44.83\\
           J$1031+4234$   &     43.54     &   43.85    &    44.02\\
           J$1037+0036$   &     44.00     &   44.40    &    44.55\\
           J$1038+4227$   &     42.57     &   42.68    &    42.93\\
           J$1047+4725$   &     44.92     &   45.50    &    45.60\\
           J$1048+2222$   &     42.59     &   42.71    &    42.95\\
           J$1102+2239$   &     42.98     &   43.18    &    43.40\\
           J$1110+3653$   &     43.92     &   44.31    &    44.46\\
           J$1138+3653$   &     43.41     &   43.70    &    43.88\\
           J$1146+3236$   &     43.64     &   43.97    &    44.14\\
           J$1159+2838$   &     42.51     &   42.61    &    42.86\\
           J$1227+3214$   &     42.58     &   42.70    &    42.95\\
           J$1238+3942$   &     43.72     &   44.06    &    44.22\\
           J$1246+0238$   &     43.34     &   43.61    &    43.79\\
           J$1333+4141$   &     42.51     &   42.61    &    42.87\\
           J$1346+3121$   &     42.42     &   42.50    &    42.76\\
           J$1348+2622$   &     43.38     &   43.65    &    43.84\\
           J$1358+2658$   &     42.76     &   42.91    &    43.14\\
           J$1421+2824$   &     43.77     &   44.12    &    44.28\\
           J$1505+0326$   &     44.57     &   45.08    &    45.20\\
           J$1548+3511$   &     44.10     &   44.52    &    44.66\\
           J$1612+4219$   &     42.74     &   42.88    &    43.12\\
           J$1629+4007$   &     43.73     &   44.07    &    44.24\\
           J$1633+4718$   &     42.76     &   42.91    &    43.14\\
           J$1634+4809$   &     43.46     &   43.75    &    43.93\\
           J$1644+2619$   &     43.31     &   43.57    &    43.76\\
           J$1709+2348$   &     42.42     &   42.58    &    42.81\\
           J$2007-4434$   &     43.96     &   44.36    &    44.50\\
           J$2021-2235$   &     42.99     &   43.19    &    43.40\\
\hline
\end{tabular}
\label{tab:jetpow}
\end{table*}

There are some cases with only a few data points, so that it is not possible to draw useful inferences (e.g.  J0100$-$0200, J0706+3901, J1333+4141, J1346+3121, J1358+2658, J1612+4219, J1709+2348). While we have added significantly
to the multiwavelength database for many of these objects, the radio observations remained limited to only  1.4~GHz. It is therefore desirable that future observations focus on radio frequencies (e.g.  Richards et al. 2014, L\"ahteenm\"aki et al., in preparation). 

\section{Jet power}
To estimate the jet power, we adopted the relationships based on the radio core measurements at 15~GHz by Foschini (2014),
\begin{equation}
\log P_{\rm jet,radiative} = (12\pm2) + (0.75\pm 0.04)\log L_{\rm radio,core}
\end{equation}
and
\begin{equation}
\log P_{\rm jet,kinetic} = (6\pm2) + (0.90\pm 0.04)\log L_{\rm radio,core}.
\end{equation}
From the values calculated in the Sect.~6, we derived the radiative, kinetic (protons, electrons, magnetic field), and the total jet power for each source. The results are given in Table~\ref{tab:jetpow}. 

In some cases, it is possible to test the present results with calculations performed by modelling the SED, with the caveat that we are comparing different epochs of strongly variable sources. For example, J0948+0022 --- the first RLNLS1 to be detected at gamma rays --- had a radiative jet power of $\log P_{\rm radiative} =45.5$, while the kinetic part was estimated as $\log P_{\rm kinetic} = 46.9$ (Abdo et al. 2009a). During the 2009 MW campaign, these values ranged from 44.9 to 45.54 for the radiative power, and from 45.67 to 46.2 for the kinetic power (Abdo et al. 2009b). During more than three years of monitoring, $\log P_{\rm rad}$ spanned the interval 44.55--45.97, while $\log P_{\rm kinetic}$ was in the range 46.19--47.61 (Foschini et al. 2012). The present estimate (Table~\ref{tab:jetpow}) is an average of several measurements done directly at 15~GHz (mostly by the MOJAVE project and Effelsberg), and is reasonably consistent with the previously published values (see also Angelakis et al., in preparation). The greater values were recorded during the exceptional 2010 outburst, when J0948+0022 reached an observed luminosity of about $10^{48}$~erg~s$^{-1}$ (Foschini et al. 2011a). 

In other cases:
\begin{itemize}
\item J0324+3410: $\log P_{\rm rad}= 42.8$, $\log P_{\rm kin}= 44.3$ (averaged over one year, Abdo et al. 2009c), and $\log P_{\rm rad}= 41.29$--41.74, $\log P_{\rm kin}= 44.06$--45.14 (different states over five years monitoring, Paliya et al. 2014).
\item J0849+5108: $\log P_{\rm rad}= 45.6$ (peak during an outburst), $\log P_{\rm kin}= 45.3$(\footnote{Electrons and magnetic field only.}) (D'Ammando et al. 2012).
\item J1505+0326: $\log P_{\rm rad}= 44.0$, $\log P_{\rm kin}= 46.2$ (averaged over one year, Abdo et al. 2009c).
\item J2007$-$4434: $\log P_{\rm rad}= 42.9$, $\log P_{\rm kin}= 44.1$ (averaged over one year, Abdo et al. 2009c).
\end{itemize}

A comparison with the jet power of FSRQs and BL Lac objects (Fig.~\ref{fig:powdistr}) shows that RLNLS1s have values comparable to BL Lac objects but lower than FSRQs. The mean values are $\log P_{\rm rad}=43.35$ and $\log P_{\rm kin}=43.62$ for RLNLS1s, $\log P_{\rm rad}=45.49$ and $\log P_{\rm kin}=46.78$ for FSRQs, and $\log P_{\rm rad}=44.14$ and $\log P_{\rm kin}=45.01$ in the case of BL Lac objects. Taking into account a mean value for the masses of the central black holes of the three populations ($M_{\rm RLNLS1}=6.8\times 10^{7}M_{\odot}$, $M_{\rm FSRQs}=1.5\times 10^{9}M_{\odot}$, and $M_{\rm BL\,Lacs}=7.2\times 10^{8}M_{\odot}$) and renormalizing by $M^{1.4}$, we obtained $\log P_{\rm rad}=32.38$ and $\log P_{\rm kin}=32.65$ for RLNLS1s, $\log P_{\rm rad}=32.64$ and $\log P_{\rm kin}=33.93$ for FSRQs, and $\log P_{\rm rad}=31.74$ and $\log P_{\rm kin}=32.61$ in the case of BL Lac objects. Thus, the normalised jet power is almost the same for all the three types of AGNs, as expected (see Sect.~6), and it is also consistent with the jets from Galactic binaries (Foschini 2014).

\section{Discussion}
Since the discovery of NLS1s, there has been debate as to whether they are an intrinsically separate AGN class, or simply the low-mass tail of the distribution of Seyferts (Osterbrock \& Pogge 1985). Many authors favoured the latter hypothesis (e.g.  Grupe 2000, Mathur 2000, Botte et al. 2004). The same question has been proposed in the case of RLNLS1s (Yuan et al. 2008). The first studies following the detections at $\gamma$ rays suggested a simple mass difference (Abdo et al. 2009a,c, Foschini 2011a, 2012a, Foschini et al. 2011a, 2013). The unification of relativistic jets provided further support for this point of this view (Foschini 2011b, 2012b,c, 2014). On the basis of what we have found in this survey, with more sources and data, we can confirm that, although RLNLS1s show some peculiar observational differences with respect to the other radio-loud AGNs (the optical spectrum and the possible starburst activity), the physical characteristics inferred from the data (mass of the central black hole, Eddington ratio, spectrum, jet power) favour the hypothesis that RLNLS1s are the low-mass tail of AGNs with jets. This is one more point favouring the Livio (1997) conjecture, according to which the jet engine is the same, but the observational features are different, depending on a number of variables, such as the mass of the central accreting body, the accretion flow, and the local environment. 

In the case of RLNLS1s, the relatively lower mass of the central black hole implies variability on very short timescales, much smaller than expected only from Doppler boosting, which is exactly what is seen when the observational coverage allows it. It is known that the power spectral densities of AGNs show a break timescale, $t_{\rm b}$, separating long-term timescales from the shorter ones (McHardy et al. 2006). There are some relationships linking $t_{\rm b}$ with the mass of the central black hole, the bolometric luminosity, or the FWHM of the H$\beta$ (McHardy et al. 2006, Gonz\'alez-Mart\'in \& Vaughan 2012). By taking as representative values the averages of the selected quantities, it is possible to estimate $t_{\rm b}$, which is expected to be around minutes to hours for RLNLS1s, and hours to a few days for blazars. Indeed, hour timescales at high energies are exceptional events for blazars (e.g.  Foschini et al. 2011b,c), but are quite common for RLNLS1s as there are sufficient statistics to allow a meaningful detection (Table~10). As stated in Sect. 4.2, the claim of minute timescale X-ray variability requires further detailed study, but it is worth noting the 2--3 minute timescale variability in the optical polarisation reported by Itoh et al. (2013). 

We have observed not only flux variability, but also spectral changes, suggesting the interplay of jet and disk components (see the case of J0948+0022 in Sect.~7). At a first look, the SEDs suggest two different classes of RLNLS1s, depending on the slope of the optical/UV spectra. However, the spectral variability of some sources (e.g.  J0849+5108, J0945+1915, J0948+0022, J1159+2838) simply indicates that we are observing different states of activity of the same central engine. Indeed, the two classes do not show any difference in the mass, disk, and jet parameters. 

The lower mass of the central black holes in RLNLS1s has an important implication: the observed jet luminosity is lower than that of quasars, but comparable to that of BL Lac objects. Therefore, one could wonder why the RLNLS1s are more difficult to discover than BL Lacs? The latter are generally more luminous at X-rays than RLNL1s because the synchrotron radiation peaks in the UV/X-rays (see Fig.~\ref{fig:monochromo}, {\it bottom panel}), and indeed, BL Lac objects are more easily found in X-ray surveys (Padovani \& Giommi 1995). At $\gamma$ rays, {\it Fermi}/LAT discovered many BL Lac objects  because the instrumental characteristics of LAT favour hard sources at low fluxes: this made it easier to detect BL Lacs ($\alpha_{\rm \gamma}<1$), but not RLNLS1s ($\alpha_{\rm \gamma}>1$) (see Sect.~4.1). At radio frequencies, both RLNLS1s and BL Lac objects are weak (see Fig.~\ref{fig:monochromo}, {\it top panel}). However, Giroletti et al. (2012) noted that BL Lacs have extended radio emission, which is almost missing in RLNLS1s (e.g,. Doi et al. 2012). One possible explanation, advanced by Doi et al. (2012), is that in the case of RLNLS1s, the jet has low kinetic power because of the low mass and because it has to propagate in a gas-rich environment, while in BL Lacs the jet power is slightly greater and develops in a more rarified medium. Another possibility is to invoke the young age of RLNLS1s (Mathur 2000, Mathur et al. 2012) and, indeed, many authors made the hypothesis of a link with GPS/CSS sources, which in turn are believed to be very young radio galaxies (Oshlack et al. 2001, Komossa et al. 2006a, Gallo et al. 2006, Yuan et al. 2008, Caccianiga et al. 2014). Yet another option has been proposed by Gu \& Chen (2010): the jet activity could be intermittent, as observed in other Seyferts (e.g. Brunthaler et al. 2005, Mundell et al. 2009). Therefore, as the technological improvement of radio surveys allows better monitoring of these sources (e.g.  Square Kilometer Array, SKA), the rate of detection should increase. 

The intermittent jet should not be confused with the outburst/flare activity as observed in blazars. In the case of RLNLS1s, the periods of activity/inactivity might be separated by dramatic changes in flux. Indeed, in addition to the episodes of strong variability already described (see Sect.~7), we also note two sources where the X-ray flux was three-to-four orders of magnitude lower than the optical/UV emission (J0100$-$0200, J0706+3901). In other cases, although the SED displayed the double-humped shape typical of a domination of a relativistic jet, the lack of $\gamma$-ray detection (no new detection was reported to date) set very stringent constraints (e.g.  J0814+5609, J1031+4234, J1421+2824, J1629+4007). 
This can be compared with the lowest-known state of the BL Lac object PKS~2155$-$304 ($z=0.116$) where the changes in the X-ray flux were of about an order of magnitude and there was a shift of the synchrotron peak at lower frequencies (Foschini et al. 2008). This indicates a jet with a continuous background of emitted radiation, with superimposed outbursts and flares, as new blobs are ejected. The more dramatic changes of three-to-four orders of magnitudes observed in RLNLS1s suggests that the central engine changes its level of activity significantly: not only the jet, but also the corona seems to be strongly reduced.

Czerny et al. (2009), supported by Wu (2009), proposed a radiative instability in the accretion disk to explain the intermittent activity in young radio sources. RLNLS1s have all accretion luminosities sufficiently high to be in the radiation-pressure dominated regime (Moderski \& Sikora 1996, Ghosh \& Abramowicz 1997; see Foschini 2011b for the application to RLNLS1s), where Czerny's theory applies. The timescale of the active phase in the case of low-mass AGNs, such as NLS1s, could be very small, of the order of tens-to-hundreds of years (Czerny et al. 2009). Therefore, the low kinetic power of the jet due to the low mass of the central black hole, the short periods of activity, and a frustrating nearby environment rich in interstellar gas and photons, are the sufficient ingredients to explain the lack of extended radio relics. As suggested by Doi et al. (2012), such structures might appear only in the sources with greatest black hole masses, which in turn might be in the final stages of their cosmological evolution before changing into broad-line AGNs. 

Another possibility is the aborted jet model proposed by Ghisellini et al. (2004), which in turn could also explain the difference between radio-loud and radio-quiet AGNs. In this case, the jet has insufficient power to escape from the central black hole and falls back. The spectral characteristics in the X-ray band could be an index generally steeper than usual for Seyferts (that is $\alpha_{\rm x}\sim 1$), significant equivalent-width fluorescent iron lines, and a steeper-when-brighter behaviour of the light curves. J0324+3410 might be a good candidate, also because it is the only one with a detected Fe\,K$\alpha$ line. However, the X-ray flux and spectral index values (Online Material Table 5) do not reveal any significant trend. We note that high-flux periods have both harder and steeper indices. We can speculate that a jet might sometimes be aborted (steeper when brighter) or launched (harder when brighter). The rather obvious question is then what determines one or the other? 
 
\section{Conclusions}
We have presented a survey of 42 RLNLS1s observed from radio to $\gamma$ rays, the largest MW sample to date of this type of source. In addition to previously published data, we present here new analyses of data obtained with {\it Swift} and {\it XMM--Newton} specifically to address these sources. 
The main results of the analyses are:
\begin{itemize}
\item $\gamma$ rays: 7/42 sources (17\%) were detected at high-energy $\gamma$ rays. The average spectral index is $\alpha_{\rm \gamma}\sim 1.6$, consistent with that of FSRQs. Intraday variability has been reported in three sources.
\item X-rays: We detected 38/42 sources (90\%), with an average spectral index $\alpha_{\rm x}\sim 1.0$ and median $0.8$. We also detected variability on timescales of hours in 6 sources.
\item Intraday variability was observed also at ultraviolet/optical wavelengths in those few sources which were targets of MW campaigns. Dramatic changes both in fluxes and spectra were also observed when comparing observations on timescales of years. Infrared colours indicate that RLNLS1s are basically on the line expected from synchrotron emission, but with a significant spread towards the starburst region.
\item We observed in some sources changes of the EVPA corresponding to $\gamma$ ray activity. We detected significant changes of radio flux density only in the VLBI-cores, suggesting that the emission of $\gamma$ rays should occur close to the central black hole.
\item Comparison of monochromatic luminosities at 15~GHz, 203~nm, 1~keV, and 100~MeV with a sample of blazars (FSRQs, and BL Lac objects) suggest that RLNLS1s are the low-power tail of the quasar distribution. 
\item Some SEDs confirm the dramatic variability already apparent from the single band analysis. We modelled one case (J0948+0022) to show how the observed spectral variability can be interpreted as the interplay of the jet and accretion disk emission.
\end{itemize}

The radio coverage are still deficient, but some programs are ongoing (Richards et al. 2014, Angelakis et al. in preparation, L\"ahteenm\"aki et al. in preparation).

The main results calculated from the data are:
\begin{itemize}
\item The estimated masses of the central black holes ($10^{6-8}M_{\odot}$) and Eddington ratios (0.01--$0.49L_{\rm Edd}$) are in the range typical of NLS1s. The masses are lower than those of blazars ($10^{8-10}M_{\odot}$), indicating that we are studying a new different regime of the mass-accretion parameter space.
\item The calculated jet powers ($10^{42.6-45.6}$~erg~s$^{-1}$) are generally lower than those of FSRQs and partially overlapping, but still slightly lower than those of BL Lac objects. Once normalised by the mass of the central black holes, the jet powers of the three populations are consistent with each other, indicating the scalability of the jet.
\end{itemize}

The inferences that can be drawn from this study are that, despite the observational differences, the central engine of RLNLS1s is quite similar to that of blazars, as indicated by the scalability of the jet emission. The difficulties in finding this type of source might be due to the low observed power and an intermittent activity of the jet. Large monitoring programs with high-performance instruments (e.g.  SKA) should allow us  to greatly improve our understanding of these sources, 
which will lead to a better understanding of the more general issue of the physics of relativistic jets and how they are generated.

\onltab{
\begin{table*}
\caption{Gamma-ray spectral characteristics. Columns: (1) Name of the source; (2) Flux in the $0.1-100$~GeV energy band [$10^{-8}$~ph~cm$^{-2}$~s$^{-1}$]; (3) Photon index of the power-law model ($\Gamma=\alpha+1$); (4) Test Statistic (Mattox et al. 1996); (5) Time period; (6) Reference.}\label{tab:gray}
\centering
\begin{tabular}{lcccll}
\hline
Name & Flux & Photon Index & TS & Time Period & Reference\\
\hline
J$0324+3410$ & $6.0\pm0.7$ & $2.87\pm0.09$ & $164$ & 2008 Aug - 2011 Feb & Foschini (2011a)\\
J$0849+5108$ & $0.51\pm0.15$ & $2.0\pm0.1$ & $52$ & 2008 Aug - 2011 Feb & Foschini (2011a)\\
{}           & $2.6\pm0.2$   & $2.18\pm0.05$ & $658$ & 2008 Aug - 2012 Aug & D'Ammando et al. (2013d)\\
J$0948+0022$ & $13.7\pm0.7$ & $2.85\pm0.04$ & $1081$ & 2008 Aug - 2011 Feb & Foschini (2011a)\\
{}           & $13.6\pm0.3$ & $2.67\pm0.03$ & $2015$ & 2008 Aug - 2011 Dec & Foschini et al. (2012)\\
J$1102+2239$ & $2.0\pm0.6$ & $3.1\pm0.2$ & $32$  & 2008 Aug - 2011 Feb & Foschini (2011a)\\
J$1246+0238$ & $1.7\pm0.7$ & $3.1\pm0.3$ & $15$ & 2008 Aug - 2011 Feb & Foschini (2011a)\\
J$1505+0326$ & $7.0\pm0.6$ & $2.71\pm0.07$ & $411$ & 2008 Aug - 2011 Feb & Foschini (2011a)\\
{}           & $5.1\pm0.4$ & $2.67\pm0.06$ & $419$ & 2008 Aug - 2012 Jul & Paliya et al. (2013b)\\
{}           & $4.0\pm0.4$ & $2.60\pm0.06$ & $305$ & 2008 Aug - 2012 Nov & D'Ammando et al. (2013a)\\
J$2007-4434$ & $1.2\pm0.3$ & $2.3\pm0.1$ & $44$ & 2008 Aug - 2011 Feb & Foschini (2011a)\\
{}           & $1.7\pm0.3$ & $2.6\pm0.1$ & $68$ & 2008 Aug - 2012 Jul & Palyia et al. (2013b)\\
{}           & $1.4\pm0.3$ & $2.6\pm0.1$ & $49$ & 2008 Aug - 2012 Aug & D'Ammando et al. (2013d)\\
\hline
\end{tabular}
\end{table*}
}

\onllongtab{
\begin{landscape}
\label{tab:xray}
\begin{longtable}{lclccccrcc}
\caption{X-ray characteristics. Columns: (1) Name of the source; (2) Satellite used: C, {\it Chandra}; S, {\it Swift}; X, {\it XMM--Newton}; R, {\it ROSAT}; (3) Observing date (start, YYYY/MM/DD HH:MM; or interval of dates, if it is the result of an integration of different snapshots); (4) Exposure [ks]; (5) Photon index (power-law model, $\Gamma=\alpha+1$) or low-energy photon index (broken power-law model); (6) Break energy [keV]; (7) High-energy photon index (broken power-law model); (8) Statistics/Value/Degrees of freedom, where statistics can be $\chi^2$, or likelihood (Cash 1979), or CXA, XSS, or RASS, if the detection is extracted from the {\it Chandra}, {\it XMM--Newton}, or {\it ROSAT} catalogues; (9) Observed flux in the $0.3-10$~keV [$10^{-12}$~erg~cm$^{-2}$~s$^{-1}$]; (10) Intrinsic luminosity [$10^{44}$~erg~s$^{-1}$].}\\
\hline
Name         & Sat. & Date & Exp. & $\Gamma_1$ & $E_{\rm break}$ & $\Gamma_2$ & stat/val/dof & $F_{\rm 0.3-10\ keV}$ & $L_{\rm 0.3-10\ keV}$\\
\hline
J$0100-0200$ & C & 2003 Sep 04 11:46 & 9.2 & $2.0$(f) & {} & {} & CXA & $0.0051\pm0.0023$ & $0.0090$\\
J$0134-4258$ & S & 2007 Nov 29 07:04 & 2.3 & $2.4\pm0.4$     & {}              & {}         & $\chi^2$/4.3/3 & $2.2\pm0.5$ & $4.6$\\
{}           & S & 2008 Mar 13 01:29 & 4.0 & $1.9\pm0.2$     & {}              & {}         & $\chi^2$/6.5/7 & $2.7\pm0.5$ & $4.8$\\
{}           & S & 2008 Mar 25 04:27 & 4.3 & $2.1\pm0.2$     & {}              & {}         & $\chi^2$/1.4/4 & $1.5\pm0.3$ & $2.5$\\
{}           & S & 2008 Mar 27 04:38 & 5.0 & $1.5\pm0.4$     & {}              & {}         & $\chi^2$/3.8/3 & $1.7\pm0.6$ & $2.7$\\
{}           & S & 2008 Mar 29 01:24 & 3.3 & $2.3\pm0.2$     & {}              & {}         & $\chi^2$/3.4/6 & $2.9\pm0.4$ & $5.7$\\
{}           & S & 2008 Mar 31 06:29 & 3.2 & $1.7\pm0.2$     & {}              & {}         & $\chi^2$/8.5/9 & $4.6\pm0.7$ & $7.8$\\
{}           & X & 2008 Dec 11 20:03 & 32.4 & $1.94_{-0.09}^{+0.03}$ & $1.4_{-0.4}^{+0.1}$ & $2.29_{-0.08}^{+0.05}$ & $\chi^2$/540.4/541 & $2.9\pm0.8$ & $5.3$\\
{}           & S & 2010 Mar 21 04:38 & 2.3 & $2.4\pm0.3$     & {}              & {}         & C/48.1/64      & $1.3\pm0.3$ & $2.6$\\
{}           & S & 2010 Mar 21 06:14 & 1.7 & $2.1\pm0.3$     & {}              & {}         & C/48.2/61      & $1.7\pm0.4$ & $3.1$\\
{}           & S & 2010 Nov 21 17:06 & 1.4 & $1.6\pm0.3$     & {}              & {}         & C/60.3/64      & $2.5\pm0.9$ & $4.1$\\
{}           & S & 2010 Nov 21 21:55 & 1.0 & $1.9\pm0.3$     & {}              & {}         & C/44.3/55      & $2.7\pm0.8$ & $4.7$\\
{}           & S & 2011 Mar 20 02:01 & 1.4 & $1.7\pm0.2$     & {}              & {}         & C/86.0/107     & $4.4\pm0.9$ & $7.4$\\
{}           & S & 2011 Mar 20 03:57 & 1.3 & $1.9\pm0.2$     & {}              & {}         & C/87.2/98      & $4.0\pm0.9$ & $6.9$\\
{}           & S & 2011 Mar 25 05:32 & 3.7 & $1.9\pm0.2$     & {}              & {}         & $\chi^2$/7.0/8 & $3.5\pm0.6$ & $6.1$\\
{}           & S & 2011 Mar 26 07:06 & 2.7 & $2.0\pm0.3$     & {}              & {}         & $\chi^2$/3.8/5 & $3.4\pm0.7$ & $6.1$\\
{}           & S & 2012 Dec 06 02:38 & 2.7 & $2.3\pm0.3$     & {}              & {}         & $\chi^2$/6.9/6 & $3.6\pm0.7$ & $7.2$\\
J$0324+3410$ & S & 2006 Jul 06 00:53 & 8.3 & $2.03\pm0.06$ & {} & {} & $\chi^2$/99.1/82 & $16.1\pm0.9$ & $1.9$\\
{}           & S & 2006 Jul 09 10:51 & 2.4 & $2.0\pm0.1$ & {} & {} & $\chi^2$/38.2/27 & $15.6\pm1.4$ & $1.8$\\
{}           & S & 2006 Jul 09 23.42 & 8.8 & $2.3\pm0.1$ & $1.3_{-0.2}^{+0.4}$ & $1.8\pm0.1$ & $\chi^2$/126.3/99 & $28.2\pm9.0$ & $3.3$\\ 
{}           & S & 2007 Jul 20 16:59 & 6.4 & $2.04\pm0.05$ & {} & {} & $\chi^2$/83.9/81 & $19.2\pm0.9$ & $2.3$\\
{}           & S & 2007 Nov 04 03:08 & 1.9 & $1.9\pm0.1$ & {} & {} & $\chi^2$/20.9/12 & $12.5\pm1.4$ & $1.4$\\
{}           & S & 2007 Nov 11 02:08 & 2.2 & $2.0\pm0.1$ & {} & {} & $\chi^2$/9.9/13 & $10.5\pm1.3$ & $1.2$\\
{}           & S & 2007 Nov 25 18:02 & 2.2 & $1.9\pm0.2$ & {} & {} & $\chi^2$/22.1/10 & $7.7\pm1.2$ & $0.87$\\
{}           & S & 2007 Dec 01 04:14 & 2.4 & $2.0\pm0.1$ & {} & {} & $\chi^2$/27.1/23 & $14.9\pm1.3$ & $1.8$\\
{}           & S & 2007 Dec 06 00:18 & 6.1 & $1.92\pm0.07$ & {} & {} & $\chi^2$/50.0/47 & $11.8\pm0.6$ & $1.3$\\
{}           & S & 2007 Dec 15 09:05 & 2.5 & $2.1\pm0.1$ & {} & {} & $\chi^2$/11.0/13 & $9.6\pm1.2$ & $1.2$\\
{}           & S & 2007 Dec 23 08:18 & 2.1 & $2.3\pm0.1$ & {} & {} & $\chi^2$/13.9/24 & $15.0\pm1.1$ & $2.0$\\
{}           & S & 2007 Dec 28 06:48 & 1.3 & $2.2\pm0.2$ & {} & {} & $\chi^2$/5.4/9 & $10.8\pm1.6$ & $1.4$\\
{}           & S & 2008 Jan 04 04:25 & 2.3 & $2.0\pm0.1$ & {} & {} & $\chi^2$/21.0/21 & $13.5\pm1.3$ & $1.6$\\
{}           & S & 2008 Jan 14 08:28 & 2.3 & $2.7\pm0.2$ & $1.3_{-0.2}^{+0.4}$ & $1.8_{-0.2}^{+0.1}$ & $\chi^2$/43.2/32 & $25.1\pm8.6$ & $3.3$\\
{}           & S & 2008 Nov 16 17:35 & 5.9 & $2.5_{-0.2}^{+0.5}$ & $1.3_{-0.4}^{+0.3}$ & $1.7_{-0.1}^{+0.2}$ & $\chi^2$/38.8/30 & $13.1\pm5.7$ & $1.6$\\
{}           & S & 2009 Jul 24 03:07 & 3.2 & $1.71\pm0.08$ & {} & {} & $\chi^2$/53.2/38 & $19.2\pm1.4$ & $2.0$\\
{}           & S & 2009 Jul 27 08:13 & 2.8 & $2.1\pm0.1$ & {} & {} & $\chi^2$/55.5/30 & $14.4\pm1.3$ & $1.7$\\
{}           & S & 2009 Jul 30 08:29 & 2.9 & $1.87\pm0.07$ & {} & {} & $\chi^2$/47.3/44 & $22.5\pm1.5$ & $2.5$\\
{}           & S & 2009 Aug 02 00:46 & 3.2 & $2.4\pm0.2$ & $1.8_{-0.3}^{+1.2}$ & $1.6_{-0.7}^{+0.2}$ & $\chi^2$/50.0/32 & $16.8\pm13.5$ & $2.0$\\
{}           & S & 2009 Aug 05 00:50 & 3.3 & $2.03\pm0.08$ & {} & {} & $\chi^2$/28.0/33 & $14.0\pm0.9$ & $1.7$\\
{}           & S & 2009 Aug 08 12:30 & 2.9 & $2.2\pm0.1$ & {} & {} & $\chi^2$/28.6/31 & $14.3\pm1.1$ & $1.8$\\
{}           & S & 2010 Oct 28 21:36 & 2.9 & $1.8\pm0.1$ & {} & {} & $\chi^2$/24.1/29 & $16.0\pm1.6$ & $1.8$\\
{}           & S & 2010 Oct 29 02:16 & 2.9 & $1.9\pm0.1$ & {} & {} & $\chi^2$/18.4/19 & $11.6\pm1.0$ & $1.3$\\
{}           & S & 2010 Oct 30 02:21 & 3.1 & $2.4_{-0.2}^{+0.3}$ & $1.5\pm0.4$ & $1.6\pm0.2$ & $\chi^2$/36.6/29 & $16.4\pm8.0$ & $2.0$\\
{}           & S & 2010 Oct 31 13:40 & 3.1 & $1.90\pm0.09$ & {} & {} & $\chi^2$/33.7/33 & $15.8\pm1.1$ & $1.8$\\
{}           & S & 2010 Nov 01 07:20 & 3.3 & $2.14\pm0.09$ & {} & {} & $\chi^2$/66.4/34 & $15.1\pm1.2$ & $1.9$\\
{}           & S & 2010 Nov 02 04:18 & 3.4 & $2.0\pm0.1$ & {} & {} & $\chi^2$/46.0/29 & $12.7\pm1.0$ & $1.5$\\
{}           & S & 2010 Nov 03 09:07 & 3.3 & $2.1\pm0.1$ & {} & {} & $\chi^2$/32.9/31 & $12.8\pm0.9$ & $1.6$\\ 
{}           & S & 2010 Nov 04 20:42 & 2.9 & $1.9\pm0.1$ & {} & {} & $\chi^2$/25.1/25 & $13.1\pm1.1$ & $1.1$\\
{}           & S & 2010 Nov 05 20:46 & 2.8 & $2.01\pm0.08$ & {} & {} & $\chi^2$/39.8/39 & $19.6\pm1.4$ & $2.3$\\
{}           & S & 2010 Nov 06 20:46 & 3.0 & $1.97\pm0.09$ & {} & {} & $\chi^2$/37.9/31 & $15.8\pm1.4$ & $1.8$\\
{}           & S & 2010 Nov 07 20:57 & 1.4 & $2.1\pm0.2$ & {} & {} & $\chi^2$/9.8/5 & $9.2\pm1.7$ & $1.1$\\
{}           & S & 2010 Nov 08 21:00 & 2.9 & $2.0\pm0.1$ & {} & {} & $\chi^2$/26.8/20 & $10.9\pm0.9$ & $1.3$\\
{}           & S & 2010 Nov 09 19:32 & 2.9 & $2.1\pm0.1$ & {} & {} & $\chi^2$/29.7/29 & $14.4\pm1.2$ & $1.8$\\
{}           & S & 2010 Nov 10 21:12 & 2.4 & $2.0\pm0.1$ & {} & {} & $\chi^2$/27.0/23 & $16.3\pm1.4$ & $1.9$\\
{}           & S & 2010 Nov 11 13:00 & 3.1 & $2.07\pm0.07$ & {} & {} & $\chi^2$/51.5/45 & $22.0\pm1.3$ & $2.7$\\
{}           & S & 2010 Nov 12 01:54 & 2.8 & $2.1\pm0.1$ & {} & {} & $\chi^2$/34.4/29 & $18.3\pm1.4$ & $2.2$\\
{}           & S & 2010 Nov 13 13:10 & 3.0 & $2.0\pm0.1$ & {} & {} & $\chi^2$/30.2/29 & $22.7\pm1.9$ & $2.6$\\
{}           & S & 2010 Nov 14 05:16 & 3.0 & $1.93\pm0.07$ & {} & {} & $\chi^2$/64.3/47 & $23.3\pm1.5$ & $2.7$\\
{}           & S & 2010 Nov 15 00:47 & 1.9 & $1.9\pm0.2$ & {} & {} & $\chi^2$/6.2/8 & $11.5\pm2.1$ & $1.3$\\
{}           & S & 2010 Nov 16 13:40 & 3.1 & $1.80\pm0.08$ & {} & {} & $\chi^2$/52.6/38 & $22.3\pm1.7$ & $2.4$\\
{}           & S & 2010 Nov 17 11:54 & 3.0 & $1.9\pm0.1$ & {} & {} & $\chi^2$/29.4/24 & $20.9\pm1.9$ & $2.4$\\
{}           & S & 2010 Nov 18 07:14 & 2.8 & $1.97\pm0.09$ & {} & {} & $\chi^2$/49.9/31 & $16.3\pm1.4$ & $1.9$\\
{}           & S & 2010 Nov 19 15:50 & 3.0 & $1.9\pm0.1$ & {} & {} & $\chi^2$/27.1/23 & $11.5\pm1.0$ & $1.3$\\
{}           & S & 2010 Nov 23 04:25 & 2.6 & $2.29\pm0.08$ & {} & {} & $\chi^2$/52.7/45 & $22.6\pm1.4$ & $3.0$\\
{}           & S & 2010 Nov 24 18:57 & 2.7 & $2.09\pm0.08$ & {} & {} & $\chi^2$/33.1/43 & $22.7\pm1.6$ & $2.8$\\
{}           & S & 2010 Nov 25 19:03 & 2.9 & $2.4\pm0.1$ & $2.0_{-0.4}^{+0.7}$ & $1.6_{-0.3}^{+0.2}$ & $\chi^2$/42.2/43 & $25.5\pm10.3$ & $3.2$\\
{}           & S & 2010 Nov 26 06:18 & 3.4 & $2.13\pm0.07$ & {} & {} & $\chi^2$/59.2/51 & $19.6\pm1.2$ & $2.4$\\
{}           & S & 2010 Nov 27 17:39 & 3.0 & $2.4\pm0.2$ & $1.4_{-0.2}^{+0.4}$ & $1.6\pm0.2$ & $\chi^2$/33.3/30 & $20.8\pm6.9$ & $2.5$\\
{}           & S & 2010 Nov 28 03:17 & 3.3 & $2.4\pm0.1$ & $2.3_{-0.6}^{+0.8}$ & $1.4_{-0.7}^{+0.4}$ & $\chi^2$/35.1/40 & $22.2\pm13.6$ & $2.7$\\
{}           & S & 2010 Nov 29 03:23 & 3.3 & $2.01\pm0.08$ & {} & {} & $\chi^2$/54.3/46 & $19.8\pm1.4$ & $2.4$\\
{}           & S & 2010 Nov 30 11:46 & 3.3 & $2.13\pm0.08$ & {} & {} & $\chi^2$/35.1/38 & $17.9\pm1.2$ & $2.2$\\
{}           & S & 2011 Jul 06 23:59 & 2.0 & $2.07\pm0.08$ & {} & {} & $\chi^2$/52.2/42 & $28.2\pm1.8$ & $3.4$\\
{}           & S & 2011 Aug 07 05:40 & 2.0 & $2.05\pm0.09$ & {} & {} & $\chi^2$/45.7/32 & $22.0\pm1.7$ & $2.6$\\
{}           & S & 2011 Sep 04 00:07 & 1.9 & $2.0\pm0.1$ & {} & {} & $\chi^2$/17.0/19 & $18.3\pm1.7$ & $2.2$\\
{}           & S & 2011 Oct 03 18:10 & 2.3 & $2.0\pm0.1$ & {} & {} & $\chi^2$/15.0/15 & $10.7\pm1.2$ & $1.3$\\
{}           & S & 2011 Oct 07 01:15 & 1.9 & $1.9\pm0.1$ & {} & {} & $\chi^2$/33.9/29 & $24.1\pm2.1$ & $2.7$\\
{}           & S & 2011 Nov 05 04:50 & 2.0 & $2.0\pm0.1$ & {} & {} & $\chi^2$/11.4/22 & $17.4\pm1.7$ & $2.1$\\
{}           & S & 2011 Dec 04 15:18 & 0.2 & $1.9\pm0.4$ & {} & {} & C/45.6/52 & $14.3\pm1.9$ & $1.6$\\
{}           & S & 2011 Dec 07 01:10 & 1.5 & $1.8\pm0.2$ & {} & {} & $\chi^2$/10.8/7 & $9.5\pm1.8$ & $1.0$\\
{}           & S & 2011 Dec 28 08:48 & 1.9 & $2.1\pm0.1$ & {} & {} & $\chi^2$/17.3/20 & $21.1\pm2.1$ & $2.5$\\
{}           & S & 2012 Jan 02 01:03 & 1.3 & $2.0\pm0.2$ & {} & {} & $\chi^2$/10.3/9 & $11.1\pm1.9$ & $1.3$\\
{}           & S & 2012 Jan 30 15:51 & 2.0 & $1.8\pm0.1$ & {} & {} & $\chi^2$/18.6/15 & $12.7\pm1.6$ & $1.4$\\
{}           & S & 2012 Mar 03 08:07 & 1.5 & $1.9\pm0.2$ & {} & {} & $\chi^2$/6.3/7 & $8.2\pm1.5$ & $0.94$\\
{}           & S & 2013 Jan 13 10:57 & 3.6 & $1.9\pm0.1$ & {} & {} & $\chi^2$/30.8/27 & $11.6\pm1.1$ & $1.3$\\
{}           & S & 2013 Jan 14 07:47 & 2.8 & $2.0\pm0.1$ & {} & {} & $\chi^2$/24.2/23 & $12.7\pm1.3$ & $1.5$\\
{}           & S & 2013 Jan 15 09:18 & 3.7 & $2.0\pm0.1$ & {} & {} & $\chi^2$/31.7/30 & $11.9\pm1.0$ & $1.4$\\
{}           & S & 2013 Feb 15 07:42 & 4.0 & $1.8\pm0.1$ & {} & {} & $\chi^2$/15.6/19 & $9.5\pm1.1$ & $1.0$\\
{}           & S & 2013 Feb 15 18:50 & 3.8 & $2.0\pm0.1$ & {} & {} & $\chi^2$/23.2/20 & $9.1\pm0.9$ & $1.1$\\
{}           & S & 2013 Mar 02 16:04 & 1.9 & $2.1\pm0.1$ & {} & {} & $\chi^2$/9.7/14 & $11.4\pm1.2$ & $1.4$\\
{}           & S & 2013 Mar 05 14:35 & 2.3 & $2.6_{-0.2}^{+0.3}$ & $2.2_{-0.8}^{+0.6}$ & $1.3_{-0.7}^{+0.6}$ & $\chi^2$/20.6/18 & $18.8\pm12.5$ & $2.4$\\
J$0706+3901$ & C & 2003 Jan 22 22:33 & 9.6 &  $2.0$(f) & {} & {} & CXA & $0.0082\pm0.0025$ & $0.0020$\\
J$0713+3820$ & S & 2011-2012 & 2.0 & $2.0\pm0.2$ & {} & {} & $\chi^2$/1.0/3 & $3.4\pm0.7$ & $1.6$\\
J$0744+5149$ & S & 2011 Dec 16-19 & 4.7 & $2.4\pm0.3$ & {} & {} & C/75.3/73 & $0.68\pm0.19$ & $8.0$\\
J$0804+3853$ & S & 2010-2012 & 4.1 & $3.0\pm0.3$ & {} & {} & C/76.3/58 & $0.68\pm0.15$ & $1.6$\\
J$0814+5609$ & S & 2011 Dec 15-16 & 3.1 & $2.0\pm0.4$ & {} & {} & C/31.4/36 & $0.62\pm0.29$ & $7.1$\\
J$0849+5108$ & S & 2011-2013 & 42.1 & $1.7\pm0.1$ & {} & {} & $\chi^2$/14.8/14 & $0.59\pm0.09$ & $8.0$\\
J$0902+0443$ & S & 2012 Jun 04-06 & 4.6 & $2.0$(f) & {} & {} & UL $3\sigma$ & $<0.10$ & $<1.3$\\
J$0937+3615$ & S & 2012 Dec 18 04:04 & 3.4 & $1.3\pm0.6$ & {} & {} & C/13.2/20 & $0.46\pm0.34$ & $3.7$\\
J$0945+1915$ & S & 2012-2013 & 6.4 & $1.3\pm0.4$ & {} & {} & C/42.7/29 & $0.38\pm0.23$ & $8.5$\\
J$0948+0022$ & X & 2008 Apr 29 14:42 & 0.05 & $1.4\pm 0.3$ & {} & {} & C/49.2/52 & $9.3\pm4.3$ & $111.6$\\
{}           & S & 2008 Dec 05 02:21 & 4.2 & $1.8\pm0.2$ & {} & {} & $\chi^2$/4.0/8 & $3.6\pm0.8$ & $54.0$\\
{}           & S & 2009 Mar 26 06:21 & 4.8 & $1.7\pm0.1$ & {} & {} & $\chi^2$/24.3/23 & $7.5\pm0.9$ & $107.4$\\
{}           & S & 2009 Apr 15 08:16 & 4.4 & $1.7\pm0.1$ & {} & {} & $\chi^2$/11.3/13 & $5.2\pm0.9$ & $73.3$\\
{}           & S & 2009 May 05 10:06 & 4.8 & $1.6\pm0.1$ & {} & {} & $\chi^2$/17.6/22 & $7.6\pm1.1$ & $99.6$\\
{}           & S & 2009 May 10 12:21 & 4.9 & $1.8\pm0.1$ & {} & {} & $\chi^2$/12.8/12 & $4.2\pm0.8$ & $62.7$\\
{}           & S & 2009 May 15 03:04 & 1.4 & $1.8\pm0.3$ & {} & {} & C/51.2/57 & $2.1\pm0.4$ & $31.2$\\
{}           & S & 2009 May 25 10:25 & 5.0 & $1.7\pm0.1$ & {} & {} & $\chi^2$/6.9/11 & $4.0\pm0.6$ & $58.1$\\
{}           & S & 2009 Jun 04 03:56 & 4.5 & $1.7\pm0.2$ & {} & {} & $\chi^2$/13.9/10 & $4.1\pm1.0$ & $56.8$\\
{}           & S & 2009 Jun 14 01:04 & 3.9 & $1.6\pm0.1$ & {} & {} & $\chi^2$/11.0/11 & $5.3\pm1.0$ & $70.7$\\
{}           & S & 2009 Jun 23 10:03 & 7.7 & $1.6\pm0.1$ & {} & {} & $\chi^2$/10.5/15 & $3.5\pm0.4$ & $46.0$\\
{}           & S & 2009 Jun 24 19:29 & 4.7 & $1.5\pm0.2$ & {} & {} & $\chi^2$/5.4/7 & $3.3\pm0.9$ & $40.6$\\
{}           & S & 2009 Jul 03 12:41 & 4.2 & $1.7\pm0.3$ & {} & {} & $\chi^2$/4.2/5 & $3.0\pm1.0$ & $41.2$\\
{}           & S & 2010 Jul 03 19:30 & 1.6 & $1.4\pm0.3$ & {} & {} & $\chi^2$/1.1/2 & $5.6\pm2.2$ & $65.0$\\
{}           & S & 2011 Apr 29 04:10 & 2.0 & $1.6\pm0.4$ & {} & {} & $\chi^2$/0.82/2 & $4.0\pm1.7$ & $52.5$\\
{}           & S & 2011 May 15 03:53 & 4.7 & $1.8\pm0.2$ & {} & {} & $\chi^2$/15.8/10 & $4.0\pm0.9$ & $58.0$\\
{}           & S & 2011 May 28 01:54 & 3.6 & $1.6\pm0.2$ & {} & {} & $\chi^2$12.5/9 & $4.6\pm1.1$ & $62.1$\\
{}           & X & 2011 May 28 11:21 & 34.9 & $2.33\pm0.03$ & $1.21\pm0.06$ & $1.61\pm0.03$ & $\chi^2$/943.5/932 & $3.9\pm0.2$ & $78.5$\\
{}           & S & 2011 Jun 04 08:54 & 2.0 & $1.7\pm0.3$ & {} & {} & $\chi^2$/0.95/2 & $3.3\pm1.3$ & $45.7$\\
{}           & S & 2011 Jun 14 03:17 & 5.2 & $1.7\pm0.2$ & {} & {} & $\chi^2$/5.7/9 & $3.2\pm0.6$ & $45.6$\\
{}           & S & 2011 Jul 02 17:30 & 2.0 & $1.4\pm0.2$ & {} & {} & $\chi^2$/5.8/7 & $5.0\pm1.5$ & $59.3$\\
{}           & S & 2011 Oct 09 00:12 & 0.2 & $1.4\pm0.5$ & {} & {} & C/39.5/33 & $10.2\pm8.6$ & $117.4$\\
{}           & S & 2011 Oct 12 21:11 & 1.9 & $1.6\pm0.3$ & {} & {} & $\chi^2$/1.5/4 & $10.0\pm3.2$ & $132.7$\\
{}           & S & 2011 Nov 05 01:57 & 2.0 & $1.6\pm0.2$ & {} & {} & $\chi^2$/3.8/6 & $7.1\pm1.8$ & $95.6$\\
{}           & S & 2011 Dec 08 23:59 & 1.9 & $1.5\pm0.2$ & {} & {} & C/80.7/95 & $3.3\pm1.1$ & $41.7$\\
{}           & S & 2012 Jan 02 11:08 & 2.0 & $1.3\pm0.4$ & {} & {} & $\chi^2$/0.72/1 & $3.6\pm1.1$ & $40.5$\\
{}           & S & 2012 Jan 05 23:59 & 0.5 & $1.7\pm0.5$ & {} & {} & C/22.1/21 & $2.8\pm2.1$ & $39.5$\\
{}           & S & 2012 Jan 31 05:21 & 0.8 & $1.7\pm0.3$ & {} & {} & C/42.5/49 & $3.4\pm1.6$ & $46.4$\\
{}           & S & 2012 Feb 27 14:56 & 2.1 & $1.8\pm0.5$ & {} & {} & $\chi^2$/0.34/1 & $2.6\pm1.6$ & $38.9$\\
{}           & S & 2012 Mar 26 03:38 & 2.0 & $1.8\pm0.2$ & {} & {} & $\chi^2$/6.0/5 & $5.3\pm1.3$ & $78.9$\\
{}           & S & 2012 Mar 30 02:16 & 2.2 & $1.8\pm0.3$ & {} & {} & $\chi^2$/3.8/4 & $4.4\pm1.4$ & $66.0$\\
{}           & S & 2012 Apr 23 05:46 & 0.9 & $1.4\pm0.3$ & {} & {} & C/66.7/75 & $5.1\pm2.0$ & $61.8$\\
{}           & S & 2012 Apr 28 13:43 & 2.0 & $1.4\pm0.3$ & {} & {} & C/75.0/65 & $2.1\pm1.0$ & $25.7$\\
{}           & S & 2012 May 21 11:54 & 0.7 & $1.7\pm0.3$ & {} & {} & C/46.5/64 & $4.7\pm1.8$ & $65.4$\\
{}           & S & 2012 Jun 18 12:03 & 2.1 & $1.6\pm0.3$ & {} & {} & $\chi^2$/0.34/2 & $3.8\pm1.7$ & $50.8$\\
{}           & S & 2012 Jun 30 18:57 & 2.0 & $1.8\pm0.4$ & {} & {} & $\chi^2$/2.0/2 & $4.8\pm2.1$ & $70.6$\\
{}           & S & 2012 Nov 05 01:51 & 2.1 & $1.8_{-0.3}^{+0.4}$ & {} & {} & $\chi^2$/2.6/2 & $3.8\pm2.0$ & $55.4$\\
{}           & S & 2012 Dec 03 04:56 & 2.0 & $1.8_{-0.5}^{+0.6}$ & {} & {} & $\chi^2$/0.23/1 & $2.3\pm0.8$ & $33.4$\\
{}           & S & 2012 Dec 25 07:15 & 1.9 & $1.4\pm0.4$ & {} & {} & $\chi^2$/0.11/1 & $4.9\pm3.4$ & $58.6$\\
{}           & S & 2012 Dec 30 15:34 & 0.7 & $1.8\pm0.2$ & {} & {} & $\chi^2$/1.6/4 & $13.1\pm3.8$ & $198.5$\\
{}           & S & 2013 Jan 03 10:52 & 3.0 & $1.5\pm0.1$ & {} & {} & $\chi^2$/8.1/10 & $6.9\pm1.2$ & $85.8$\\
{}           & S & 2013 Jan 11 08:07 & 2.9 & $1.6\pm0.2$ & {} & {} & $\chi^2$/6.4/9 & $6.6\pm1.6$& $85.2$\\
{}           & S & 2013 Jan 17 08:17 & 3.3 & $1.7\pm0.1$ & {} & {} & $\chi^2$/7.1/13 & $9.2\pm1.6$ & $126.4$\\
J$0953+2836$ & S & 2012-2013 & 11.0 & $1.9\pm0.3$ & {} & {} & C/51.6/61 & $0.28\pm0.10$ & $5.8$\\
J$1031+4234$ & S & 2012 Jan-Oct & 9.2 & $1.6\pm0.3$ & {} & {} & C/56.2/41 & $0.23\pm0.10$ & $1.1$\\
J$1037+0036$ & S & 2013 Jan-Feb & 6.9 & $2.0\pm0.3$ & {} & {} & C/60.3/50 & $0.43\pm0.16$ & $7.5$\\
J$1038+4227$ & S & 2013 Jan-Feb & 7.4 & $1.7\pm0.5$ & {} & {} & C/31.4/24 & $0.16\pm0.09$ & $0.23$\\
J$1047+4725$ & S & 2012 Feb-Apr & 6.7 & $1.6\pm0.5$ & {} & {} & C/20.3/17 & $0.18\pm0.16$ & $4.6$\\
J$1048+2222$ & S & 2013 Jan-Feb & 3.1 & $2.0$(f) & {} & {} & C/10.1/16 & $0.25\pm0.11$ & $0.95$\\
J$1102+2239$ & S & 2012 Jan-Jul & 25.1 & $1.6\pm0.2$(*) & {} & {} & $\chi^2$/0.94/3 & $0.40\pm0.19$ & $1.5$\\
{}           & X & 2012 Jun 11 07:00 & 5.0 & $2.0\pm0.2$ & {} & {} & $\chi^2$/7.8/11 & $0.20\pm0.04$ & $1.7$\\
J$1110+3653$ & S & 2012 Mar-Oct & 9.0 & $1.0\pm0.5$ & {} & {} & C/21.2/17 & $0.16\pm0.12$ & $1.8$\\
J$1138+3653$ & S & 2012-2013 & 7.4 & $1.5\pm0.5$ & {} & {} & C/23.4/15 & $0.33\pm0.22$ & $1.3$\\
J$1146+3236$ & S & 2013 Jan-Apr & 7.4 & $2.1\pm0.3$ & {} & {} & C/68.7/72 & $0.44\pm0.11$ & $4.0$\\
J$1159+2838$ & S & 2012 Mar-Oct & 4.3 & $1.8\pm0.6$ & {} & {} & C/17.8/12 & $0.16\pm0.11$ & $0.21$\\
J$1227+3214$ & S & 2012 Feb-Aug & 4.0 & $1.3\pm0.2$ & {} & {} & $\chi^2$/8.8/6 & $2.6\pm0.7$ & $1.2$\\
J$1238+3942$ & S & 2012 Nov 10-14 & 5.0 & $2.0\pm0.3$ & {} & {} & C/43.6/62 & $0.66\pm0.23$ & $11.5$\\
J$1246+0238$ & S & 2012-2013 & 23.6 & $1.6\pm0.3$ & {} & {} & $\chi^2$/1.0/3 & $0.23\pm0.08$ & $0.95$\\
{}           & X & 2012 Dec 14 21:05 & 10.3 & $2.0\pm0.2$ & {} & {} & $\chi^2$/25.5/16 & $0.13\pm0.02$ & $0.76$\\
J$1333+4141$ & X & 2006 Dec 19 09:57 & 0.0055 & $2.0$(f) & {} & {} & XSS & $3.0\pm0.9$ & $4.7$\\
J$1346+3121$ & R & 1990-1991 & 0.714 & $2.0$(f) & {} & {} & RASS & $<0.11$ & $<0.20$\\
J$1348+2622$ & X & 2003 Jan 13 13:25 & 40.1 & $3.32\pm0.04$ & {} & {} & $\chi^2$/314.5/253 & $0.379\pm0.008$ & $58.8$\\
J$1358+2658$ & R & 1990-1991 & 0.531 & $2.0$(f) & {} & {} & RASS & $<0.16$ & $<0.63$\\
J$1421+2824$ & S & 2012 Aug 16-17 & 6.1 & $1.6\pm0.3$ & {} & {} & C/53.7/44 & $0.41\pm0.18$ & $4.1$\\
J$1505+0326$ & S & 2009-2012 & 24.2 & $1.8\pm0.2$ & {} & {} & $\chi^2$/8.5/7 & $0.32\pm0.07$ & $2.0$\\
{}           & X & 2012 Aug 07 20:12 & 10.8 & $2.1\pm0.1$ & $1.8\pm0.5$ & $1.4_{-0.3}^{+0.2}$ & $\chi^2$/47.4/49 & $0.37\pm0.13$ & $3.0$\\
J$1548+3511$ & X & 2011 Aug 08 00:55 & 13.9 & $2.64_{-0.08}^{+0.09}$ & $1.8_{-0.4}^{+0.3}$ & $1.7\pm0.2$ & $\chi^2$/95.3/94 & $0.41\pm0.11$ & $6.5$\\
{}           & X & 2011 Aug 20 00:04 & 17.6 & $2.55\pm0.08$ & $1.6\pm0.3$ & $1.7\pm0.2$ & $\chi^2$/103.7/126 & $0.48\pm0.11$ & $6.7$\\	
J$1612+4219$ & R & 1990-1991 & 1.0 & $2.0$(f) & {} & {} & RASS & $<0.077$ & $<0.13$\\
J$1629+4007$ & S & 2005 Apr 20 07:06 & 4.9 & $2.4\pm0.2$ & {} & {} & $\chi^2$/15.9/11 & $2.6\pm0.3$ & $7.2$\\
{}           & S & 2005 May 23 20:38 & 0.6 & $2.7\pm0.4$ & {} & {} & C/26.8/35 & $1.9\pm0.6$ & $5.7$\\
{}           & S & 2006 Jan 20 23:59 & 6.2 & $2.2\pm0.1$ & {} & {} & $\chi^2$/17.3/13 & $2.3\pm0.2$ & $5.9$\\
{}           & S & 2007 Apr 22 05:25 & 1.9 & $2.6\pm0.3$ & {} & {} & $\chi^2$/2.1/2 & $2.1\pm0.4$ & $6.1$\\
{}           & S & 2007 Apr 28 04:45 & 1.9 & $2.6\pm0.4$ & {} & {} & $\chi^2$/1.1/1 & $1.9\pm0.4$ & $5.7$\\
{}           & S & 2007 May 04 11:17 & 2.2 & $2.3\pm0.3$ & {} & {} & C/27.8/43 & $0.79\pm0.24$ & $2.1$\\
{}           & S & 2007 May 12 08:50 & 2.1 & $2.4\pm0.4$ & {} & {} & $\chi^2$/1.1/2 & $2.2\pm0.5$ & $6.0$\\
{}           & S & 2007 May 18 09:36 & 2.2 & $2.5\pm0.5$ & {} & {} & $\chi^2$/1.4/1 & $1.8\pm0.5$ & $5.1$\\
{}           & S & 2007 May 27 05:39 & 2.2 & $1.9\pm0.3$ & {} & {} & $\chi^2$/0.91/2 & $2.7\pm0.7$ & $6.5$\\
{}           & S & 2007 Jun 26 00:32 & 3.0 & $2.4\pm0.3$ & {} & {} & $\chi^2$/0.71/2 & $2.0\pm0.4$ & $5.6$\\
{}           & S & 2008 Jan 08 01:03 & 7.6 & $2.4\pm0.1$ & {} & {} & $\chi^2$/25.1/14 & $2.8\pm0.3$ & $7.6$\\
{}           & S & 2008 Jan 11 19:09 & 6.5 & $2.5\pm0.2$ & {} & {} & $\chi^2$/4.7/8 & $1.8\pm0.2$ & $5.0$\\
{}           & S & 2008 Apr 05 01:52 & 2.0 & $2.4\pm0.4$ & {} & {} & $\chi^2$/1.8/1 & $1.6\pm0.5$ & $4.4$\\
{}           & S & 2008 May 03 10:50 & 1.0 & $2.4\pm0.3$ & {} & {} & C/48.2/54 & $2.5\pm0.6$ & $6.9$\\
{}           & S & 2012 Jul 26 08:22 & 0.9 & $1.7\pm0.4$ & {} & {} & C/35.1/41 & $3.0\pm1.3$ & $6.8$\\
{}           & S & 2012 Oct 13 02:38 & 0.6 & $2.5\pm0.6$ & {} & {} & C/19.1/26 & $2.5\pm1.0$ & $7.1$\\
{}           & S & 2012 Oct 29 16:14 & 1.4 & $3.6\pm0.4$ & {} & {} & C/19.8/40 & $1.7\pm0.5$ & $6.5$\\
{}           & S & 2012 Oct 30 11:45 & 1.1 & $2.8\pm0.4$ & {} & {} & C/25.9/38 & $1.6\pm0.5$ & $4.8$\\
J$1633+4718$ & X & 2011 Jul 09 05:50 & 17.0 & $1.58\pm0.05$ & $1.8_{-0.3}^{+0.5}$ & $1.94_{-0.08}^{+0.10}$ & $\chi^2$/522.0/368 & $1.9\pm0.5$ & $0.79$\\
{}           & X & 2011 Sep 12 22:24 & 19.0 & $1.59\pm0.03$ & $2.2_{-0.3}^{+0.4}$ & $1.96_{-0.08}^{+0.10}$ & $\chi^2$/582.0/427 & $2.0\pm0.4$ & $0.83$\\
{}           & X & 2012 Jan 14 15:56 & 12.8 & $1.40\pm0.04$ & $2.2_{-0.3}^{+0.4}$ & $1.96_{-0.09}^{+0.11}$ & $\chi^2$/483.4/370 & $2.5\pm0.5$ & $0.98$\\
{}           & X & 2012 Mar 14 10:20 & 8.7 & $1.60\pm0.03$ & {} & {} & $\chi^2$/316.0/211 & $1.96\pm0.07$ & $0.76$\\	 
{}           & S & 2012 May-Jun & 1.2 & $2.0\pm0.4$ & {} & {} & C/27.5/41 & $1.6\pm0.6$ & $0.61$\\
J$1634+4809$ & S & 2011-2012 & 6.6 & $1.8\pm0.4$ & {} & {} & C/38.7/35 & $0.26\pm0.12$ & $2.4$\\
J$1644+2619$ & S & 2011 Dec 26 09:41 & 1.3 & $2.2\pm0.3$ & {} & {} & C/73.9/63 & $2.1\pm0.6$ & $1.5$\\
J$1709+2348$ & R & 1990-1991 & 8.0 & $2.0$(f) & {} & {} & RASS & $0.65\pm0.06$ & $1.3$\\
J$2007-4434$ & X & 2004 Apr 11 18:52 & 20.7 & $2.1\pm0.2$ & $0.65_{-0.08}^{+0.10}$ & $1.53\pm0.03$ & $\chi^2$/351.1/338 & $1.4\pm0.4$ & $3.9$\\
{}           & S & 2011-2013 & 58.6 & $1.7\pm0.1$ & {} & {} & $\chi^2$/13.4/23 & $0.60\pm0.06$ & $1.0$\\
{}           & X & 2012 May 01 05:32 & 19.0 & $1.69\pm0.05$ & {} & {} & $\chi^2$/166.0/146 & $0.47\pm0.02$ & $1.2$\\
{}           & X & 2012 Oct 18 21:24 & 23.2 & $1.67\pm0.03$ & {} & {} & $\chi^2$/235.2/206 & $0.69\pm0.02$ & $1.5$\\
J$2021-2235$ & X & 2005 Apr 06 14:31 & 0.011 & $2.0$(f) & {} & {} & XSS & $2.1\pm0.6$ & $2.5$\\
\hline
\end{longtable}
\end{landscape}
}

\onllongtab{
\begin{landscape}
\label{tab:uvotintegrated}
\begin{longtable}{lcccccccccccc}
\caption{{\it Swift}/UVOT Observed average magnitudes (extracted from all the data integrated). Column Exp. indicated the resulting exposure [ks].}\\
\hline
Name           & $v$            & Exp. & $b$            & Exp. & $u$            & Exp.  & $uvw1$        & Exp. & $uvm2$         & Exp. & $uvw2$         & Exp.\\
\hline
J$0134-4258$   & $16.21\pm0.03$ & 2.5  & $16.35\pm0.03$ & 2.6  & $15.17\pm0.03$ & 5.1  & $14.73\pm0.04$ & 5.3  & $14.59\pm0.04$ & 7.0  & $14.69\pm0.04$ & 16.0\\
J$0324+3410$   & $15.70\pm0.02$ & 41.0 & $16.24\pm0.03$ & 15.5 & $15.34\pm0.03$ & 17.0 & $15.61\pm0.04$ & 31.8 & $15.91\pm0.04$ & 40.1 & $15.80\pm0.04$ & 91.7\\
J$0713+3820$   & $14.94\pm0.04$ & 0.08 & $15.38\pm0.04$ & 0.12 & $14.41\pm0.04$ & 0.12 & $14.70\pm0.04$ & 1.1  & $14.92\pm0.06$ & 0.11 & $15.12\pm0.04$ & 0.47\\
J$0744+5149$   & $18.90\pm0.03$ & 0.17 & $18.57\pm0.11$ & 0.17 & $17.63\pm0.04$ & 1.2  & $17.74\pm0.06$ & 0.76 & $17.88\pm0.06$ & 1.6  & $17.87\pm0.07$ & 0.67\\
J$0804+3853$   & {}             & {}   & {}             & {}   & $17.25\pm0.04$ & 1.7  & $17.76\pm0.08$ & 0.64 & $18.11\pm0.08$ & 0.94 & $18.21\pm0.08$ & 0.79\\
J$0814+5609$   & $18.04\pm0.17$ & 0.18 & $18.40\pm0.09$ & 0.18 & $17.47\pm0.05$ & 0.52 & $17.21\pm0.07$ & 0.36 & $17.22\pm0.06$ & 1.1  & $17.14\pm0.05$ & 0.72\\
J$0849+5108$   & $17.32\pm0.04$ & 2.4  & $17.89\pm0.03$ & 2.2  & $17.60\pm0.04$ & 2.2  & $17.60\pm0.04$ & 16.8 & $17.60\pm0.04$ & 12.9 & $17.63\pm0.04$ & 12.6\\
J$0902+0443$   & $19.31\pm0.45$ & 0.27 & $19.36\pm0.22$ & 0.27 & $18.54\pm0.10$ & 0.58 & $18.45\pm0.08$ & 1.5  & $18.36\pm0.10$ & 0.82 & $18.56\pm0.08$ & 1.1\\
J$0937+3615$   & $17.42\pm0.13$ & 0.01 & $17.97\pm0.09$ & 0.01 & $17.18\pm0.08$ & 0.01 & $17.49\pm0.04$ & 2.3  & $17.60\pm0.10$ & 0.31 & $17.83\pm0.08$ & 0.40\\
J$0945+1915$   & $16.71\pm0.06$ & 0.24 & $17.01\pm0.04$ & 0.24 & $15.93\pm0.03$ & 1.5  & $15.79\pm0.04$ & 1.5  & $15.69\pm0.04$ & 1.9  & $15.85\pm0.04$ & 1.6\\
J$0948+0022$   & $17.72\pm0.03$ & 6.6  & $18.10\pm0.03$ & 6.7  & $17.38\pm0.03$ & 40.1 & $17.11\pm0.04$ & 14.2 & $17.20\pm0.04$ & 17.4 & $17.21\pm0.04$ & 26.9\\
J$0953+2836$   & $19.44\pm0.34$ & 0.37 & $19.33\pm0.13$ & 0.37 & $18.31\pm0.04$ & 6.4  & $18.20\pm0.08$ & 0.74 & $17.93\pm0.06$ & 1.4  & $18.01\pm0.05$ & 1.5\\
J$1031+4234$   & $19.39\pm0.26$ & 0.72 & $19.82\pm0.16$ & 0.72 & $19.27\pm0.13$ & 0.72 & $19.65\pm0.13$ & 1.6  & $19.62\pm0.12$ & 2.1  & $20.02\pm0.11$ & 3.0\\
J$1037+0036$   & $>19.3$        & 0.31 & $19.97\pm0.24$ & 0.31 & $19.46\pm0.21$ & 0.31 & $19.93\pm0.16$ & 1.4  & $20.00\pm0.24$ & 0.83 & $20.90\pm0.17$ & 3.6\\
J$1038+4227$   & $17.73\pm0.07$ & 0.53 & $18.27\pm0.05$ & 0.53 & $17.92\pm0.05$ & 0.67 & $18.74\pm0.07$ & 1.6  & $19.26\pm0.10$ & 1.8  & $19.83\pm0.09$ & 2.7\\
J$1047+4725$   & $19.05\pm0.21$ & 0.35 & $19.35\pm0.12$ & 0.35 & $18.48\pm0.06$ & 1.3  & $18.11\pm0.06$ & 1.6  & $17.68\pm0.06$ & 1.5  & $18.00\pm0.05$ & 1.9\\
J$1048+2222$   & $18.86\pm0.30$ & 0.17 & $18.78\pm0.13$ & 0.17 & $17.83\pm0.09$ & 0.17 & $17.66\pm0.05$ & 1.3  & $17.42\pm0.08$ & 0.52 & $17.87\pm0.07$ & 0.69\\
J$1102+2239$   & $19.37\pm0.26$ & 0.76 & $19.85\pm0.17$ & 0.76 & $19.66\pm0.07$ & 5.0  & $20.47\pm0.15$ & 5.3  & $20.86\pm0.17$ & 5.5  & $20.75\pm0.11$ & 7.4\\
J$1110+3653$   & $>19.6$        & 0.37 & $20.51\pm0.39$ & 0.37 & $20.10\pm0.26$ & 0.84 & $19.65\pm0.10$ & 3.0  & $19.67\pm0.10$ & 3.4  & $19.77\pm0.11$ & 2.1\\
J$1138+3653$   & $19.25\pm0.20$ & 0.63 & $20.08\pm0.17$ & 0.63 & $19.64\pm0.16$ & 0.63 & $20.37\pm0.17$ & 2.3  & $21.29\pm0.38$ & 2.0  & $21.60\pm0.32$ & 2.7\\
J$1146+3236$   & $18.73\pm0.15$ & 0.49 & $19.11\pm0.49$ & 0.49 & $18.33\pm0.05$ & 1.1  & $18.34\pm0.06$ & 1.3  & $18.27\pm0.07$ & 1.3  & $18.33\pm0.05$ & 2.5\\
J$1159+2838$   & $18.13\pm0.16$ & 0.24 & $18.72\pm0.12$ & 0.24 & $17.82\pm0.05$ & 0.93 & $18.23\pm0.07$ & 1.0  & $18.44\pm0.12$ & 0.49 & $18.66\pm0.07$ & 1.4\\
J$1227+3214$   & $18.22\pm0.37$ & 0.07 & $19.01\pm0.36$ & 0.07 & $19.07\pm0.08$ & 2.3  & $19.14\pm0.20$ & 0.33 & $19.88\pm0.30$ & 0.45 & $19.48\pm0.14$ & 0.82\\
J$1238+3942$   & $>18.9$        & 0.21 & $>19.8$        & 0.21 & $18.99\pm0.24$ & 0.21 & $18.65\pm0.13$ & 0.42 & $18.54\pm0.16$ & 0.34 & $18.66\pm0.05$ & 3.5\\
J$1246+0238$   & $18.22\pm0.09$ & 0.71 & $18.46\pm0.06$ & 0.71 & $17.61\pm0.03$ & 4.3  & $17.53\pm0.04$ & 7.6  & $17.56\pm0.05$ & 2.8  & $17.64\pm0.04$ & 7.5\\
J$1421+2824$   & $17.63\pm0.13$ & 0.13 & $17.88\pm0.08$ & 0.17 & $16.87\pm0.06$ & 0.18 & $16.69\pm0.04$ & 4.7  & $16.70\pm0.07$ & 0.28 & $16.64\pm0.05$ & 0.51\\
J$1505+0326$   & $19.02\pm0.13$ & 1.0  & $19.65\pm0.11$ & 1.0  & $18.68\pm0.04$ & 4.4  & $18.42\pm0.04$ & 6.3  & $18.67\pm0.05$ & 6.6  & $18.44\pm0.05$ & 4.2\\
J$1629+4007$   & $18.18\pm0.05$ & 3.0  & $18.37\pm0.04$ & 2.6  & $17.32\pm0.04$ & 9.1  & $17.04\pm0.03$ & 7.9  & $16.82\pm0.04$ & 11.7 & $16.84\pm0.04$ & 10.8\\
J$1633+4718$   & {}             & {}   & {}             & {}   & $16.47\pm0.05$ & 0.27 & $16.83\pm0.06$ & 0.30 & $16.89\pm0.06$ & 0.59 & {}             & {}\\
J$1634+4809$   & $>19.1$        & 0.14 & $20.00\pm0.35$ & 0.14 & $19.43\pm0.30$ & 0.14 & $19.82\pm0.16$ & 1.1  & $19.96\pm0.11$ & 4.1  & $20.15\pm0.19$ & 0.90\\
J$1644+2619$   & $17.88\pm0.26$ & 0.08 & $18.22\pm0.14$ & 0.08 & $16.83\pm0.08$ & 0.08 & $16.98\pm0.08$ & 0.17 & $16.90\pm0.09$ & 0.20 & $16.95\pm0.05$ & 0.70\\
J$2007-4434$   & $18.77\pm0.06$ & 6.1  & $19.49\pm0.05$ & 6.5  & $18.80\pm0.03$ & 38.2 & $19.40\pm0.04$ & 36.1 & $19.92\pm0.06$ & 21.2 & $20.10\pm0.06$ & 14.2\\
\hline
\end{longtable}
\end{landscape}
}

\onllongtab{
\begin{landscape}
\label{tab:radiovlbi}
\begin{longtable}{clccccccccccc}
\caption{Characteristics at radio frequencies (VLBI). Columns: (1) Frequency; (2) date; (3) core size [mas]; (4) core flux density [Jy]; (5) total flux density from clean [Jy]; (6) total flux density from model fitting [Jy]; (7) rms after clean [mJy]; (8) rms after model fitting [mJy]; (9) core axial ratio; (10) core position angle; (11) jet direction [deg]; (12) jet width [mas]; (13) MOJAVE EVPA [deg] (Lister et al. 2009, 2013).}\\
\hline
Frequency & Date & Core Size & Core Flux & Total Flux C & Total Flux M & RMS C & RMS M & Axes Ratio & Pos Angle & Direction & Width & EVPA\\
\hline
\multicolumn{13}{c}{J$0324+3410$} \\
\hline
15~GHz & 2010 Oct 15 & 0.187 & 0.222 & 0.300 & 0.307 & 0.153 & 0.177 & 0.3 & -51.5 & -49.9 & 0.070 & 35.0\\
{}     & 2011 Mar 05 & 0.056 & 0.337 & 0.400 & 0.407 & 0.155 & 0.198 & 1.0 & {} & -61.6 & 0.056 & 26.0\\
{}     & 2011 May 26 & 0.029 & 0.237 & 0.336 & 0.344 & 0.171 & 0.185 & 1.0 & {} & -46.5 & 0.029 & 31.0\\
{}     & 2011 Jul 15 & {} & 0.169 & 0.286 & 0.294 & 0.156 & 0.183 & 1.0 & {} & -50.8 & {} & 31.0\\
{}     & 2011 Dec 29 & {} & 0.232 & 0.373 & 0.379 & 0.182 & 0.225 & 1.0 & {} & -53.4 & {} & 27.0\\
{}     & 2012 Jul 12 & 0.061 & 0.153 & 0.229 & 0.237 & 0.166 & 0.197 & 1.0 & {} & -50.9 & 0.061 & 33.0\\
{}     & 2012 Dec 23 & {} & 0.146 & 0.258 & 0.265 & 0.171 & 0.185 & 1.0 & {} & -57.8 & {} & 25.0\\
{}     & 2013 Jul 08 & 0.027 & 0.131 & 0.213 & 0.219 & 0.161 & 0.174 & 1.0 & {} & -53.3 & 0.027 & 33.0\\
\hline
\multicolumn{13}{c}{J$0814+5609$} \\
\hline
2~GHz & 2010 Mar 23 & 1.338 & 0.024 & 0.029 & 0.031 & 0.227 & 0.222 & 0.0 & 78.5 & -96.3 & 1.274 & {}\\
8~GHz & 2010 Mar 23 & 0.282 & 0.034 & 0.032 & 0.036 & 0.199 & 0.153 & 0.2 & -84.3 & -89.5 & 0.280 & {}\\
\hline
\multicolumn{13}{c}{J$0849+5108$} \\
\hline
15~GHz & 2011 May 26 & 0.009 & 0.183 & 0.242 & 0.244 & 0.168 & 0.206 & 1.0 & {} & -77.9 & 0.009 & 163.0\\
{}     & 2011 Jul 15 & 0.060 & 0.165 & 0.211 & 0.215 & 0.168 & 0.180 & 1.0 & {} & -82.5 & 0.060 & 105.0\\
{}     & 2012 Jan 02 & 0.064 & 0.186 & 0.229 & 0.233 & 0.155 & 0.182 & 1.0 & {} & -79.4 & 0.064 & 174.0\\
{}     & 2012 Jun 25 & 0.061 & 0.265 & 0.305 & 0.308 & 0.105 & 0.131 & 1.0 & {} & -80.9 & 0.061 & 169.0\\
{}     & 2012 Nov 02 & 0.034 & 0.281 & 0.310 & 0.314 & 0.132 & 0.151 & 1.0 & {} & -65.8 & 0.034 & 159.0\\
{}     & 2013 Jan 21 & 0.050 & 0.218 & 0.247 & 0.252 & 0.125 & 0.136 & 1.0 & {} & -78.9 & 0.050 & 168.0\\
{}     & 2013 Jul 08 & 0.045 & 0.333 & 0.357 & 0.361 & 0.153 & 0.165 & 1.0 & {} & -76.7 & 0.045 & 24.0\\
\hline
\multicolumn{13}{c}{J$0902+0443$} \\
\hline
5~GHz & 2013 Apr 08 & 1.040 & 0.071 & 0.076 & 0.109 & 4.210 & 1.014 & 0.3 & 47.7 & -110.5 & 0.575 & {}\\
8~GHz & 2013 Apr 08 & 0.780 & 0.069 & 0.051 & 0.095 & 4.269 & 0.383 & 0.5 & 18.0 & -122.1 & 0.457 & {}\\
\hline
\multicolumn{13}{c}{J$0948+0022$} \\
\hline
15~GHz & 2009 May 28 & 0.025 & 0.420 & 0.434 & 0.436 & 0.155 & 0.178 & 1.0 & {} & -157.6 & 0.025 & 131.0\\
{}     & 2009 Jul 23 & 0.022 & 0.326 & 0.338 & 0.340 & 0.174 & 0.175 & 1.0 & {} & -155.2 & 0.022 & 146.0\\
{}     & 2009 Dec 10 & {} & 0.381 & 0.383 & 0.387 & 0.141 & 0.155 & 1.0 & {} & -147.2 & {} & {}\\
{}     & 2010 Sep 17 & 0.027 & 0.531 & 0.531 & 0.537 & 0.158 & 0.188 & 1.0 & {} & -152.3 & 0.027 & 51.0\\
{}     & 2010 Nov 04 & 0.045 & 0.487 & 0.494 & 0.496 & 0.168 & 0.251 & 1.0 & {} & -153.5 & 0.045 & 49.0\\
{}     & 2010 Nov 29 & 0.029 & 0.477 & 0.478 & 0.483 & 0.147 & 0.164 & 1.0 & {} & -151.2 & 0.029 & 49.0\\
{}     & 2011 Feb 20 & 0.014 & 0.602 & 0.619 & 0.623 & 0.155 & 0.207 & 1.0 & {} & -179.0 & 0.014 & 19.0\\
{}     & 2011 May 26 & 0.029 & 0.652 & 0.664 & 0.665 & 0.181 & 0.198 & 1.0 & {} & -171.3 & 0.029 & 49.0\\
{}     & 2011 Jun 24 & 0.023 & 0.655 & 0.669 & 0.671 & 0.151 & 0.190 & 1.0 & {} & -158.3 & 0.023 & 51.0\\
{}     & 2011 Sep 12 & 0.032 & 0.447 & 0.455 & 0.458 & 0.182 & 0.195 & 1.0 & {} & -169.9 & 0.032 & 40.0\\
{}     & 2011 Dec 12 & 0.033 & 0.366 & 0.380 & 0.382 & 0.181 & 0.215 & 1.0 & {} & -162.2 & 0.033 & 77.0\\
{}     & 2012 Jul 12 & 0.056 & 0.321 & 0.328 & 0.330 & 0.200 & 0.219 & 1.0 & {} & -171.8 & 0.056 & 98.0\\
{}     & 2012 Nov 11 & 0.026 & 0.334 & 0.338 & 0.340 & 0.145 & 0.164 & 1.0 & {} & -170.1 & 0.026 & 82.0\\
{}     & 2012 Dec 10 & 0.016 & 0.379 & 0.381 & 0.388 & 0.139 & 0.157 & 1.0 & {} & -162.6 & 0.016 & 101.0\\
{}     & 2013 Jan 21 & 0.033 & 0.480 & 0.485 & 0.487 & 0.128 & 0.137 & 1.0 & {} & -172.7 & 0.033 & 125.0\\
{}     & 2013 May 05 & 0.041 & 0.862 & 0.866 & 0.867 & 0.164 & 0.194 & 1.0 & {} & -159.4 & 0.041 & 102.0\\
{}     & 2013 Jul 30 & 0.053 & 0.420 & 0.421 & 0.423 & 0.188 & 0.223 & 1.0 & {} & -171.3 & 0.053 & 92.0\\
\hline
\multicolumn{13}{c}{J$1505+0326$} \\
\hline
2~GHz & 1995 Jul 15 & 0.369 & 0.860 & 0.855 & 0.901 & 2.266 & 1.836 & 1.0 & {} & 85.9 & 0.369 & {}\\
{}    & 1997 Jan 10 & {} & 0.574 & 0.571 & 0.587 & 1.392 & 1.296 & 1.0 & {} & 82.2 & {} & {}\\
{}    & 2000 May 22 & 0.930 & 0.496 & 0.548 & 0.567 & 0.931 & 0.698 & 0.0 & -37.0 & 104.7 & 0.352 & {}\\
{}    & 2002 Sep 25 & 0.842 & 0.560 & 0.579 & 0.589 & 1.270 & 1.101 & 0.4 & -56.0 & 94.2 & 0.703 & {}\\
{}    & 2006 Jul 11 & 0.851 & 0.513 & 0.541 & 0.557 & 2.196 & 1.926 & 0.2 & -25.4 & 111.9 & 0.224 & {}\\
{}    & 2006 Sep 13 & 0.634 & 0.389 & 0.444 & 0.457 & 0.975 & 0.755 & 0.0 & -37.9 & 120.4 & 0.083 & {}\\
{}    & 2008 Dec 17 & 0.823 & 0.370 & 0.376 & 0.391 & 1.477 & 1.209 & 0.0 & -44.1 & 115.8 & 0.258 & {}\\
{}    & 2010 Oct 13 & 0.591 & 0.564 & 0.581 & 0.598 & 1.018 & 0.791 & 0.7 & -46.3 & 97.6 & 0.490 & {}\\
{}    & 2012 Oct 03 & 0.729 & 0.538 & 0.532 & 0.560 & 1.572 & 0.778 & 0.4 & -16.0 & 102.2 & 0.290 & {}\\
8~GHz & 1995 Jul 15 & 0.210 & 0.987 & 1.002 & 1.048 & 2.441 & 1.279 & 0.7 & -42.3 & 119.9 & 0.147 & {}\\
{}    & 1997 Jan 10 & 0.014 & 0.508 & 0.530 & 0.558 & 1.364 & 0.927 & 0.0 & 73.5  & 108.6 & 0.014 & {}\\
{}    & 2000 May 22 & 0.300 & 0.625 & 0.648 & 0.673 & 1.262 & 0.626 & 0.5 & -35.0 & 129.0 & 0.150 & {}\\
{}    & 2002 Sep 25 & 0.192 & 0.753 & 0.760 & 0.784 & 1.470 & 0.994 & 0.0 & -47.5 & 107.1 & 0.097 & {}\\
{}    & 2006 Jul 11 & 0.417 & 0.580 & 0.542 & 0.580 & 4.709 & 4.707 & 0.6 & 2.6 & 0.0 & 0.416 & {}\\
{}    & 2006 Sep 13 & 0.235 & 0.478 & 0.480 & 0.495 & 0.704 & 0.479 & 0.5 & -55.7 & 121.1 & 0.145 & {}\\
{}    & 2008 Dec 17 & 0.194 & 0.502 & 0.496 & 0.526 & 2.361 & 1.455 & 0.0 & 86.3 & 123.2 & 0.169 & {}\\
{}    & 2010 Oct 13 & 0.321 & 0.554 & 0.546 & 0.568 & 1.041 & 1.051 & 1.0 & {} & 143.8 & 0.321 & {}\\
{}    & 2012 Oct 03 & 0.192 & 0.561 & 0.596 & 0.616 & 1.316 & 0.652 & 0.0 & -39.8 & 106.4 & 0.076 & {}\\
{}    & 2013 May 18 & 0.409 & 0.429 & 0.502 & 0.508 & 0.427 & 0.457 & 0.5 & -38.8 & 139.3 & 0.232 & {}\\
15~GHz & 2010 Oct 15 & 0.154 & 0.372 & 0.502 & 0.504 & 0.166 & 0.201 & 0.4 & -58.1 & 119.4 & 0.090 & 19.0\\
{}    & 2010 Oct 25 & 0.178 & 0.366 & 0.485 & 0.487 & 0.143 & 0.188 & 0.5 & -49.8 & 118.3 & 0.100 & 18.0\\
{}    & 2011 May 26 & 0.159 & 0.348 & 0.427 & 0.429 & 0.187 & 0.233 & 0.0 & -61.0 & 118.3 & 0.086 & 179.0\\
{}    & 2012 Jan 02 & 0.122 & 0.399 & 0.443 & 0.447 & 0.152 & 0.164 & 0.3 & -64.1 & 121.2 & 0.075 & 50.0\\
{}    & 2012 Sep 27 & 0.203 & 0.577 & 0.584 & 0.590 & 0.151 & 0.318 & 0.0 & -63.7 & 112.0 & 0.135 & 81.0\\
{}    & 2013 May 05 & 0.111 & 0.401 & 0.498 & 0.500 & 0.160 & 0.204 & 0.6 & -76.7 & 120.1 & 0.094 & 22.0\\
22~GHz & 2002 Dec 26 & {} & 0.456 & 0.559 & 0.593 & 0.838 & 1.972 & 1.0 & {} & 125.2 & {} & {}\\
{}    & 2003 May 22 & 0.124 & 0.421 & 0.437 & 0.452 & 0.849 & 2.162 & 0.0 & 79.5 & 112.7 & 0.122 & {}\\
43~GHz & 2002 Dec 26 & 0.042 & 0.397 & 0.423 & 0.447 & 1.074 & 1.826 & 0.0 & 85.1 & 107.3 & 0.041 & {}\\
\hline
\multicolumn{13}{c}{J$2007-4434$} \\
\hline
8~GHz & 2002 Jan 31 & 0.425 & 0.110 & 0.072 & 0.110 & 10.647 & 9.018 & 1.0 & {} & 0.0 & 0.425 & {}\\
\hline
\end{longtable}
\end{landscape}
}

\onllongtab{
\begin{landscape}
\label{tab:spectralindices}
\begin{longtable}{lccccccccc}
\caption{Spectral indices ($S_{\rm \nu}\propto \nu^{-\alpha}$) at different frequencies. The index $\alpha_{\rm r}$ was calculated by using flux densities at 1.4 and 5~GHz when possible; otherwise, we used the available frequencies (mostly 325 MHz and 1.4 GHz, but also 150, 153, 352, and 843~MHz, 8.35 and 37~GHz). An asterisk on the name indicated the detection at frequencies $<1.4$~GHz.}\\
\hline
Name & $\alpha_{\rm r}$ & $\alpha_{\rm 5-15\,GHz}$ & $\alpha_{\rm 15-37\,GHz}$ & $\alpha_{\rm 37-142\,GHz}$ & $\alpha_{\rm IR}$   & $\alpha_{\rm o}$ & $\alpha_{\rm uv}$ & $\alpha_{\rm x}$ & $\alpha_{\rm \gamma}$\\
\hline
J$0100-0200$ & {}            & {}            & {}             & {}            & $1.37\pm0.04$ & $1.18\pm0.06$  & {}             & {}            & {}\\
J$0134-4258^{*}$ & $0.10\pm0.04$ & {}            & {}             & {}            & $1.14\pm0.03$ & $-0.49\pm0.02$  & $0.36\pm0.32$  & $0.98\pm0.05$ & {}\\
J$0324+3410^{*}$ & $0.26\pm0.03$ & $0.24\pm0.01$ & $-0.57\pm0.06$ & $0.10\pm0.06$ & $0.96\pm0.02$ & $0.59\pm0.07$ & $-0.41\pm0.32$ & $0.83\pm0.02$ & $1.87\pm0.09$\\
J$0706+3901$ & {}            & {}            & {}             & {}            & $1.44\pm0.04$ & $1.01\pm0.15$ & {}             & {}            & {}\\
J$0713+3820^{*}$ & $0.58\pm0.12$ & {}            & {}             & {}            & $1.06\pm0.02$ & $0.18\pm0.02$ & $1.43\pm0.31$  & $1.0\pm0.2$   & {}\\
J$0744+5149$ & {}            & {}            & {}             & {}            & $1.30\pm0.05$ & $0.35\pm0.17$ & $0.51\pm0.51$  & $1.4\pm0.3$   & {}\\
J$0804+3853$ & {}            & {}            & {}             & {}            & $1.21\pm0.03$ & $1.75\pm0.01$ & $1.71\pm0.65$ & $2.0\pm0.3$   & {}\\
J$0814+5609^{*}$ & $0.38\pm0.01$ & {}            & {}             & {}            & $0.92\pm0.08$ & $-0.40\pm0.02$  & $-0.22\pm0.47$ & $1.0\pm0.4$  & {}\\
J$0849+5108^{*}$ & $0.21\pm0.01$ & $-0.25\pm0.01$& $-0.64\pm0.05$ & {}            & $0.70\pm0.09$ & $1.49\pm0.02$ & $0.42\pm0.31$ & $0.7\pm0.1$ & $1.1\pm0.1$\\
J$0902+0443$ & $0.07\pm0.01$ & {}            & {}             & {}            & $1.45\pm0.06$ & $1.20\pm0.05$ & $0.71\pm0.65$ & {}         & {}\\
J$0937+3615$ & {}            & {}            & {}             & {}            & $1.36\pm0.03$ & $1.78\pm0.02$ & $1.89\pm0.47$ & $0.3\pm0.6$ & {}\\
J$0945+1915$ & {}            & {}            & {}             & {}            & $1.19\pm0.03$ & $0.94\pm0.07$ & $0.66\pm0.31$ & $0.3\pm0.4$ & {}\\
J$0948+0022$ & $-0.28\pm0.01$& $-0.71\pm0.01$& $-0.33\pm0.01$ & $0.10\pm0.03$ & $1.33\pm0.06$ & $0.39\pm0.03$ & $0.29\pm0.32$ & $0.56\pm0.05$ & $1.7\pm0.1$\\
J$0953+2836$ & {}            & {}            & {}             & {}            & $0.67\pm0.18$ & $0.62\pm0.01$ & $-0.16\pm0.53$ & $0.9\pm0.3$ & {}\\
J$1031+4234^{*}$ & $-0.40\pm0.06$  & {}            & {}             & {}            & $1.47\pm0.69$ & $1.43\pm0.04$ & $2.06\pm0.99$ & $0.6\pm0.3$ & {}\\
J$1037+0036$ & {}            & {}            & {}             & {}            & $1.00\pm0.25$ & $0.89\pm0.04$ & $3.73\pm1.41$ & $1.0\pm0.3$ & {}\\
J$1038+4227$ & {}            & {}            & {}             & {}            & $1.36\pm0.11$ & $1.34\pm0.01$ & $4.77\pm0.65$ & $0.7\pm0.5$ & {}\\
J$1047+4725^{*}$ & $0.33\pm0.01$ & {}            & {}             & {}            & $1.52\pm0.13$ & $0.89\pm0.05$ & $0.14\pm0.44$ & $0.6\pm0.5$ & {}\\
J$1048+2222$ & {}            & {}            & {}             & {}            & $1.20\pm0.07$ & $0.70\pm0.02$ & $1.33\pm0.48$ & {} & {}\\
J$1102+2239$ & {}            & {}            & {}             & {}            & $1.35\pm0.04$ & $1.96\pm0.04$ & $1.66\pm1.08$ & $0.8\pm0.2$ & $2.1\pm0.2$\\
J$1110+3653$ & {}            & {}            & {}             & {}            & $0.65\pm0.48$ & $-0.20\pm0.17$  & $0.95\pm0.87$ & $0.0\pm0.5$ & {}\\
J$1138+3653^{*}$ & $0.50\pm0.09$ & {}            & {}             & {}            & $1.05\pm0.13$ & $1.80\pm0.07$ & $5.27\pm2.01$ & $0.5\pm0.5$ & {}\\
J$1146+3236^{*}$ & $0.38\pm0.09$ & {}            & {}             & {}            & $0.79\pm0.23$ & $0.76\pm0.05$ & $0.51\pm0.44$ & $1.1\pm0.3$ & {}\\
J$1159+2838$ & {}            & {}            & {}             & {}            & $1.76\pm0.03$ & $0.04\pm0.19$ & $2.17\pm0.56$ & $0.8\pm0.6$ & {}\\
J$1227+3214$ & $-1.04\pm0.07$  & {}            & {}             & {}            & $1.32\pm0.02$ & $3.76\pm0.03$ & $1.87\pm1.49$ & $0.3\pm0.2$ & {}\\
J$1238+3942$ & {} & {}            & {}             & {}            & $0.77\pm0.19$ & $0.62\pm0.28$ & $0.59\pm0.76$ & $1.0\pm0.3$ & {}\\
J$1246+0238$ & $0.55\pm0.06$ & {}            & {}             & {}            & $1.34\pm0.11$ & $0.04\pm0.04$ & $0.87\pm0.32$ & $0.8\pm0.3$ & $2.1\pm0.3$\\
J$1333+4141$ & {}            & {}            & {}             & {}            & $1.63\pm0.03$ & $0.48\pm0.04$ & {}            & {}          & {}\\
J$1346+3121$ & {}            & {}            & {}             & {}            & $1.26\pm0.07$ & $0.60\pm0.05$ & {}            & {}          & {}\\
J$1348+2622$ & {}            & {}            & {}             & {}            & $1.10\pm0.21$ & $0.00\pm0.16$  & {}            & $2.32\pm0.04$ & {}\\
J$1358+2658$ & {}            & {}            & {}             & {}            & $1.31\pm0.04$ & $0.97\pm0.07$ & {}            & {}          & {}\\
J$1421+2824$ & $-0.21\pm0.01$& {}            & {}             & {}            & $1.26\pm0.06$ & $0.19\pm0.03$ & $0.37\pm0.36$ & $0.6\pm0.3$ & {}\\
J$1505+0326$ & $-0.31\pm0.01$& $0.05\pm0.01$ & $0.39\pm0.07$  & {}            & $1.51\pm0.06$ & $1.31\pm0.05$ & $0.21\pm0.35$ & $0.6\pm0.2$ & $1.65\pm0.06$\\
J$1548+3511^{*}$ & $0.26\pm0.01$ & {}        & {}             & {}            & $1.31\pm0.05$ & $0.06\pm0.04$ & {}            & $0.7\pm0.2$ & {}\\
J$1612+4219$ & {}            & {}            & {}             & {}            & $1.88\pm0.03$ & $0.54\pm0.07$ & {}            & {}          & {}\\
J$1629+4007$ & $-0.68\pm0.02$& {}            & {}             & {}            & $1.23\pm0.04$ & $-0.13\pm0.04$ & $-0.18\pm0.32$ & $1.26\pm0.07$ & {}\\
J$1633+4718$ & $0.42\pm0.01$ & {}            & {}             & {}            & $1.57\pm0.02$ & $0.92\pm0.01$ & {}            & $0.9\pm0.2$ & {}\\
J$1634+4809$ & {}            & {}            & {}             & {}            & $1.42\pm0.08$ & $0.78\pm0.05$ & $1.80\pm1.50$ & $0.8\pm0.4$ & {}\\
J$1644+2619$ & $-0.06\pm0.01$& {}            & {}             & {}            & $1.14\pm0.07$ & $1.33\pm0.01$ & $-0.17\pm0.53$ & $1.2\pm0.3$ & {}\\
J$1709+2348$ & {}            & {}            & {}             & {}            & $1.23\pm0.08$ & $1.24\pm0.08$             & {}            & {}          & {}\\
J$2007-4434^{*}$ & $0.41\pm0.01$ & {}            & {}             & {}            & $1.23\pm0.06$ & $1.49\pm0.03$ & $3.03\pm0.40$ & $0.65\pm0.05$ & $1.5\pm0.2$\\
J$2021-2235^{*}$ & $0.50\pm0.07$  & {}            & {}             & {}            & $1.61\pm0.03$ & {}            & {}            & {}            & {} \\
\hline
\end{longtable}
\end{landscape}
}

\onllongtab{
\begin{landscape}
\label{tab:variability}
\begin{longtable}{ccccccccc}
\caption{Shortest Variability at optical-to-$\gamma$ ray frequencies. For each source, it is indicated the $\tau$ [days] and, in the second row between parentheses, the significance of the flux change [$\sigma$].}\\
\hline
Name         & $\gamma$ rays & X-rays & $uvw2$ & $uvm2$ & $uvw1$ & $u$ & $b$ & $v$\\
\hline
J$0134-4258$ & {}            & $-0.071\pm0.025$ & {} & {} & {} & {} & {} & {}\\
{}           & {}            & (3.8) & {} & {} & {} & {} & {} & {}\\
J$0324+3410$ & $0.10\pm0.03^a$ & $-0.079\pm0.012$ & $0.09\pm0.03$  & $0.10\pm0.04$  & $<0.43$ & $7\pm3$  & $<0.7$ & $<0.28$\\
{}           & (3.8)  & (4.4) & (10.7) & (7.6) & (4.1) & (6.1) & (4.0) & (3.6)\\
J$0849+5108$ & $12\pm8^b$  & $<18$ & {}  & {} & $<0.27$ & {} & {} & {}\\
{}           & (4.7) & (3.5) & {}  & {} & (3.6) & {} & {} & {}\\
J$0948+0022$ & $<0.8^b$  & $<0.21$ & $0.12\pm0.07$ & $0.09\pm0.05$ & $0.08\pm0.03$ & $0.07\pm0.03$ & $0.06\pm0.03$ & $0.05\pm0.03$ \\
{}           & (5.4) & (5.2) & (6.3) & (6.1) & (7.3) & (7.9) & (7.8) & (5.7) \\
J$0953+2836$ & {}            & {} & {} & {} & $<18$ & {} & {} & {}\\
{}           & {}            & {} & {} & {} & (3.1) & {} & {} & {}\\
J$1031+4234$ & {}            & {} & {} & {} & {} & $<0.15$ & {} & {}\\
{}           & {}            & {} & {} & {} & {} & (3.0) & {} & {}\\
J$1038+4227$ & {}            & {} & {} & {} & {} & {} & $<12$ & {}\\
{}           & {}            & {} & {} & {} & {} & {} & (3.4) & {}\\
J$1047+4725$ & {}            & {} & $<0.83$ & {}  & {}  & {} & {} & {}\\
{}           & {}            & {} & (3.4) & {}  & {}  & {} & {} & {}\\
J$1102+2239$ & {}            & {} & {} & {} & {} & $<0.24$ & {} & {}\\
{}           & {}            & {} & {} & {} & {} & (3.1) & {} & {}\\
J$1238+3942$ & {}            & {} & $<0.58$ & {}  & {} & {} & {} & {}\\
{}           & {}            & {} & (3.4) & {}  & {} & {} & {} & {}\\
J$1505+0326$ & $1.3\pm0.5^b$   & {} & {} & {} & {} & {} & {} & $<0.04$\\
{}           & (6.6)         & {} & {} & {} & {} & {} & {} & (3.8)\\
J$1629+4007$ & {}            & $-0.12\pm0.02$ & $<0.49$ & {} & {} & $<0.17$ & $<0.23$  & $<0.09$ \\
{}           & {}            & (3.5) & (3.7) & {} & {} & (4.1) & (3.2) & (3.6)\\
J$2007-4434$ & $6\pm2^b$  & $<0.19$ & $<0.12$ & {} & {} & $<0.30$ & {} & $<0.14$ \\
{}           & (12)         & (3.3) & (3.0) & {} & {} & (3.1) & {} & (3.1)\\
\hline
\end{longtable}
\begin{list}{}{}
\item[$^{\mathrm{a}}$] Palyia et al. (2014).
\item[$^{\mathrm{b}}$] Foschini (2011a).
\end{list}
\end{landscape}
}

\onlfig{
\begin{figure*}
\centering
\includegraphics[angle=270,scale=0.35]{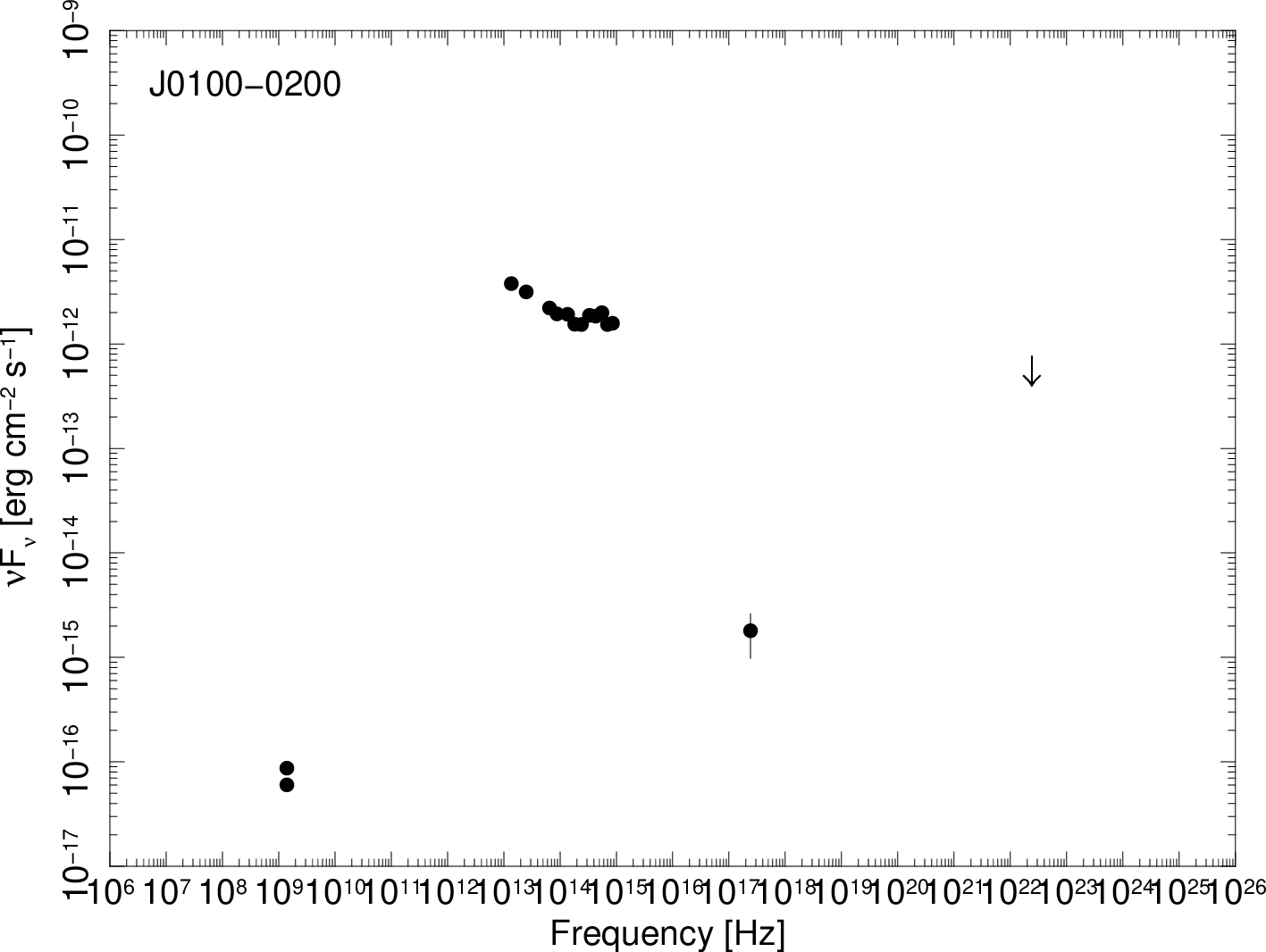}
\includegraphics[angle=270,scale=0.35]{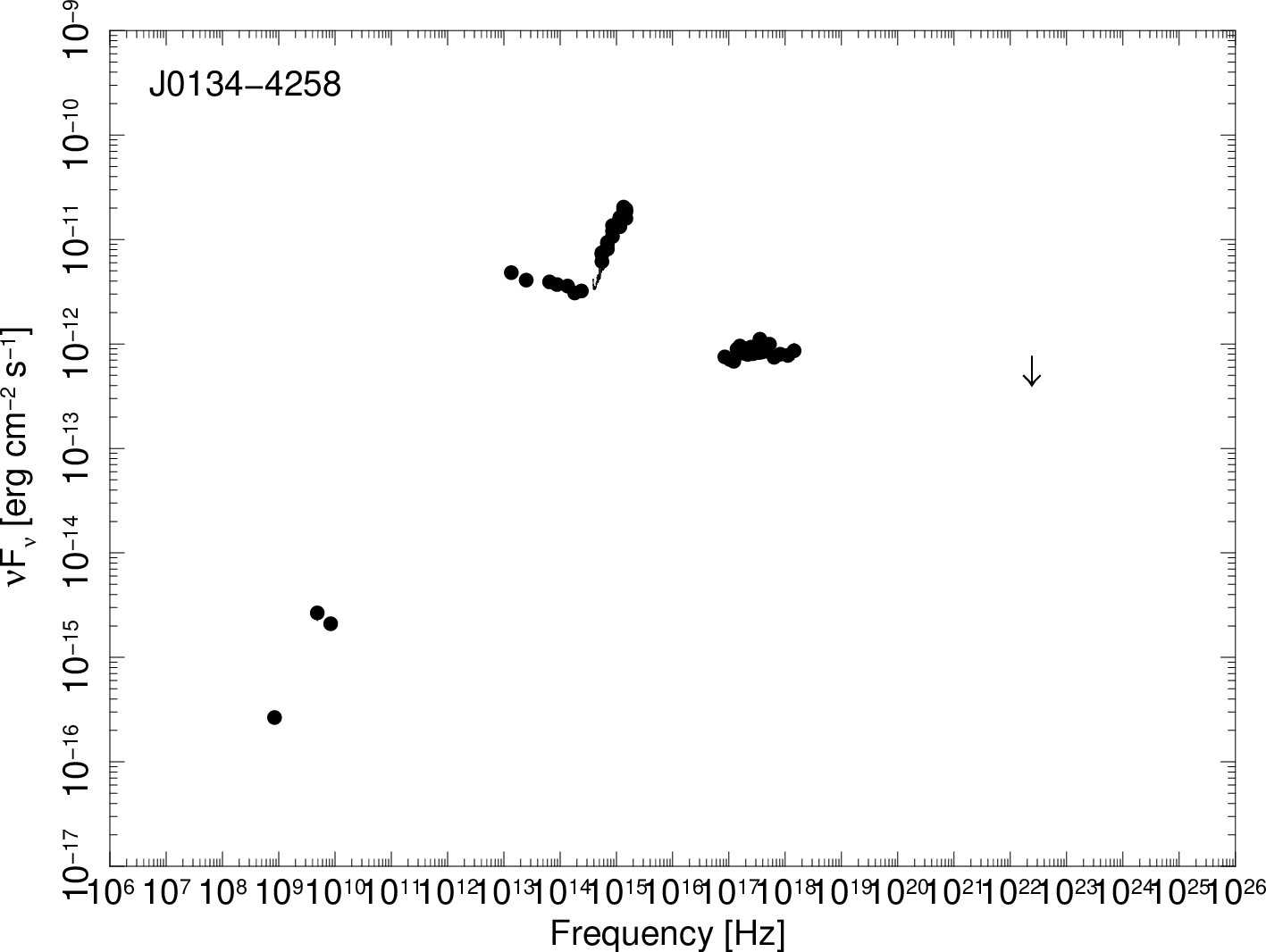}\\
\includegraphics[angle=270,scale=0.35]{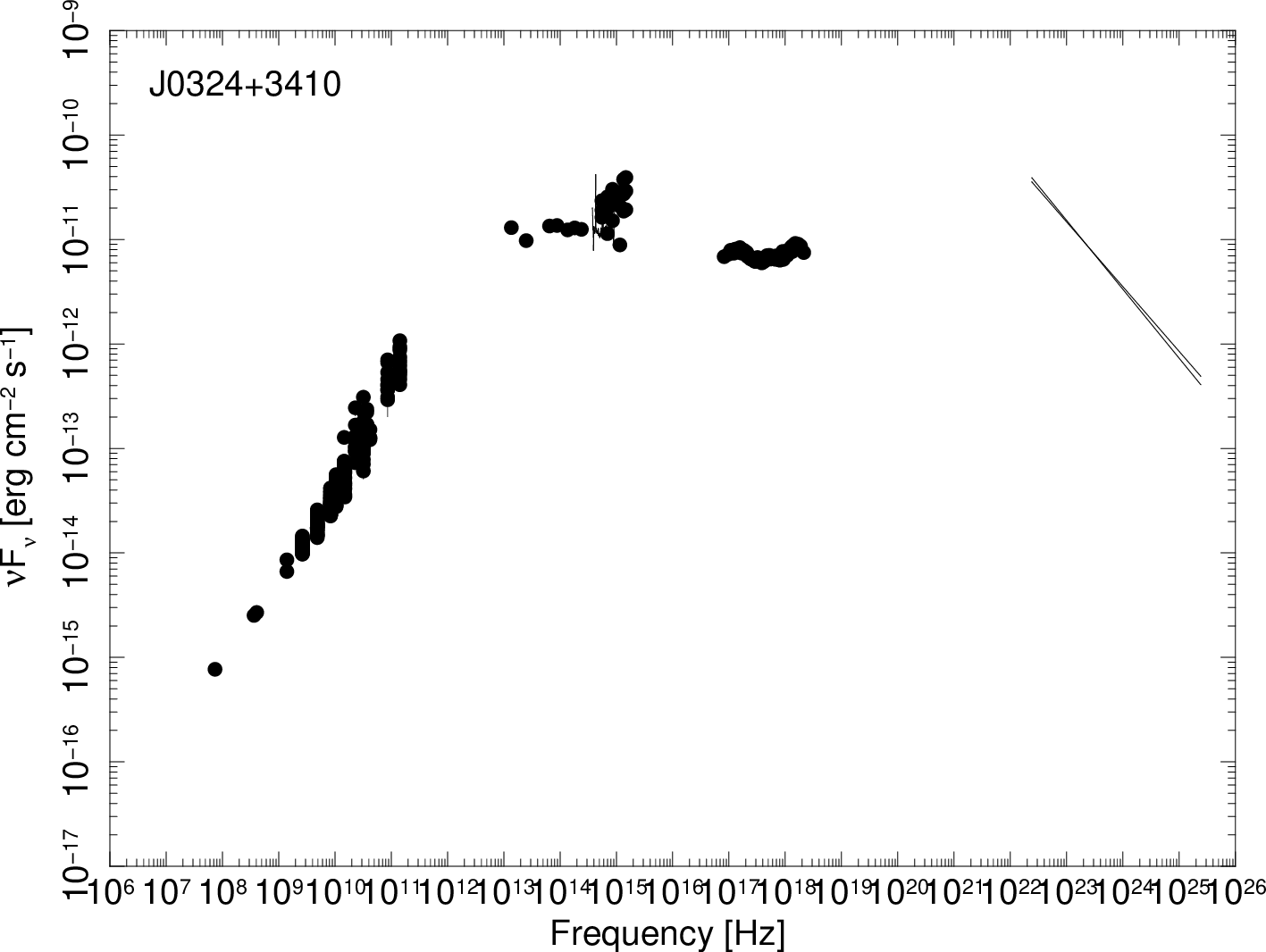}
\includegraphics[angle=270,scale=0.35]{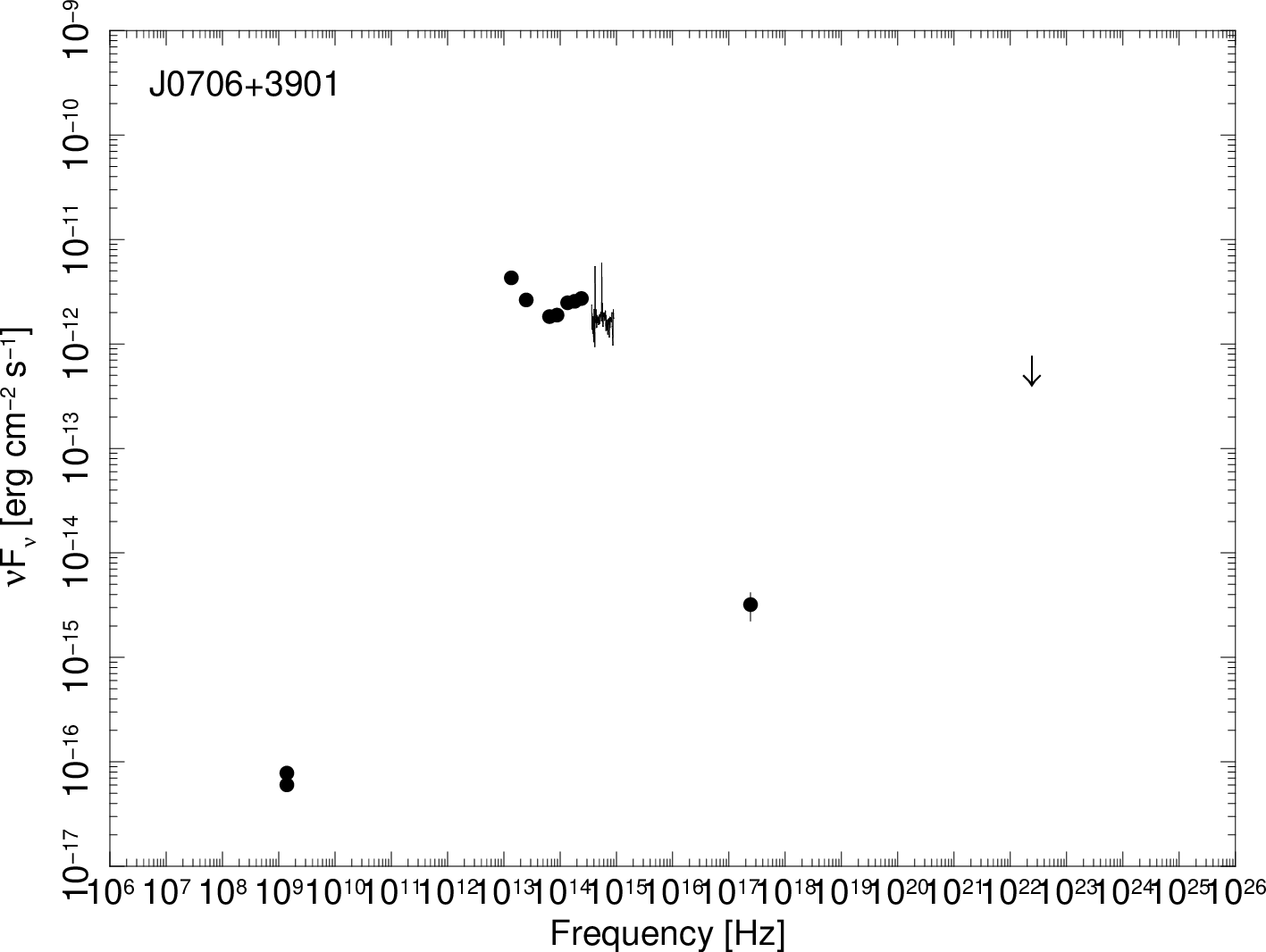}\\
\includegraphics[angle=270,scale=0.35]{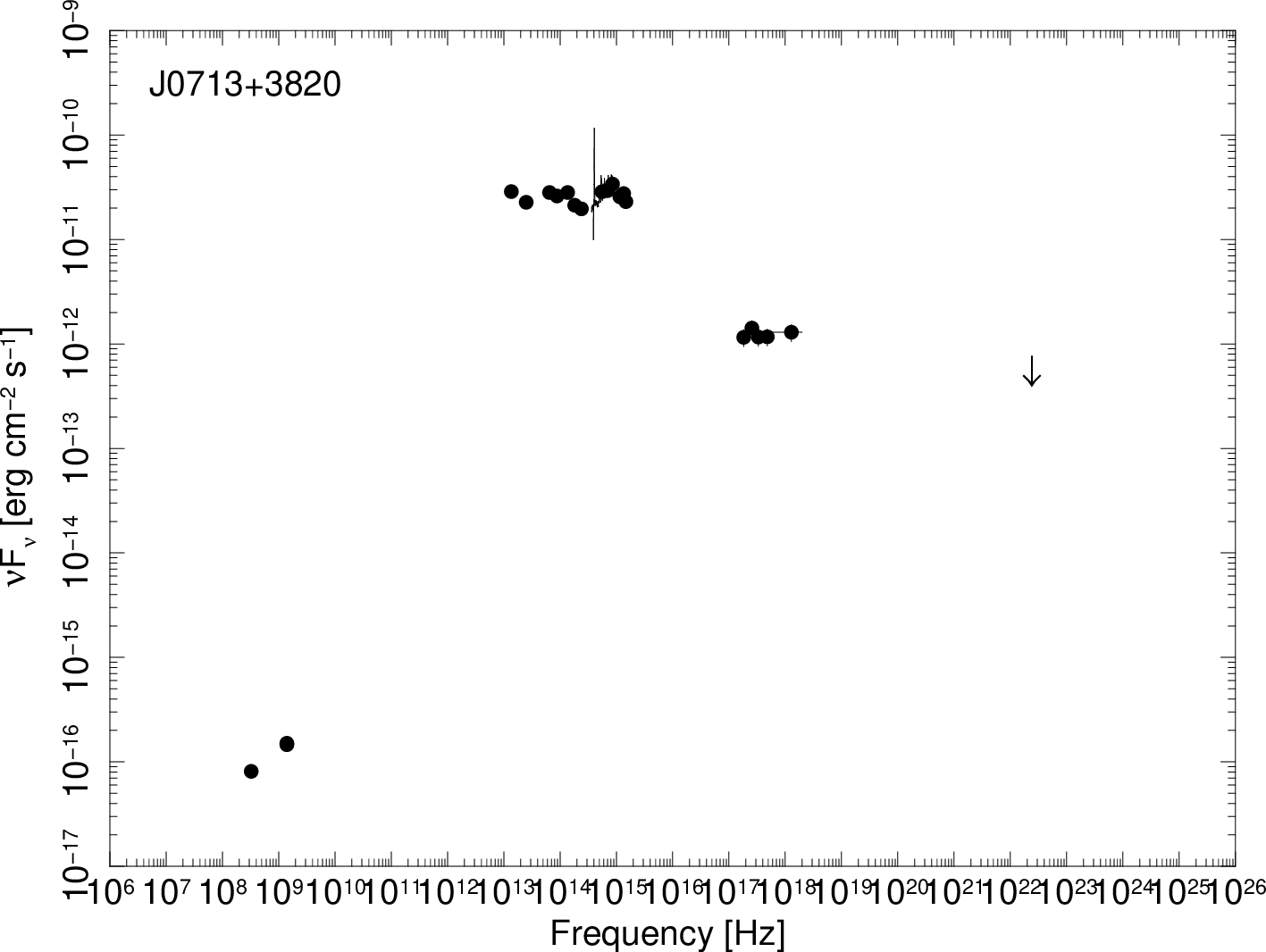}
\includegraphics[angle=270,scale=0.35]{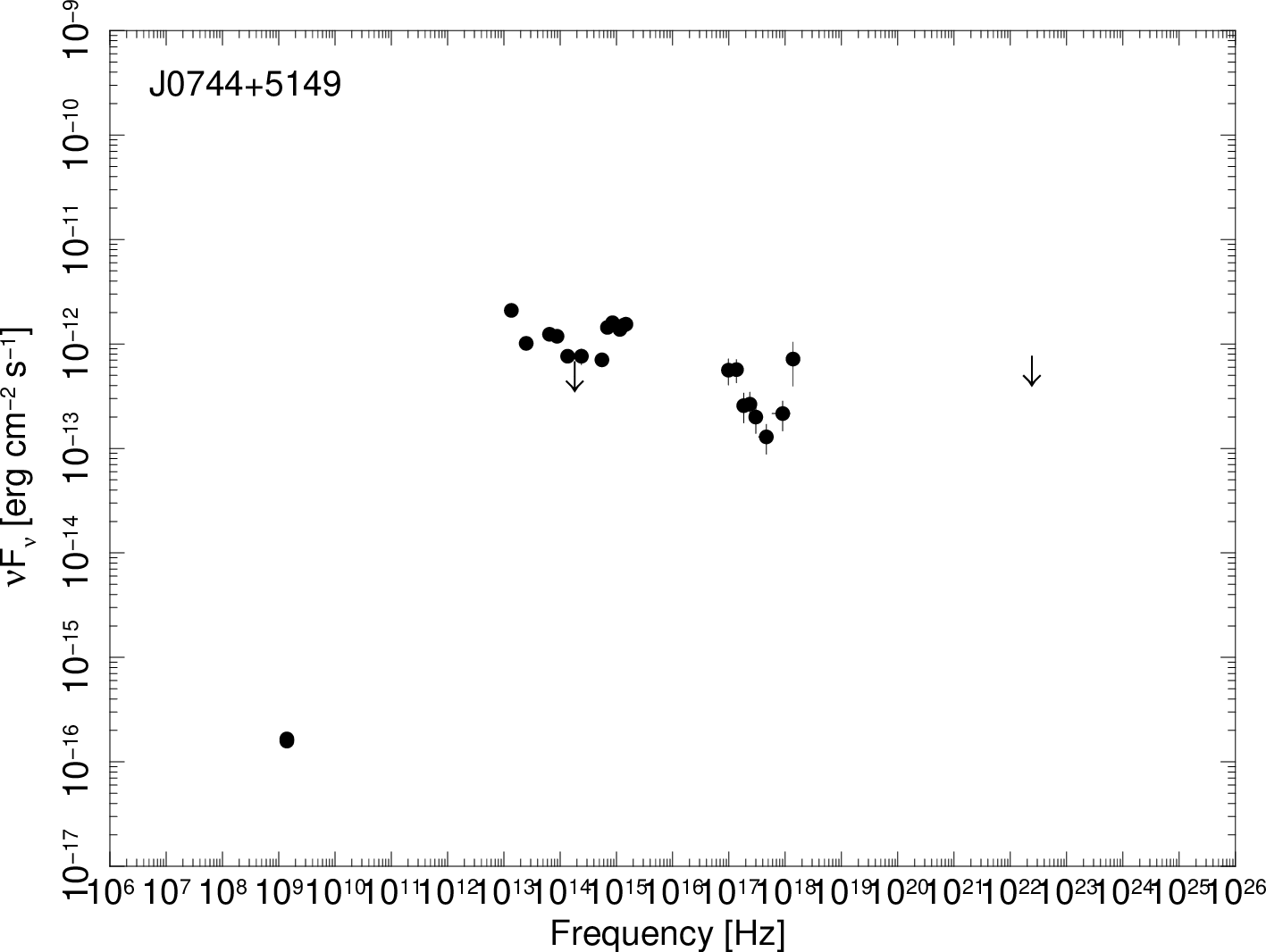}\\
\includegraphics[angle=270,scale=0.35]{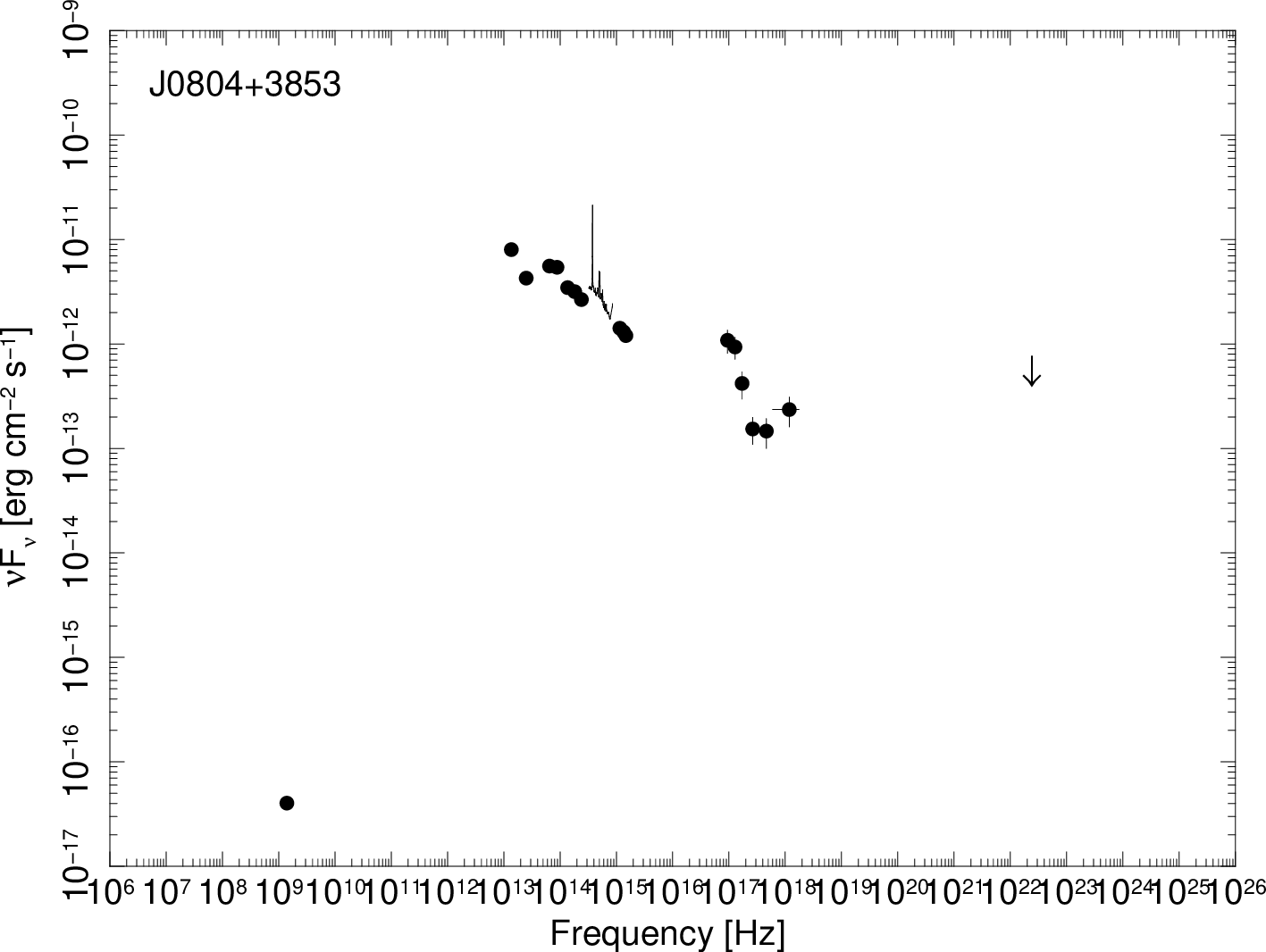}
\includegraphics[angle=270,scale=0.35]{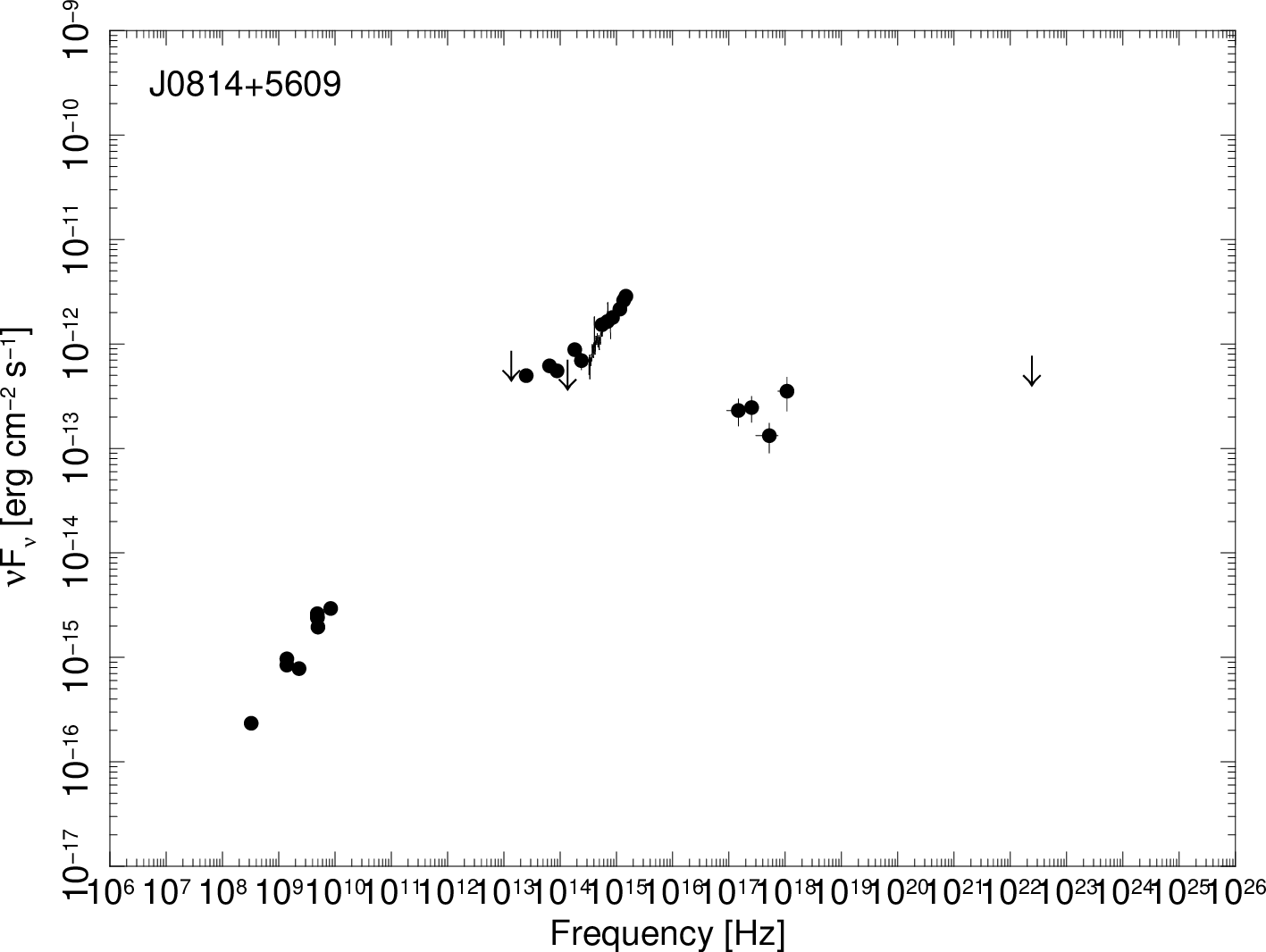}\\
\caption{Spectral Energy Distributions of the sources in the present sample. Data are corrected for the Galactic absorption. Points refer to detections; arrows are upper limits; the continuous lines are the optical spectra.} 
\label{fig:seds1}
\end{figure*}
}

\onlfig{
\begin{figure*}
\centering
\includegraphics[angle=270,scale=0.35]{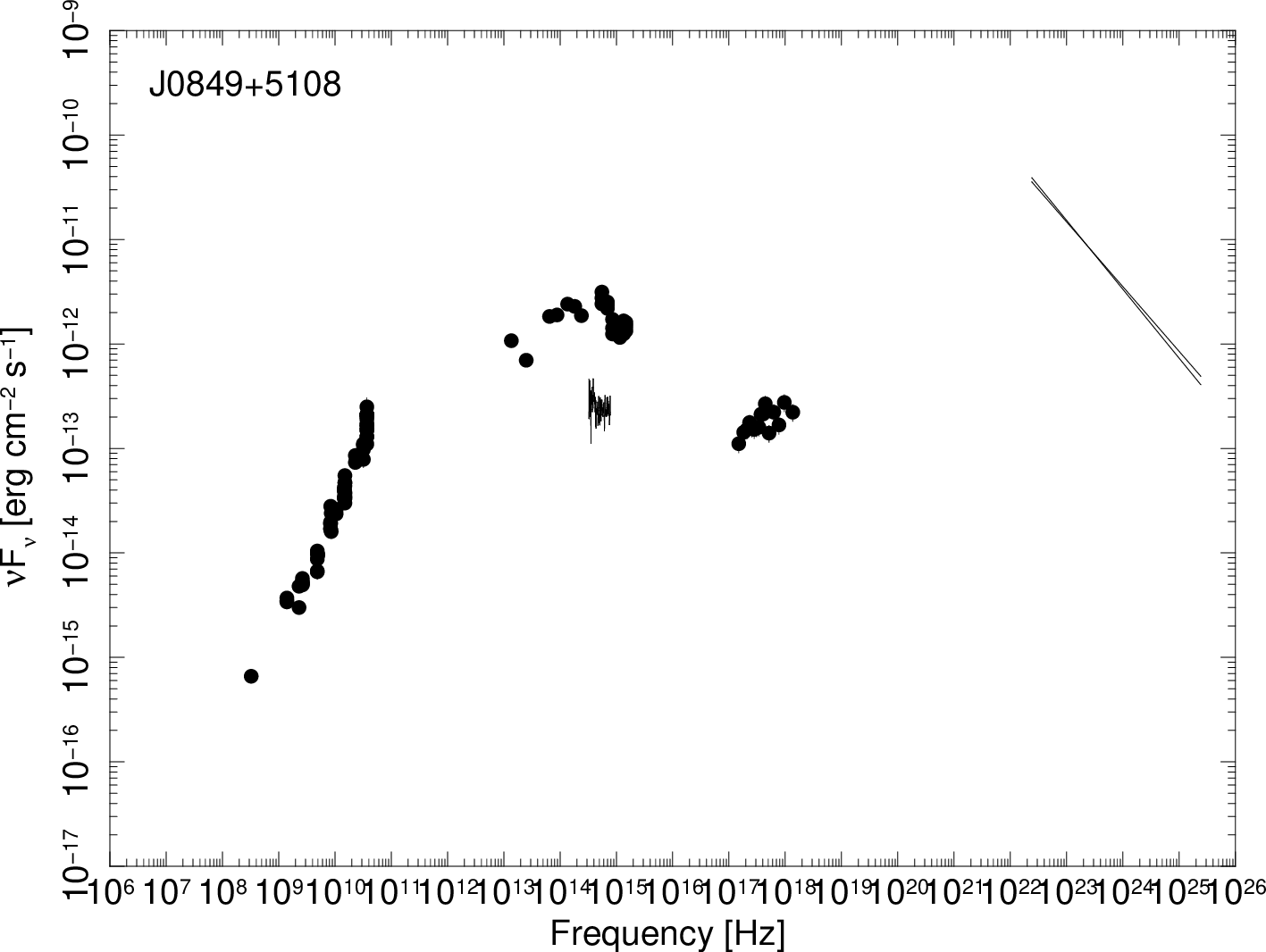}
\includegraphics[angle=270,scale=0.35]{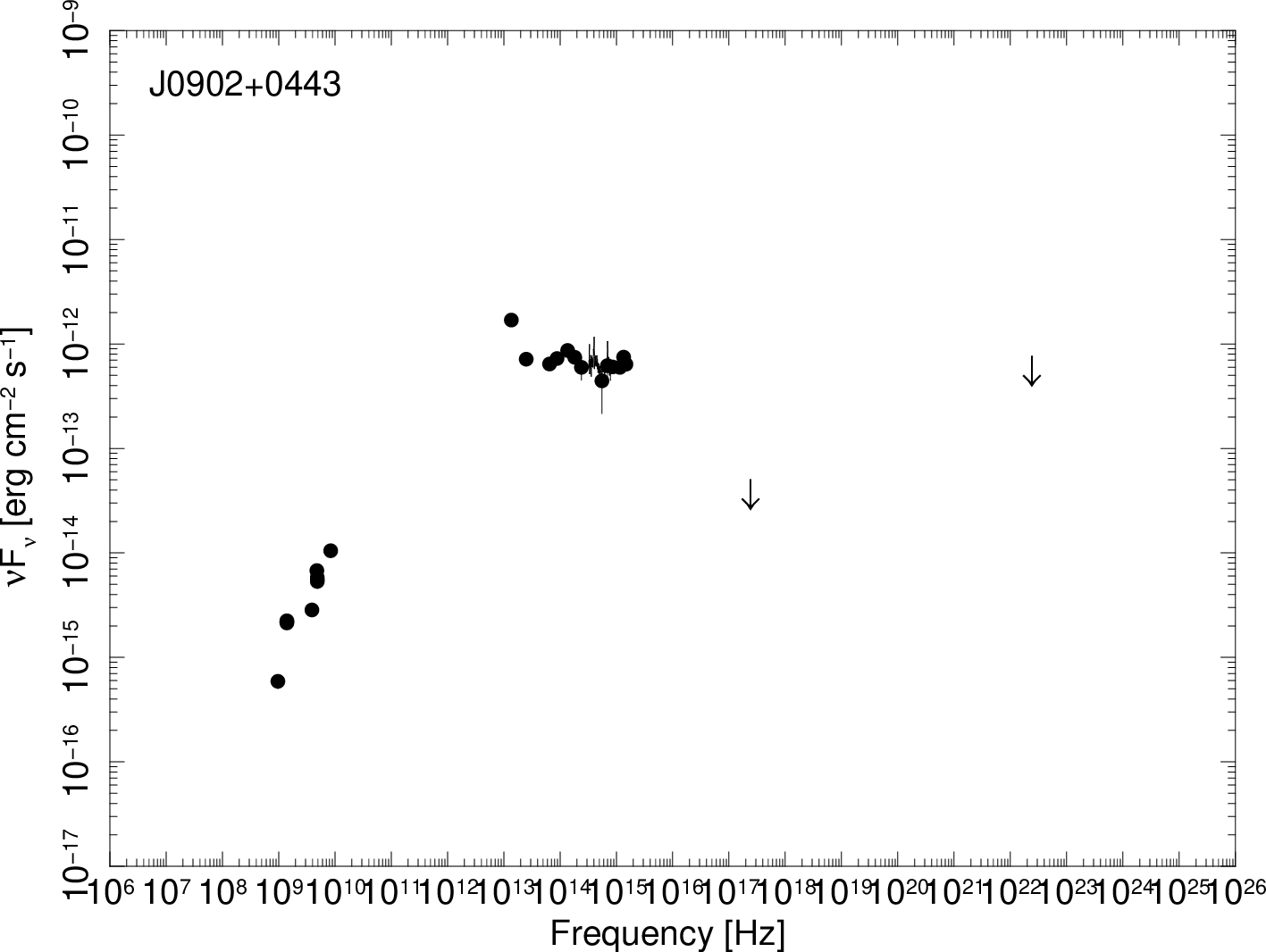}\\
\includegraphics[angle=270,scale=0.35]{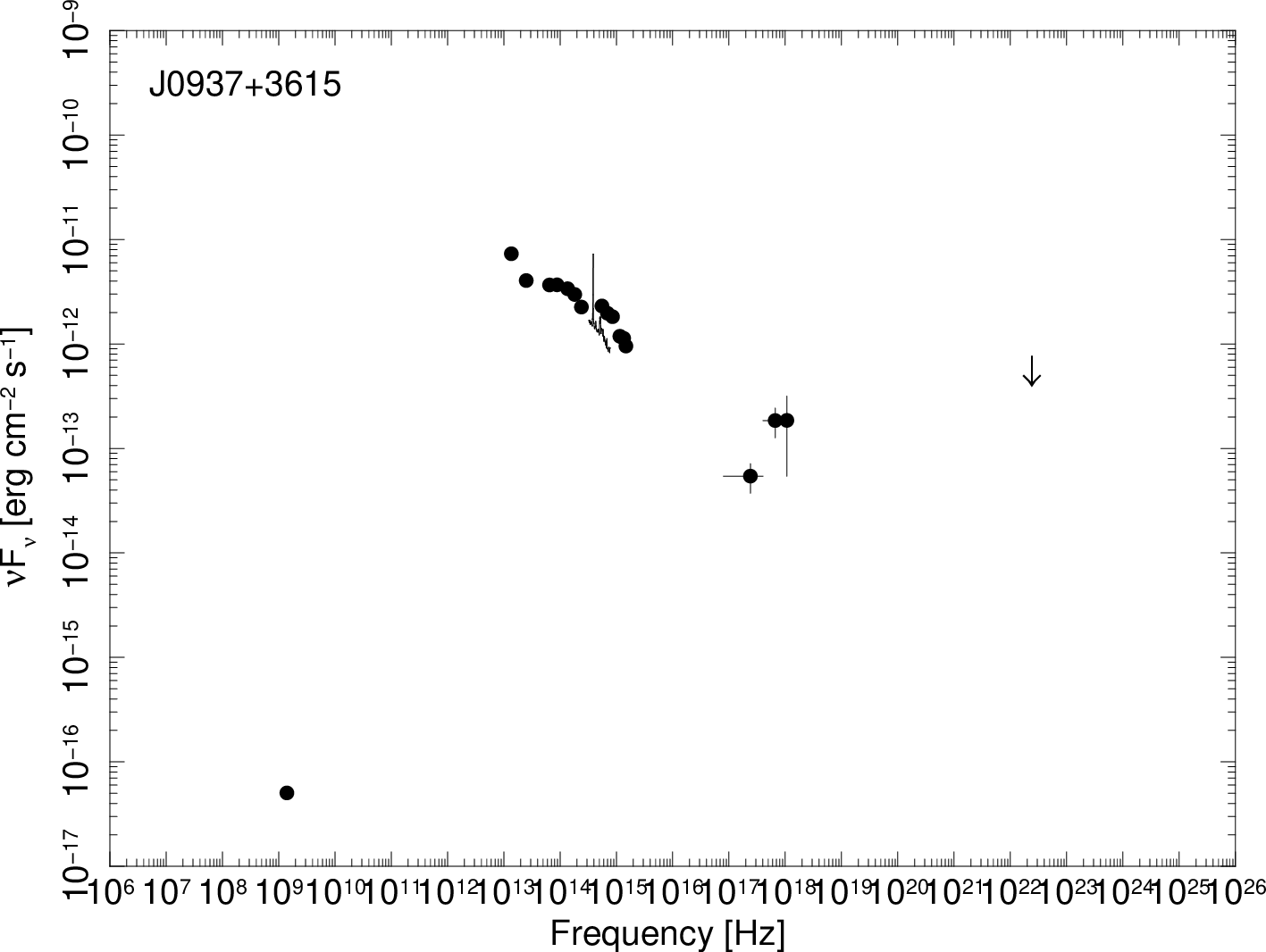}
\includegraphics[angle=270,scale=0.35]{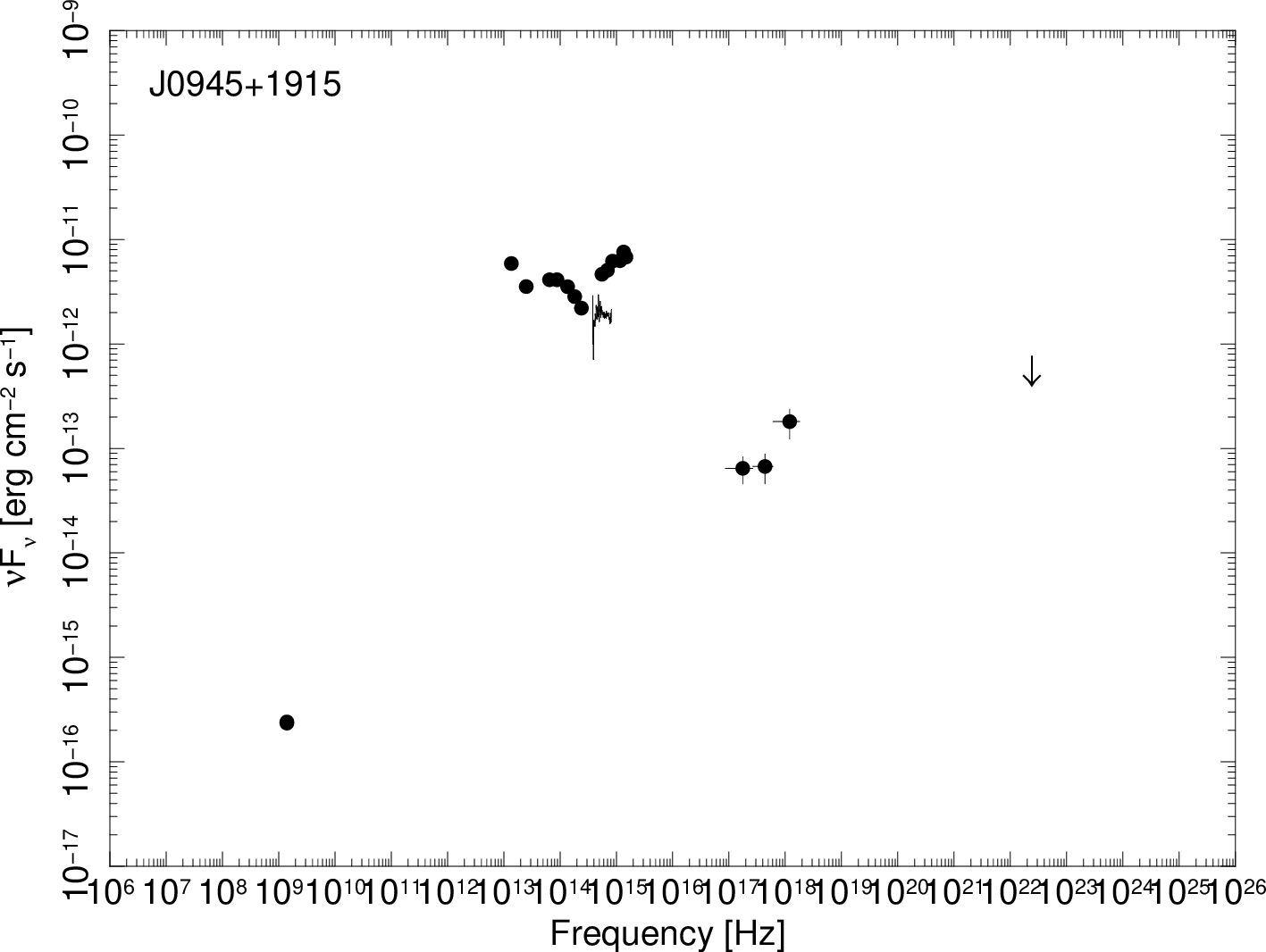}\\
\includegraphics[angle=270,scale=0.35]{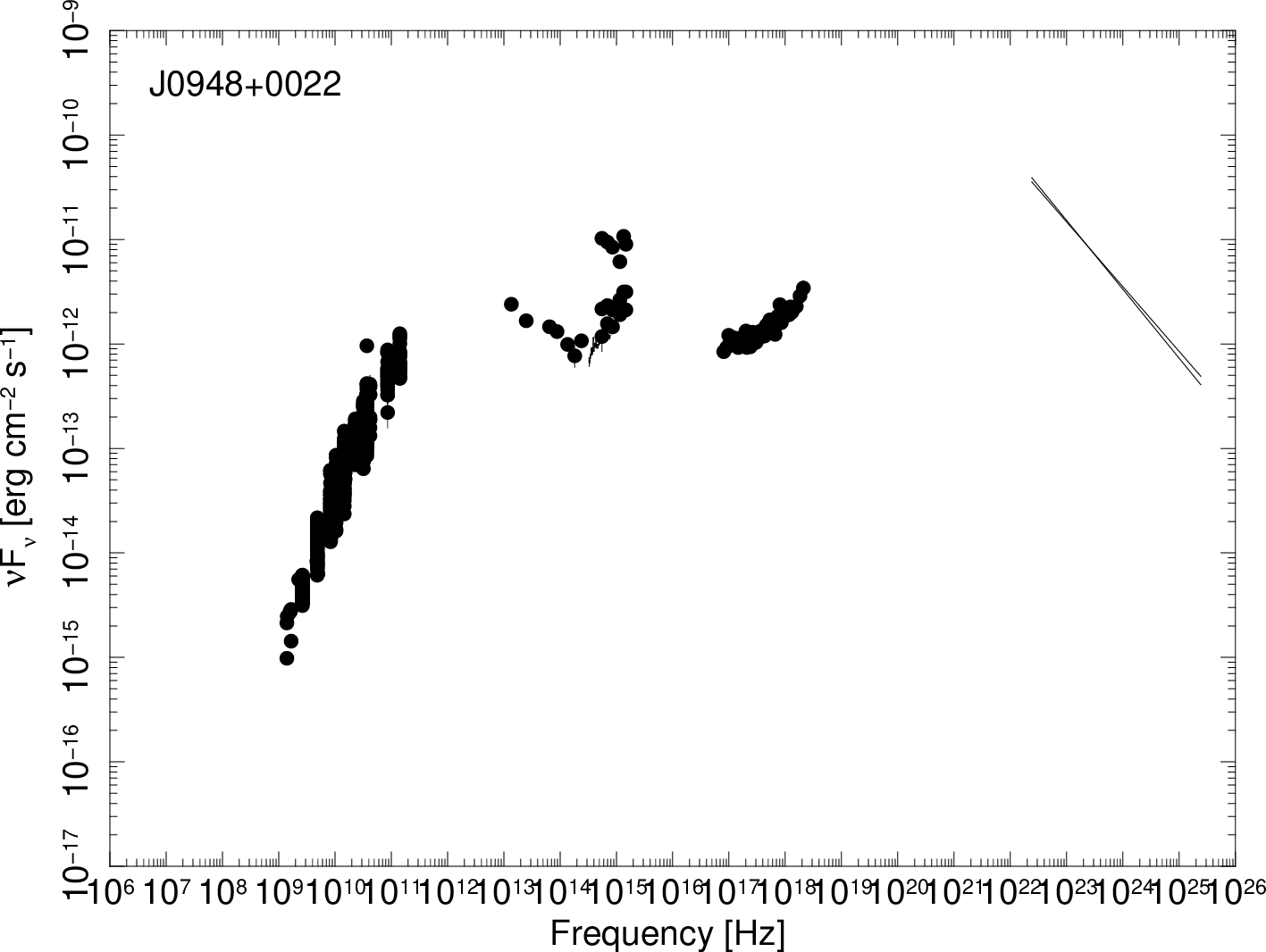}
\includegraphics[angle=270,scale=0.35]{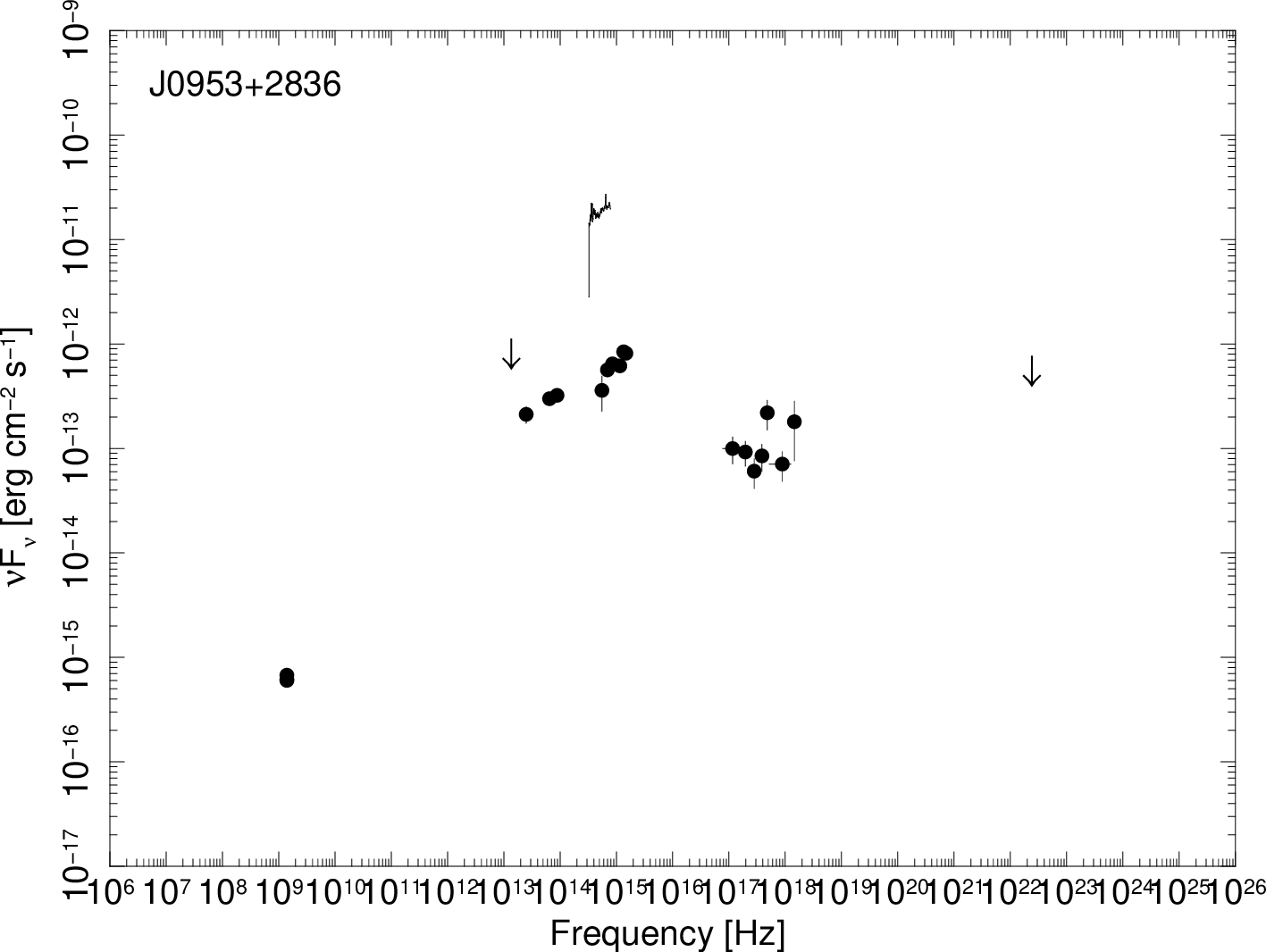}\\
\includegraphics[angle=270,scale=0.35]{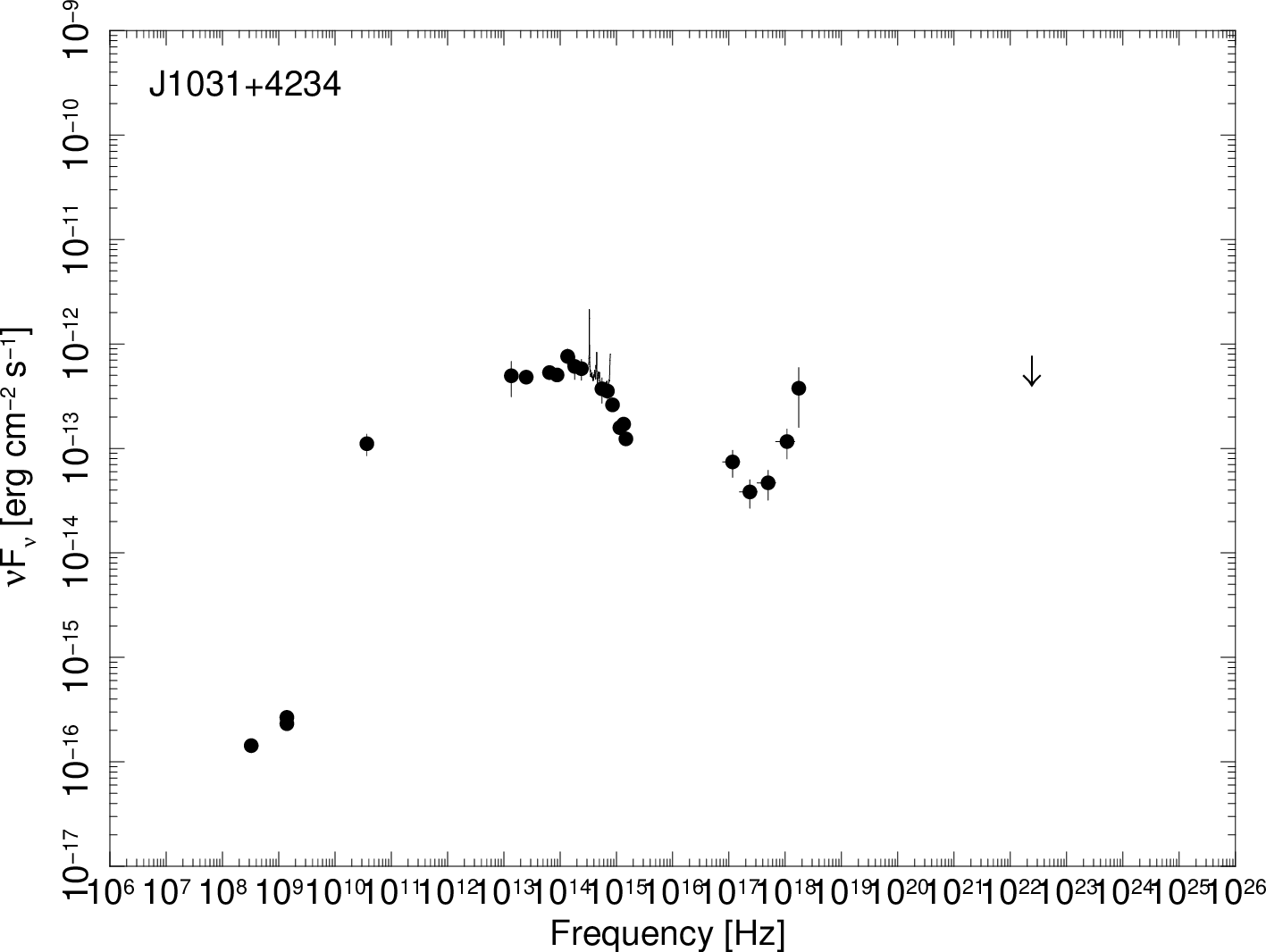}
\includegraphics[angle=270,scale=0.35]{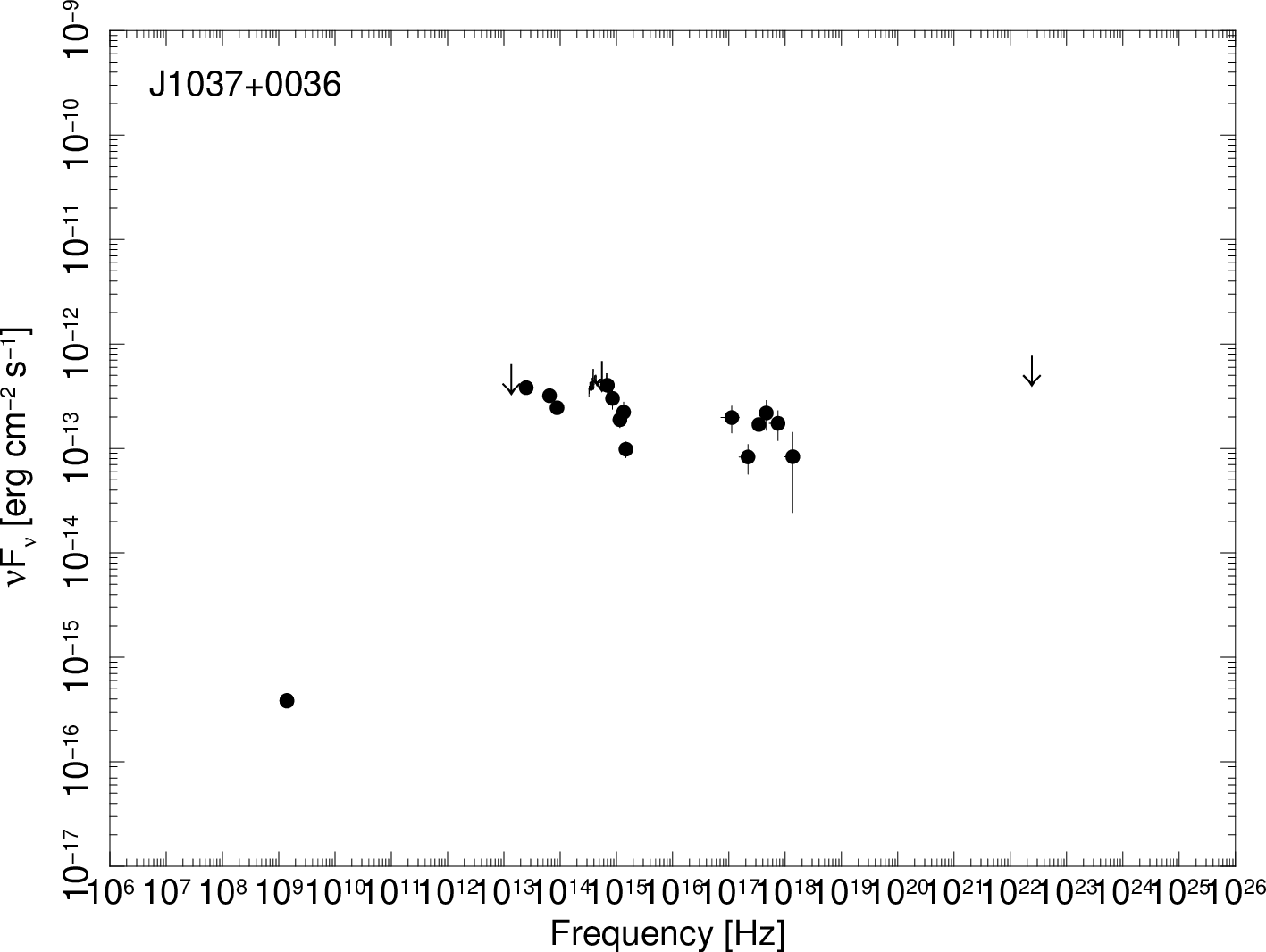}\\
\caption{Spectral Energy Distributions of the sources in the present sample. Data are corrected for the Galactic absorption. Points refer to detections; arrows are upper limits; the continuous lines are the optical spectra.} 
\label{fig:seds2}
\end{figure*}
}

\onlfig{
\begin{figure*}
\centering
\includegraphics[angle=270,scale=0.35]{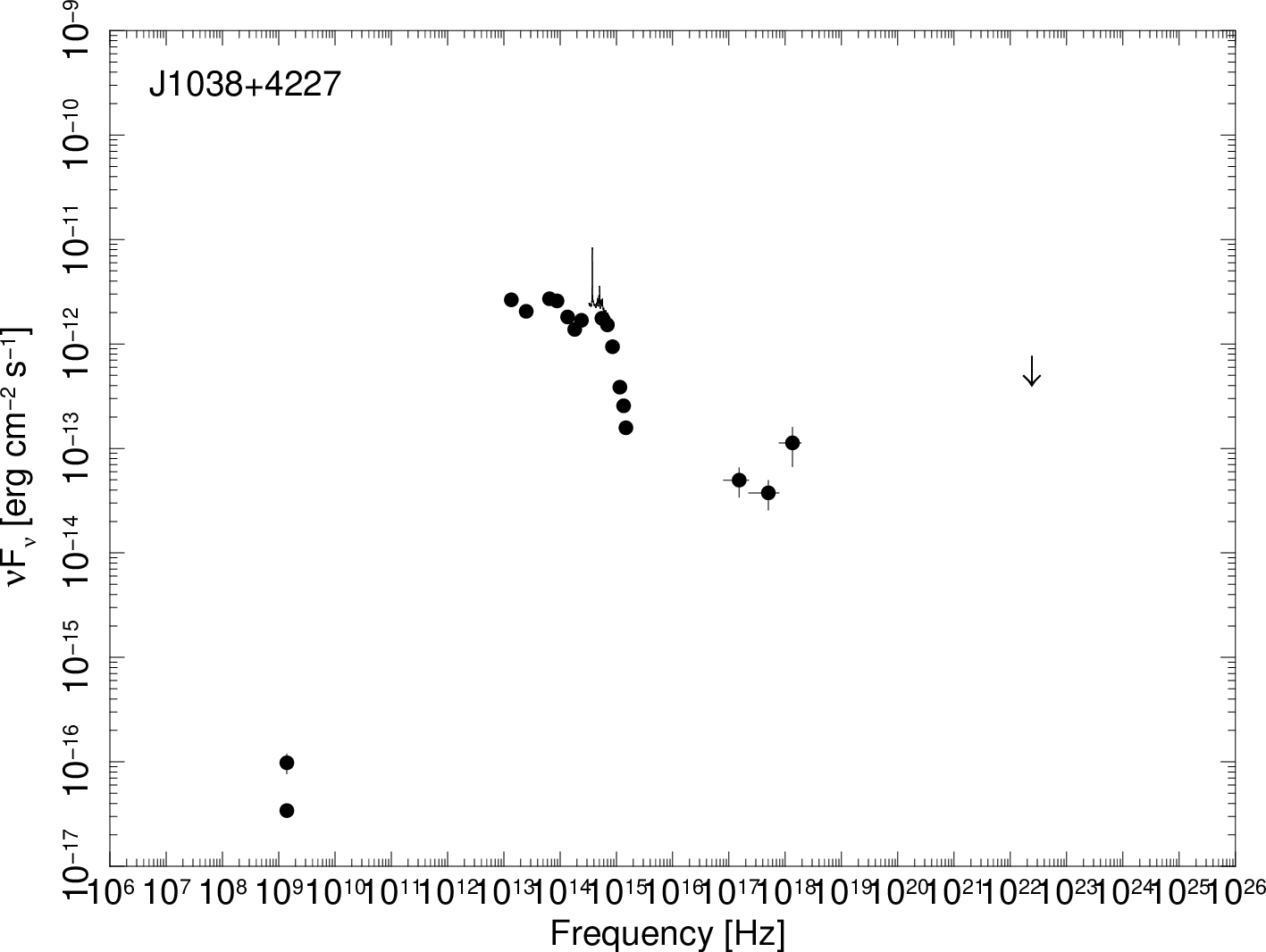}
\includegraphics[angle=270,scale=0.35]{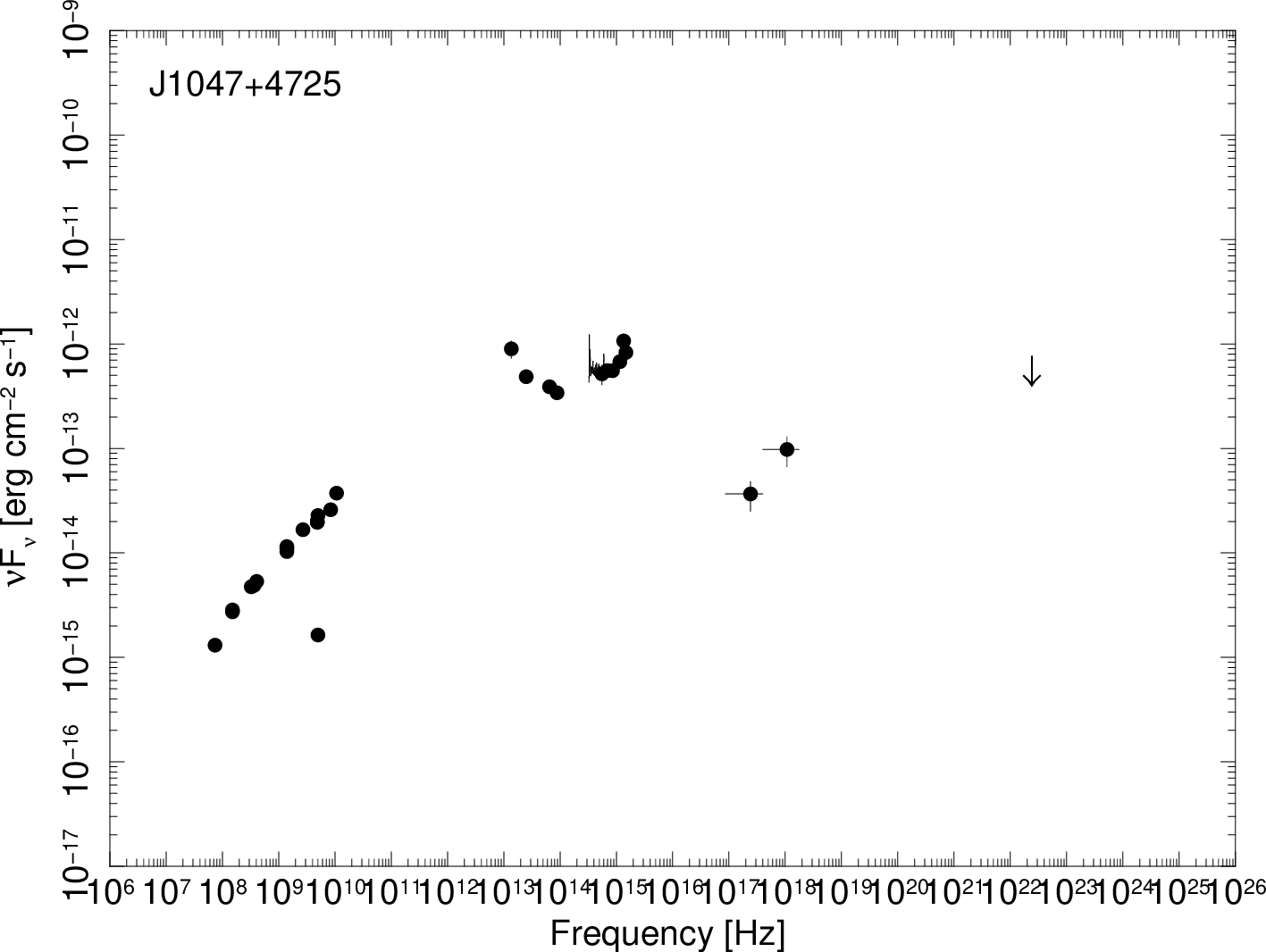}\\
\includegraphics[angle=270,scale=0.35]{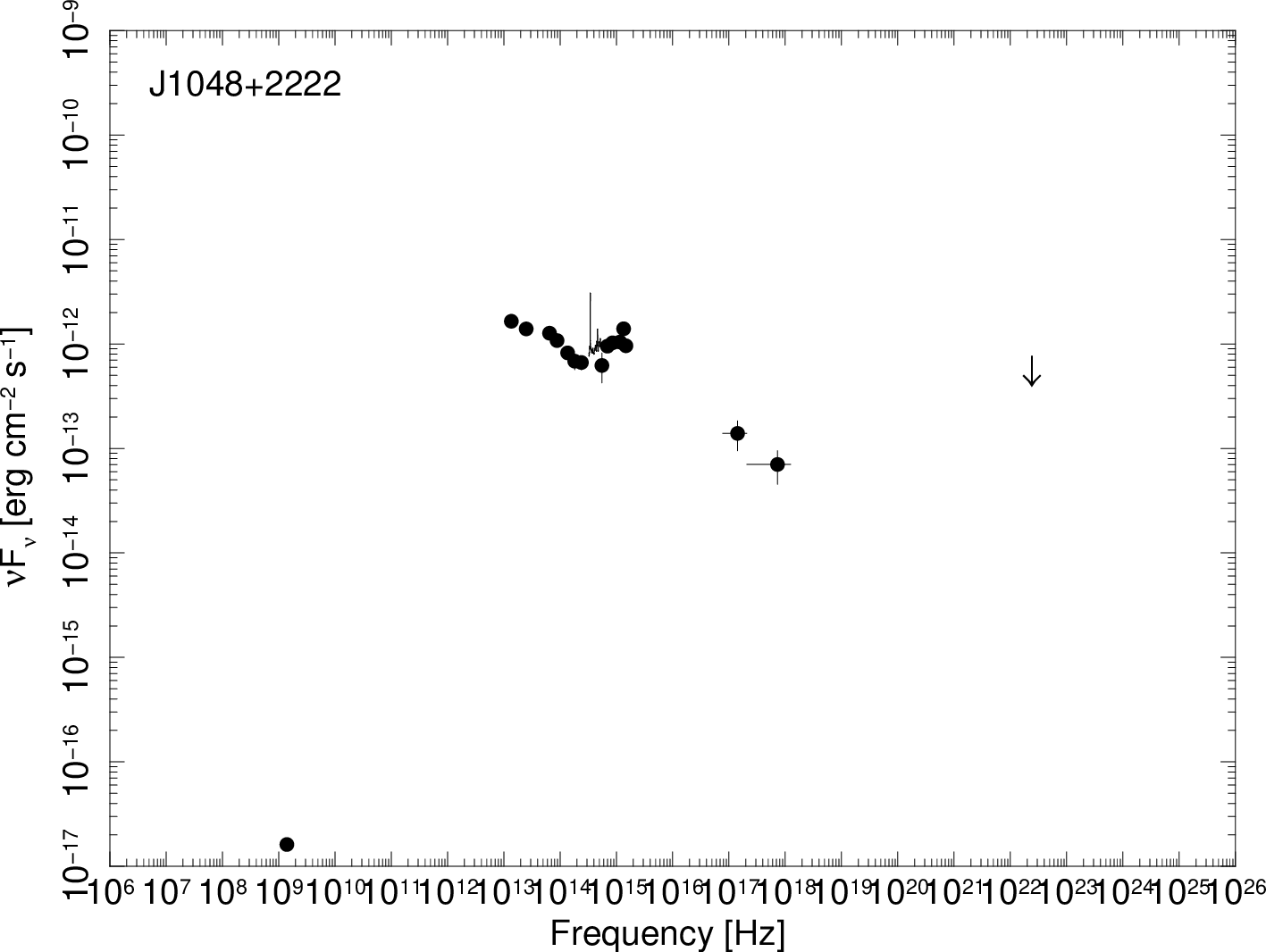}
\includegraphics[angle=270,scale=0.35]{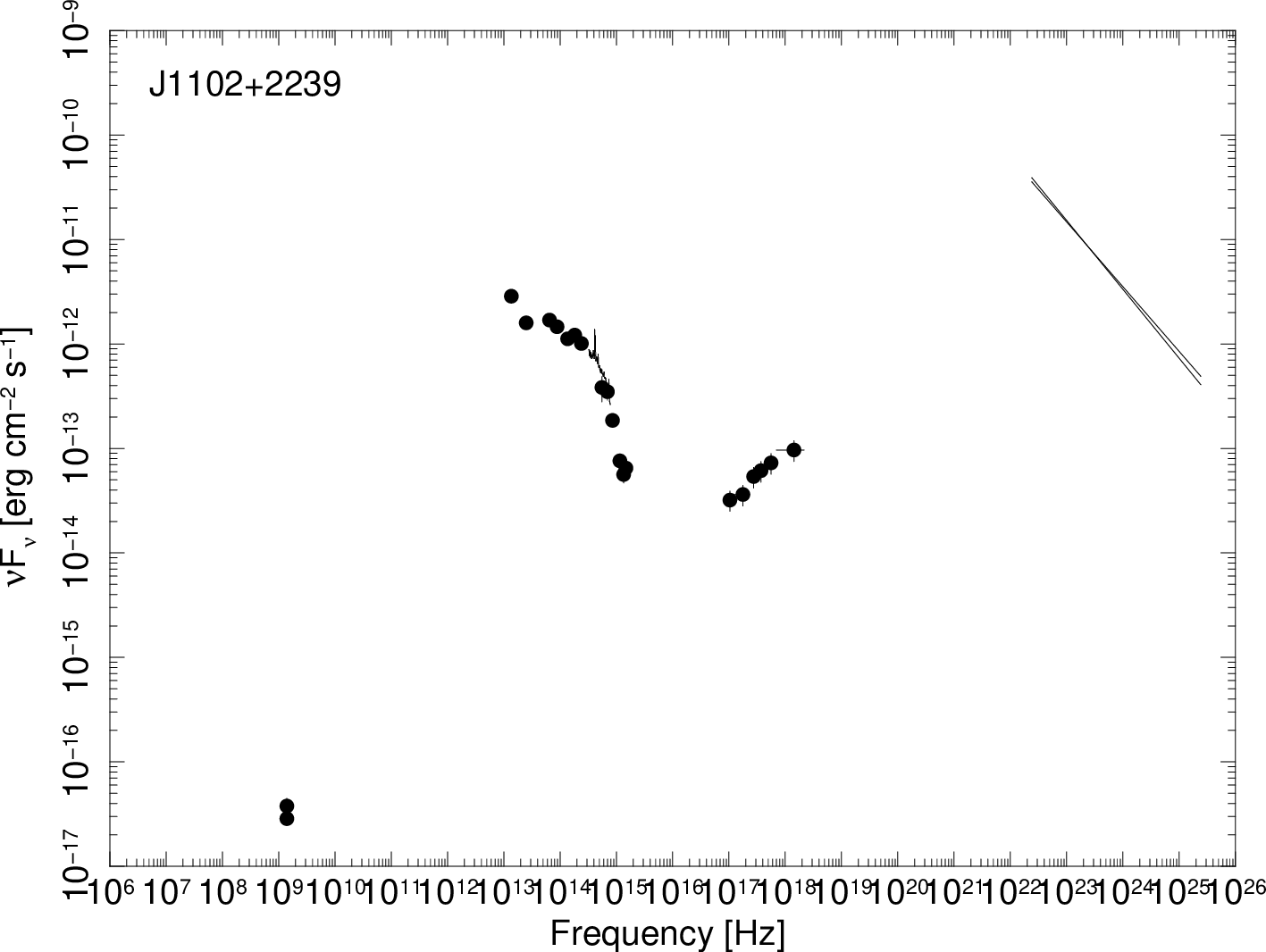}\\
\includegraphics[angle=270,scale=0.35]{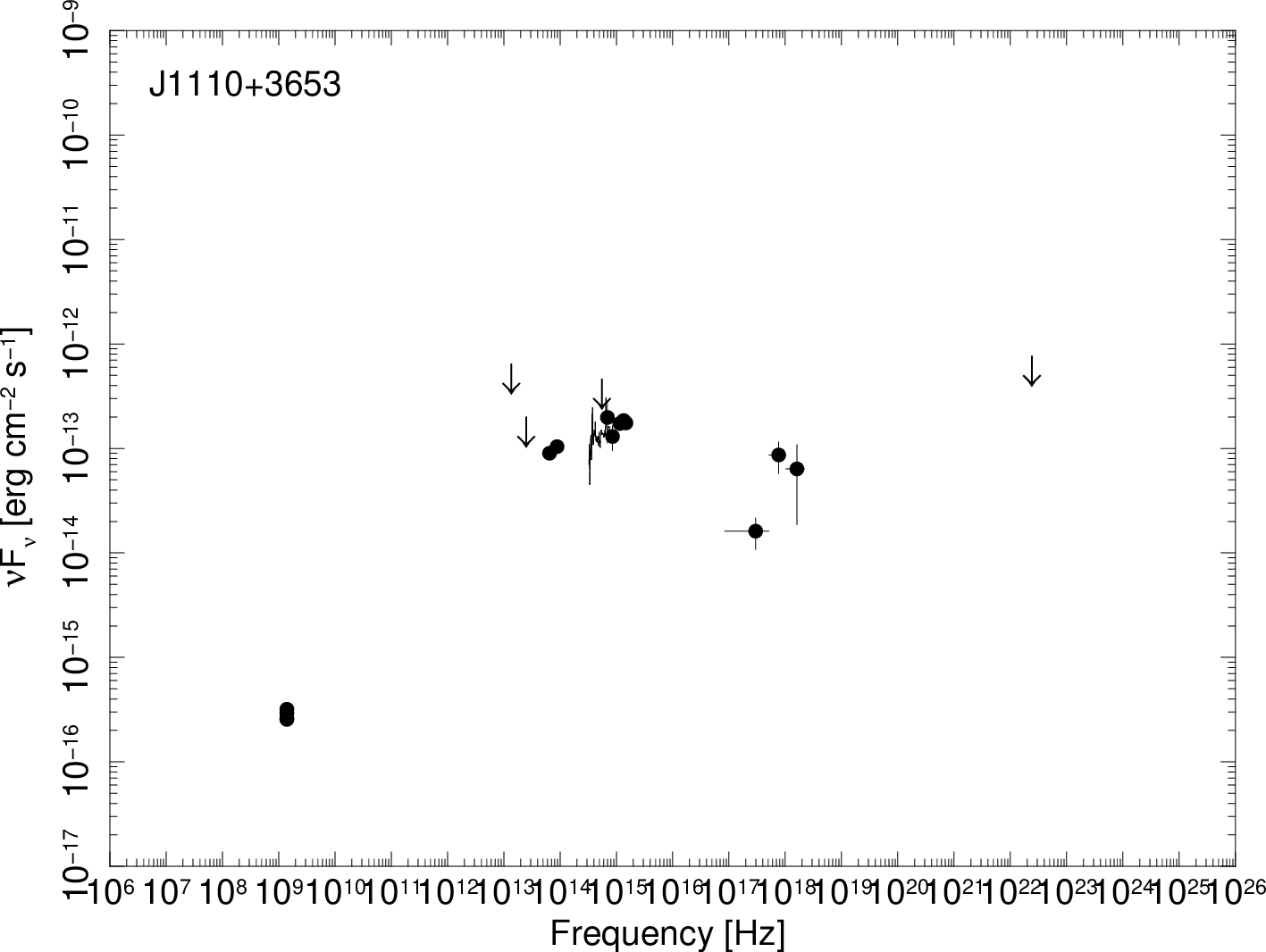}
\includegraphics[angle=270,scale=0.35]{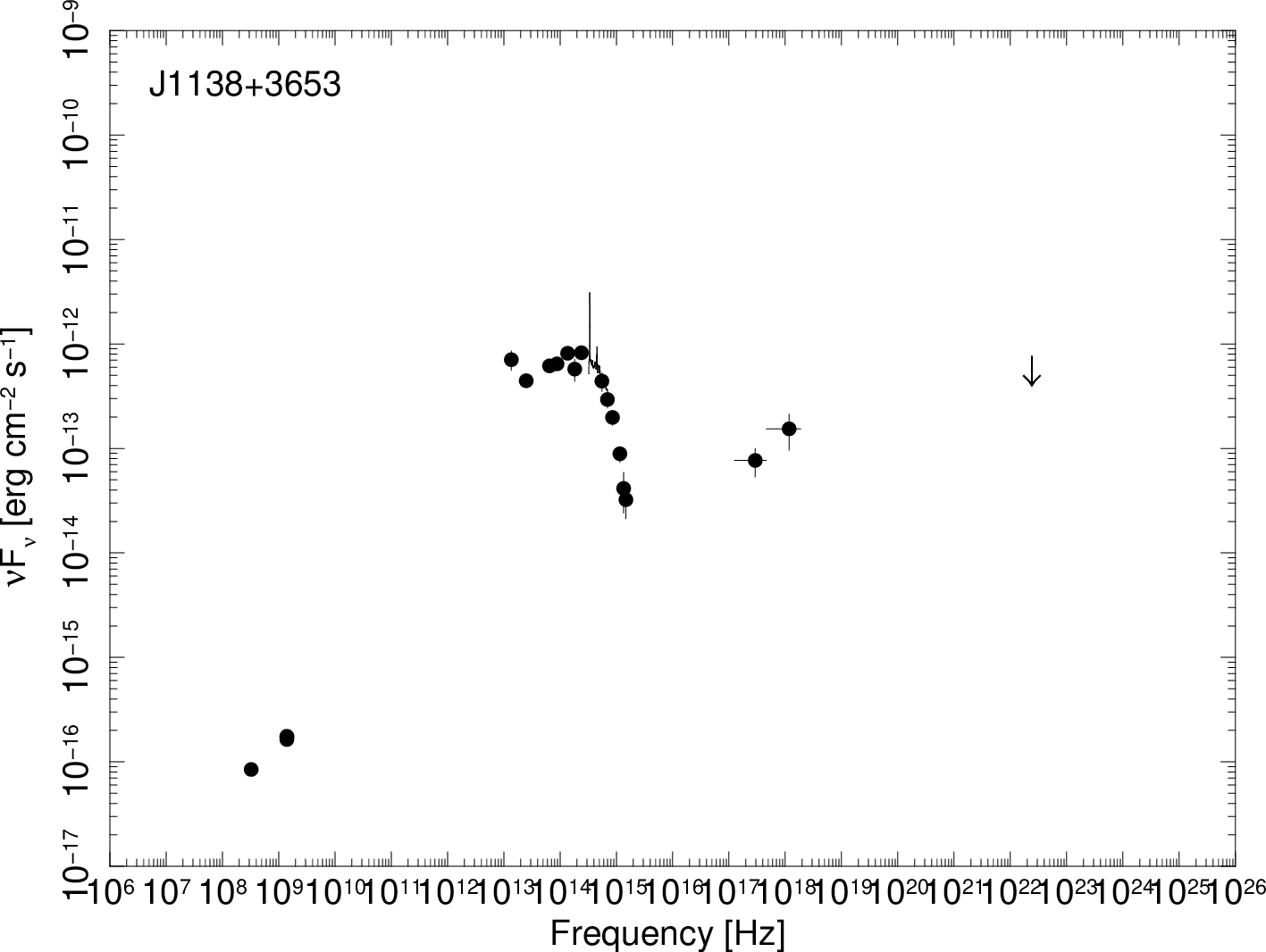}\\
\includegraphics[angle=270,scale=0.35]{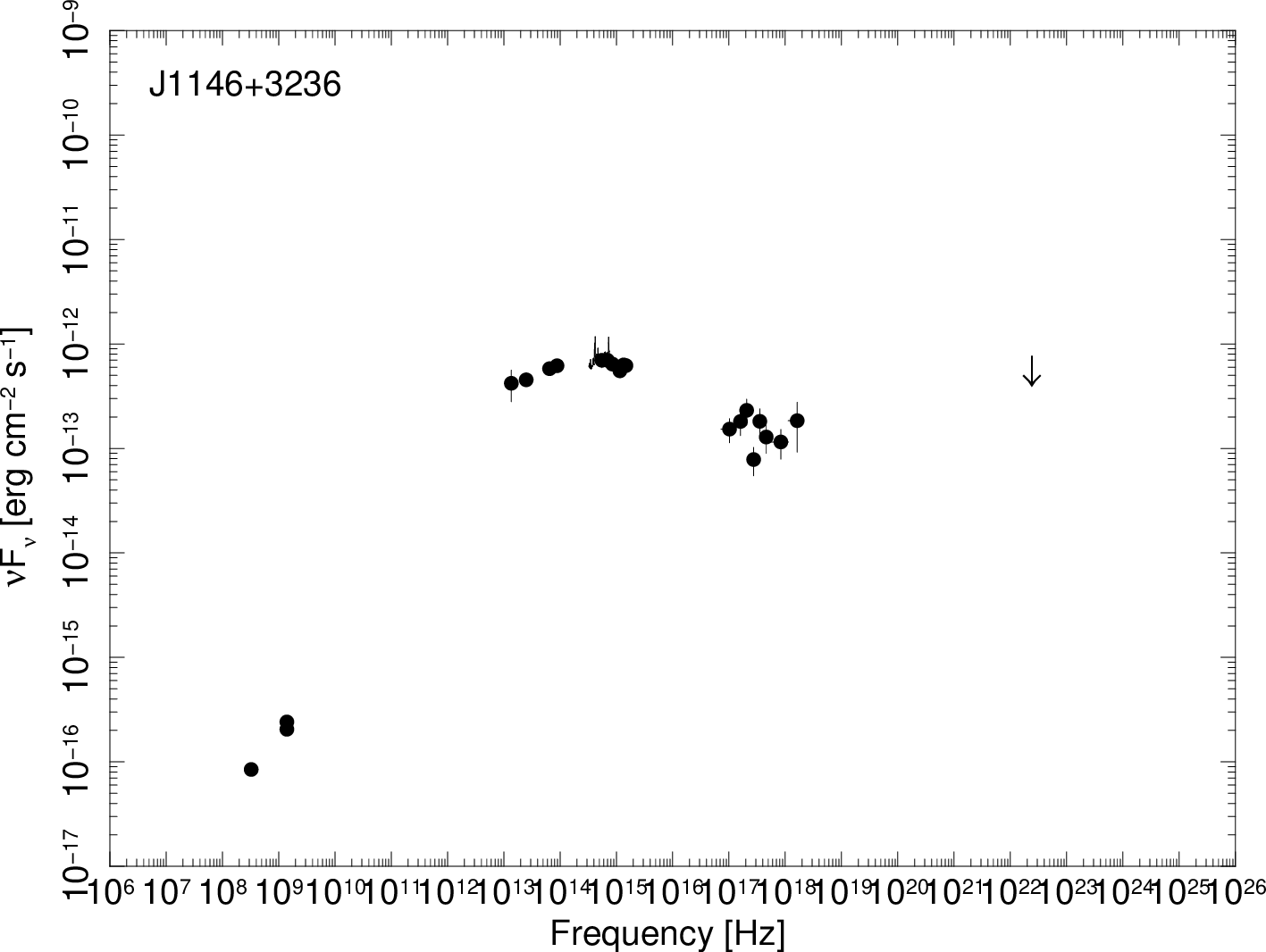}
\includegraphics[angle=270,scale=0.35]{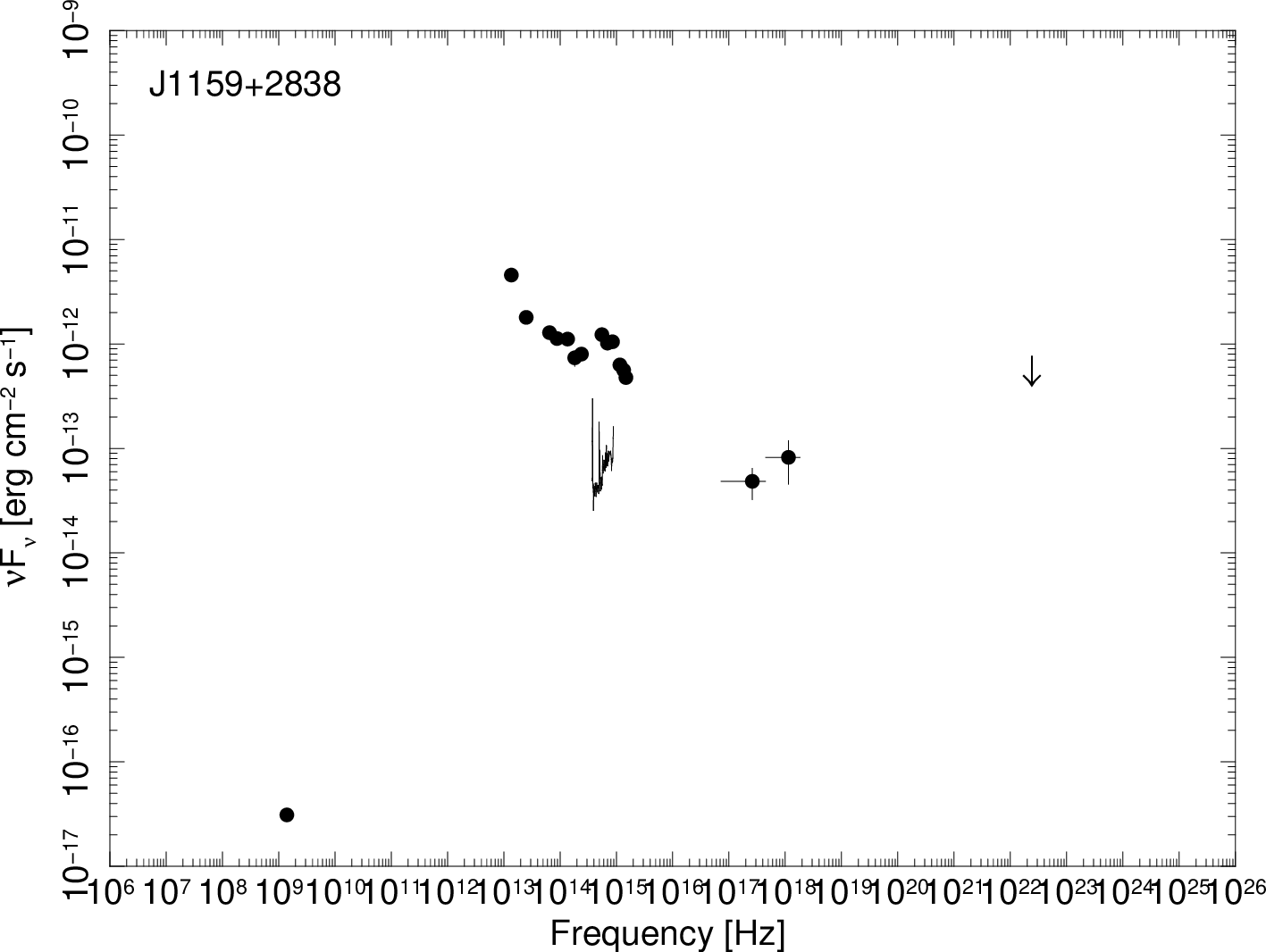}\\
\caption{Spectral Energy Distributions of the sources in the present sample. Data are corrected for the Galactic absorption. Points refer to detections; arrows are upper limits; the continuous lines are the optical spectra.} 
\label{fig:seds3}
\end{figure*}
}

\onlfig{
\begin{figure*}
\centering
\includegraphics[angle=270,scale=0.35]{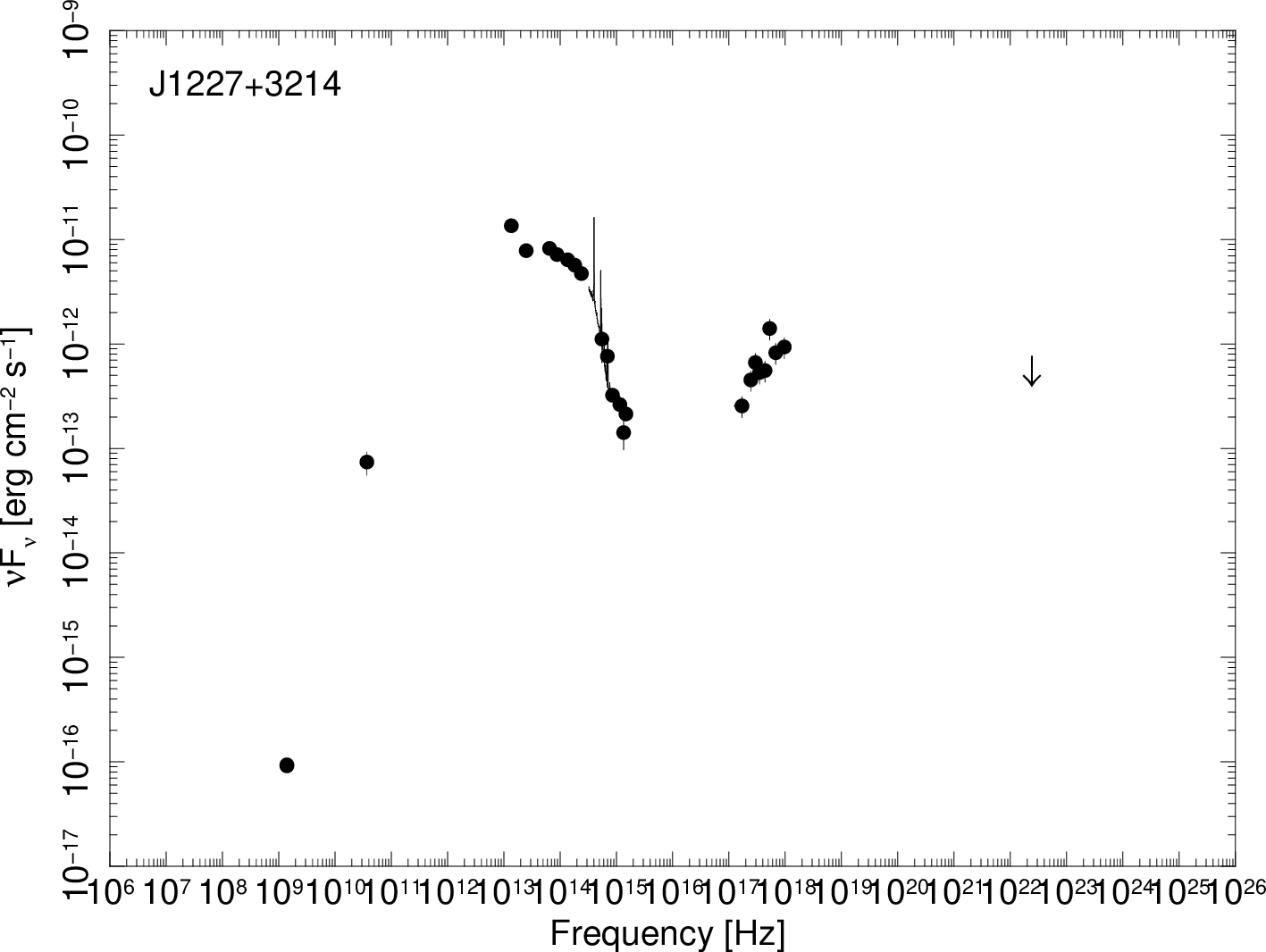}
\includegraphics[angle=270,scale=0.35]{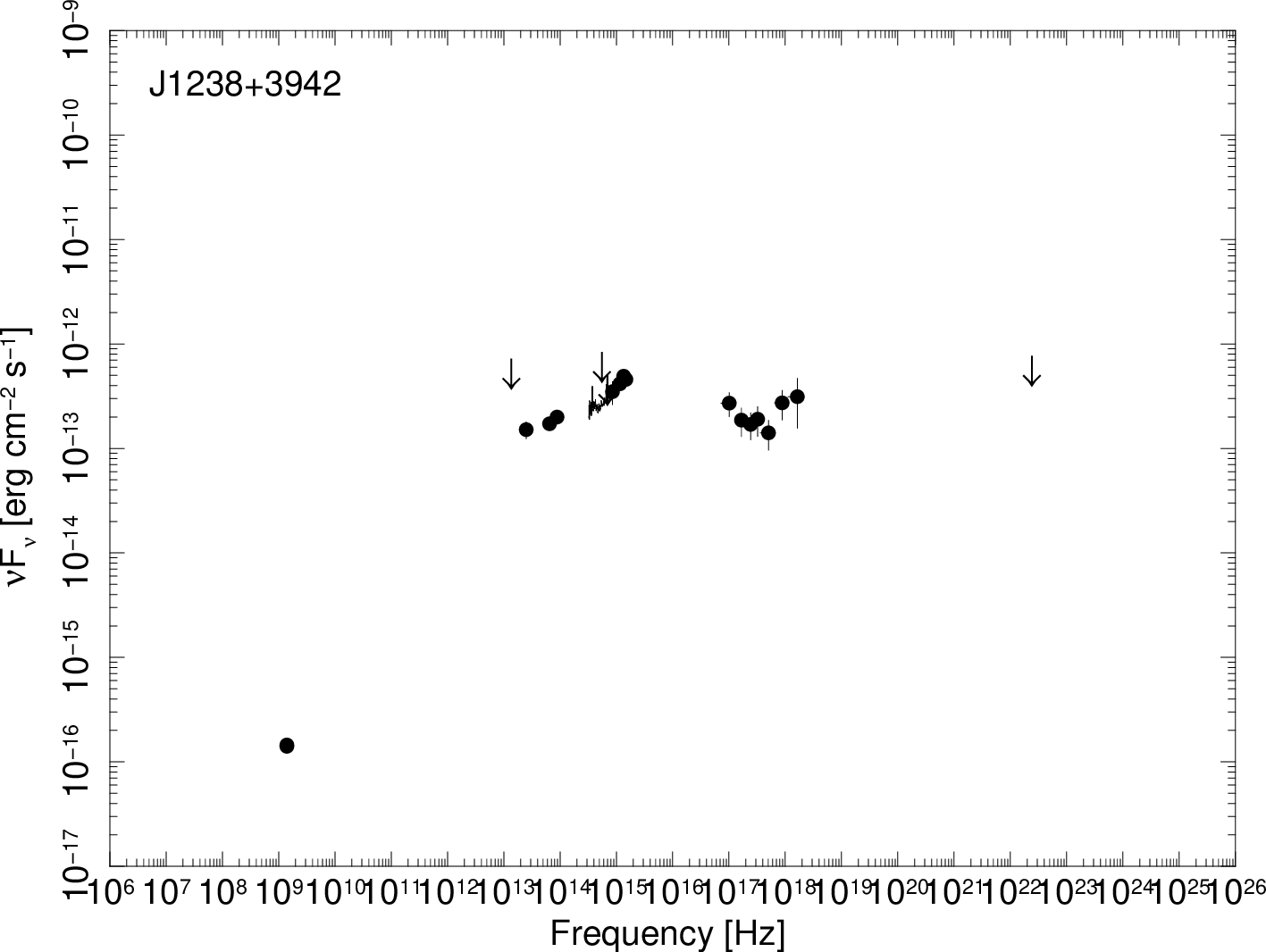}\\
\includegraphics[angle=270,scale=0.35]{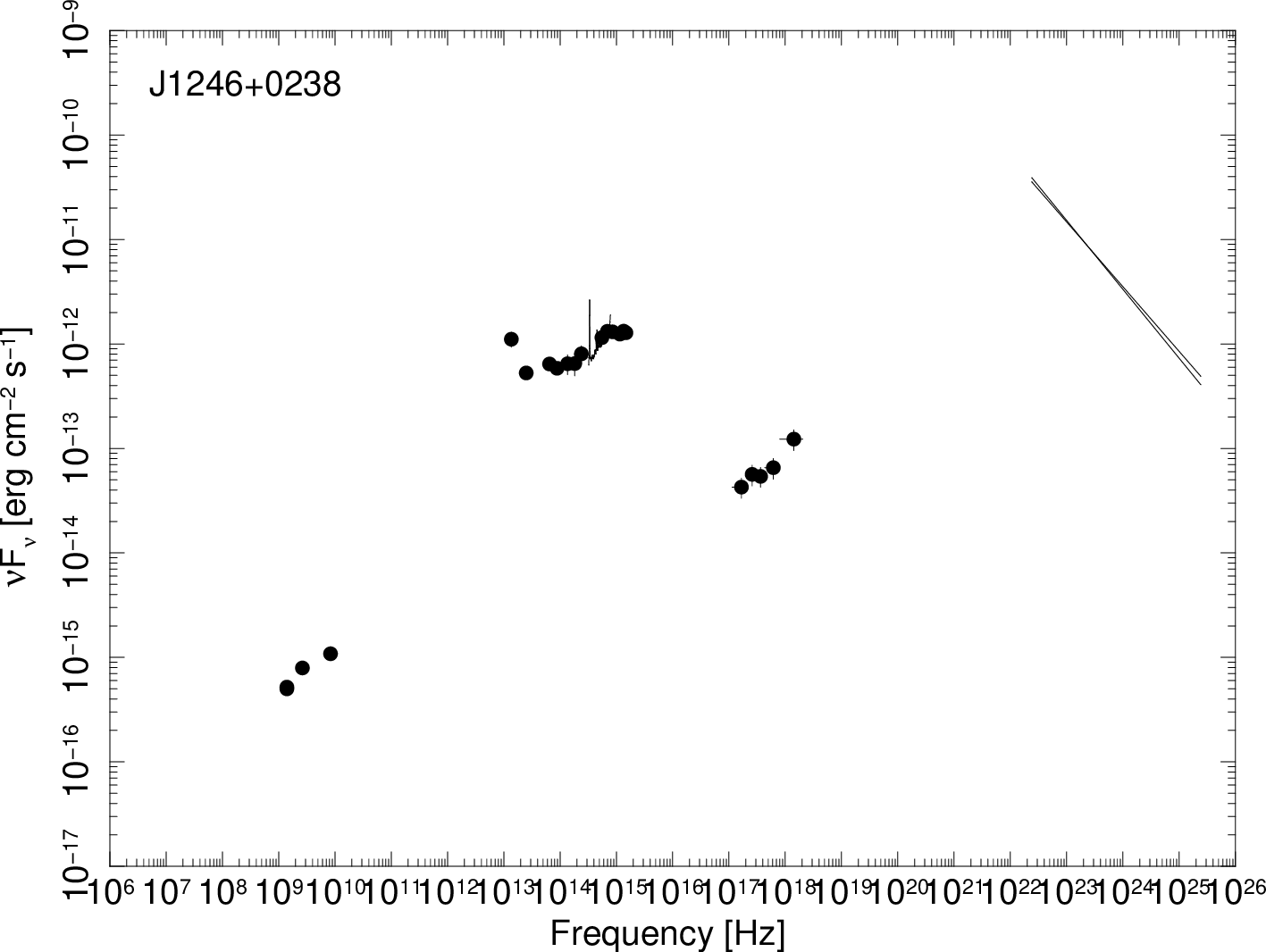}
\includegraphics[angle=270,scale=0.35]{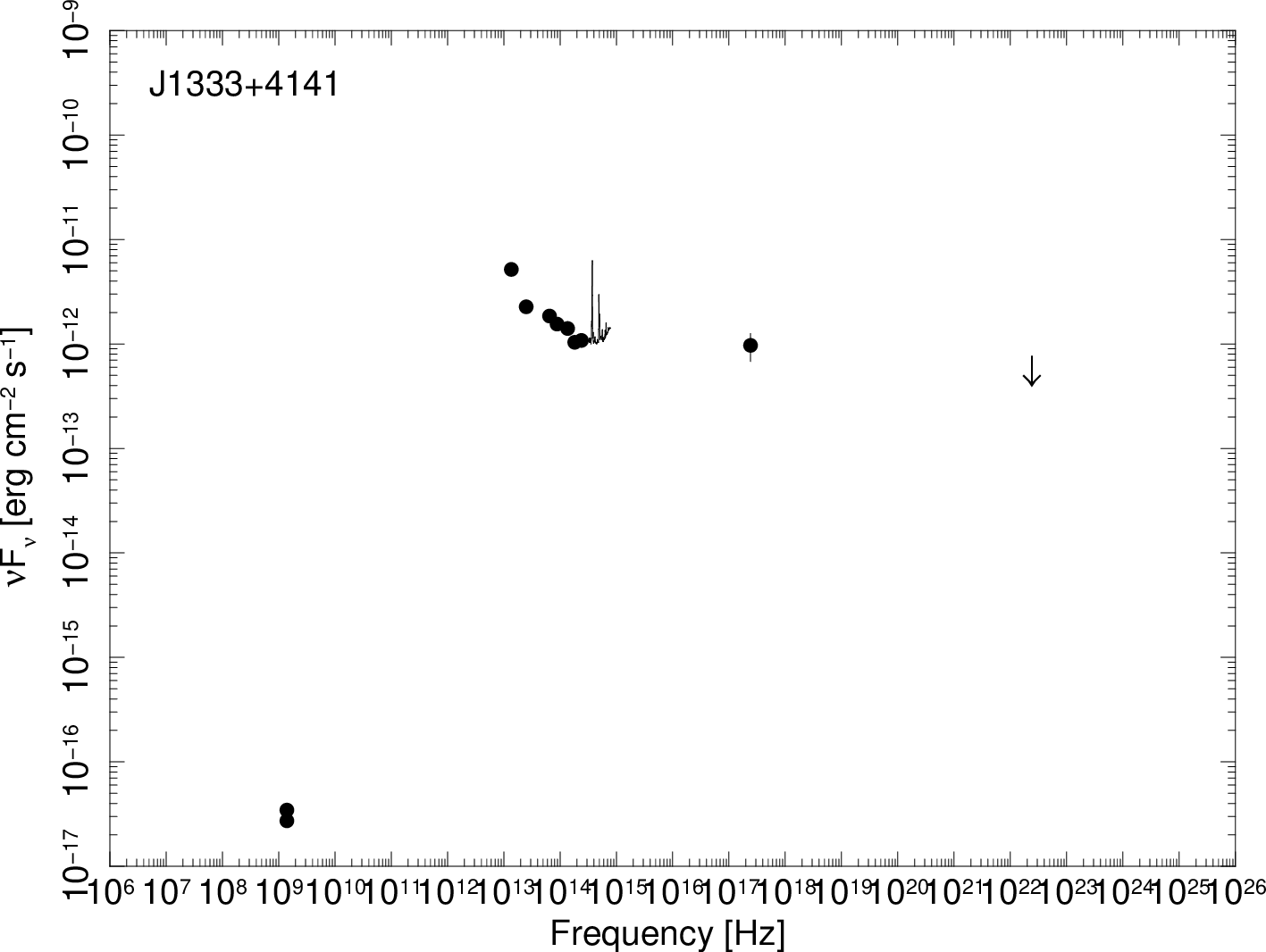}\\
\includegraphics[angle=270,scale=0.35]{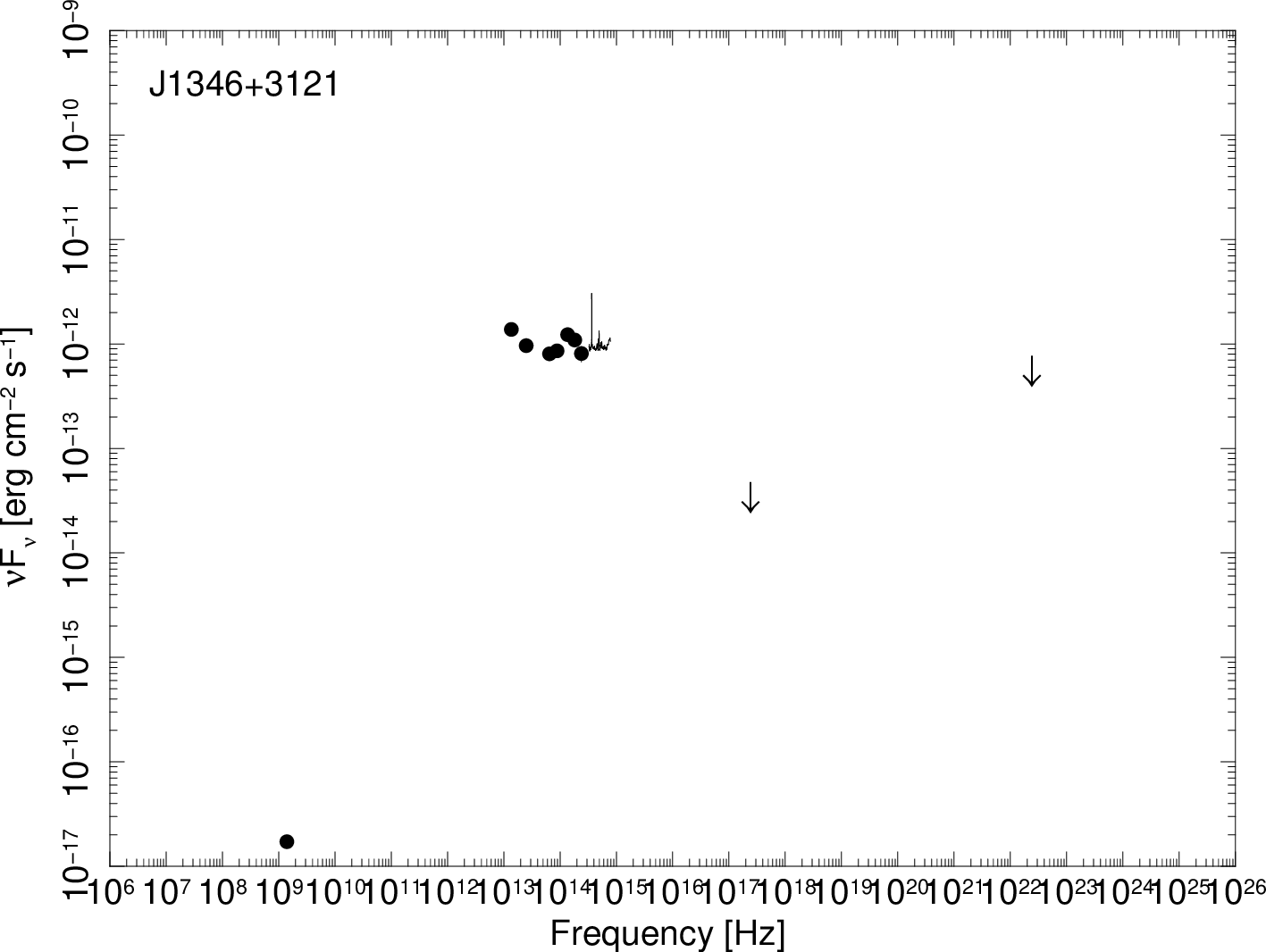}
\includegraphics[angle=270,scale=0.35]{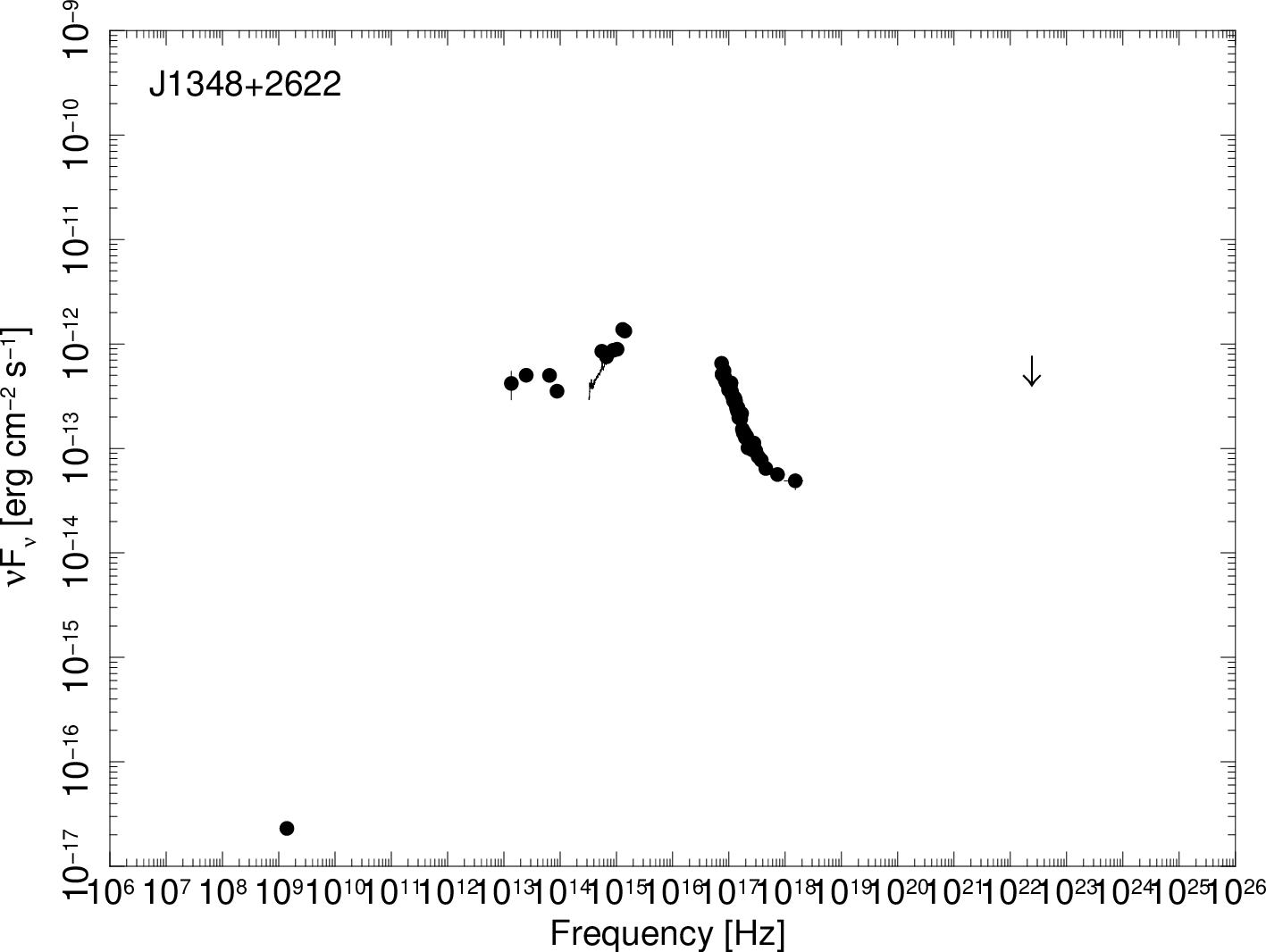}\\
\includegraphics[angle=270,scale=0.35]{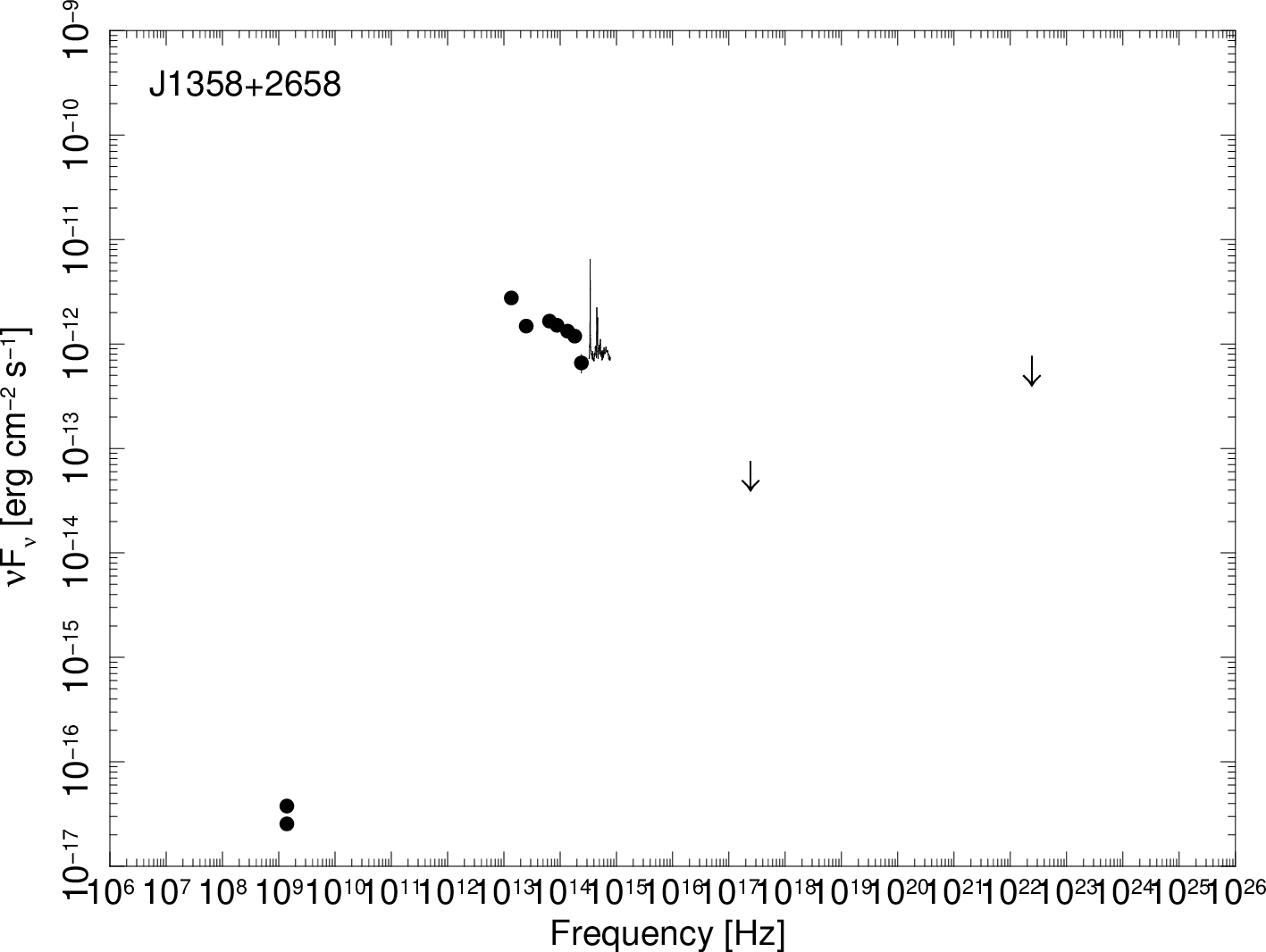}
\includegraphics[angle=270,scale=0.35]{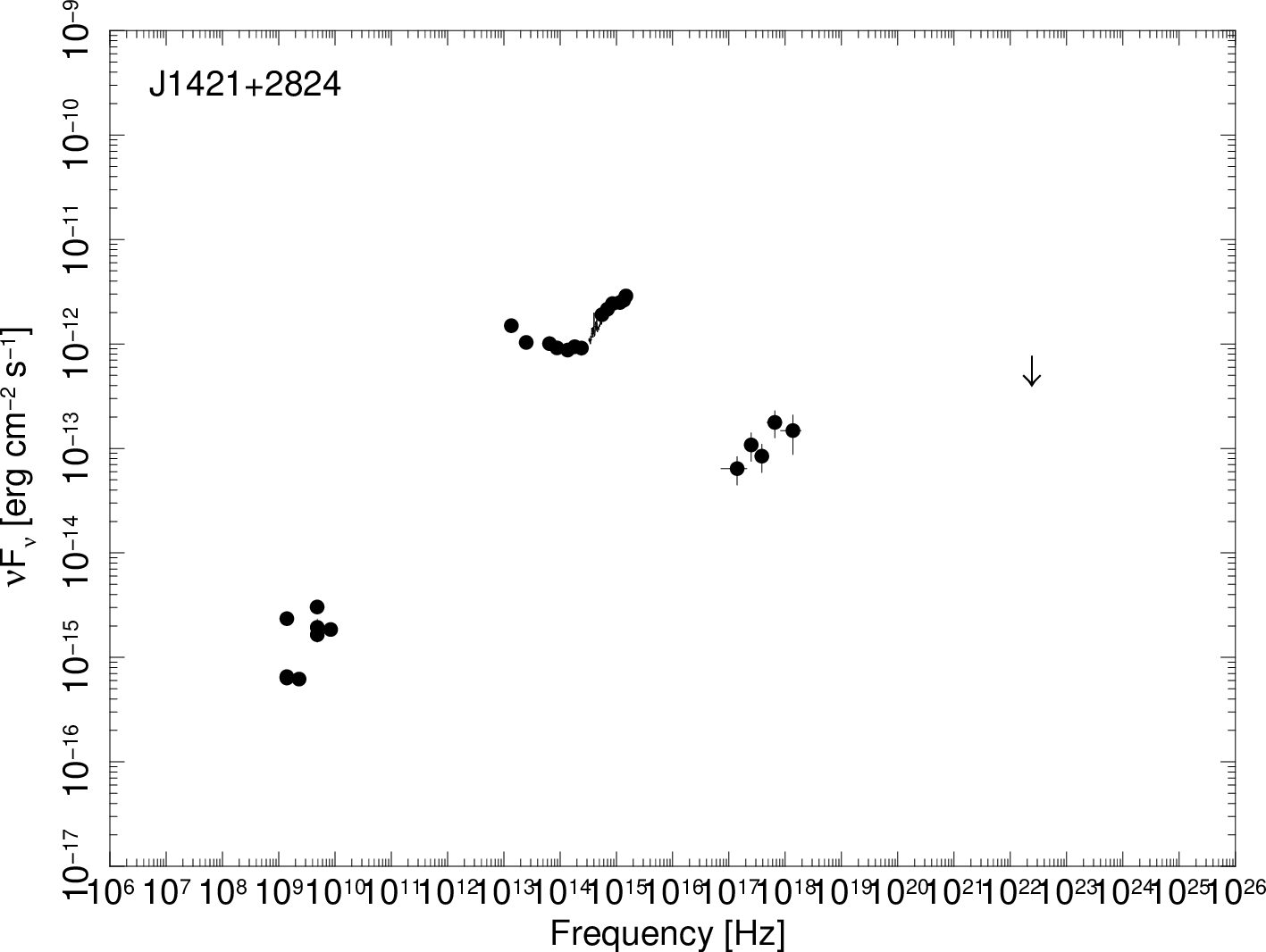}\\
\caption{Spectral Energy Distributions of the sources in the present sample. Data are corrected for the Galactic absorption. Points refer to detections; arrows are upper limits; the continuous lines are the optical spectra.} 
\label{fig:seds4}
\end{figure*}
}

\onlfig{
\begin{figure*}
\centering
\includegraphics[angle=270,scale=0.35]{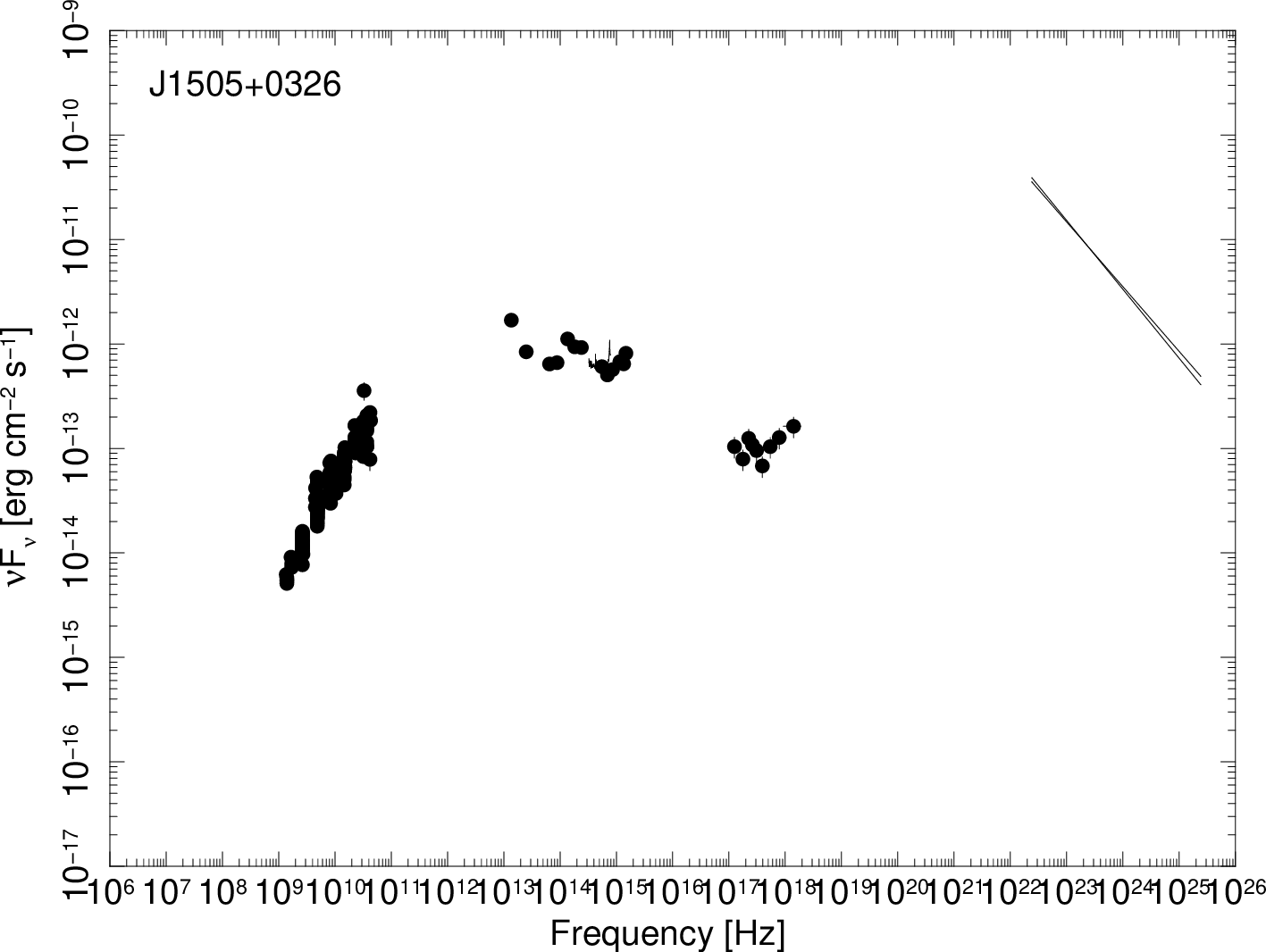}
\includegraphics[angle=270,scale=0.35]{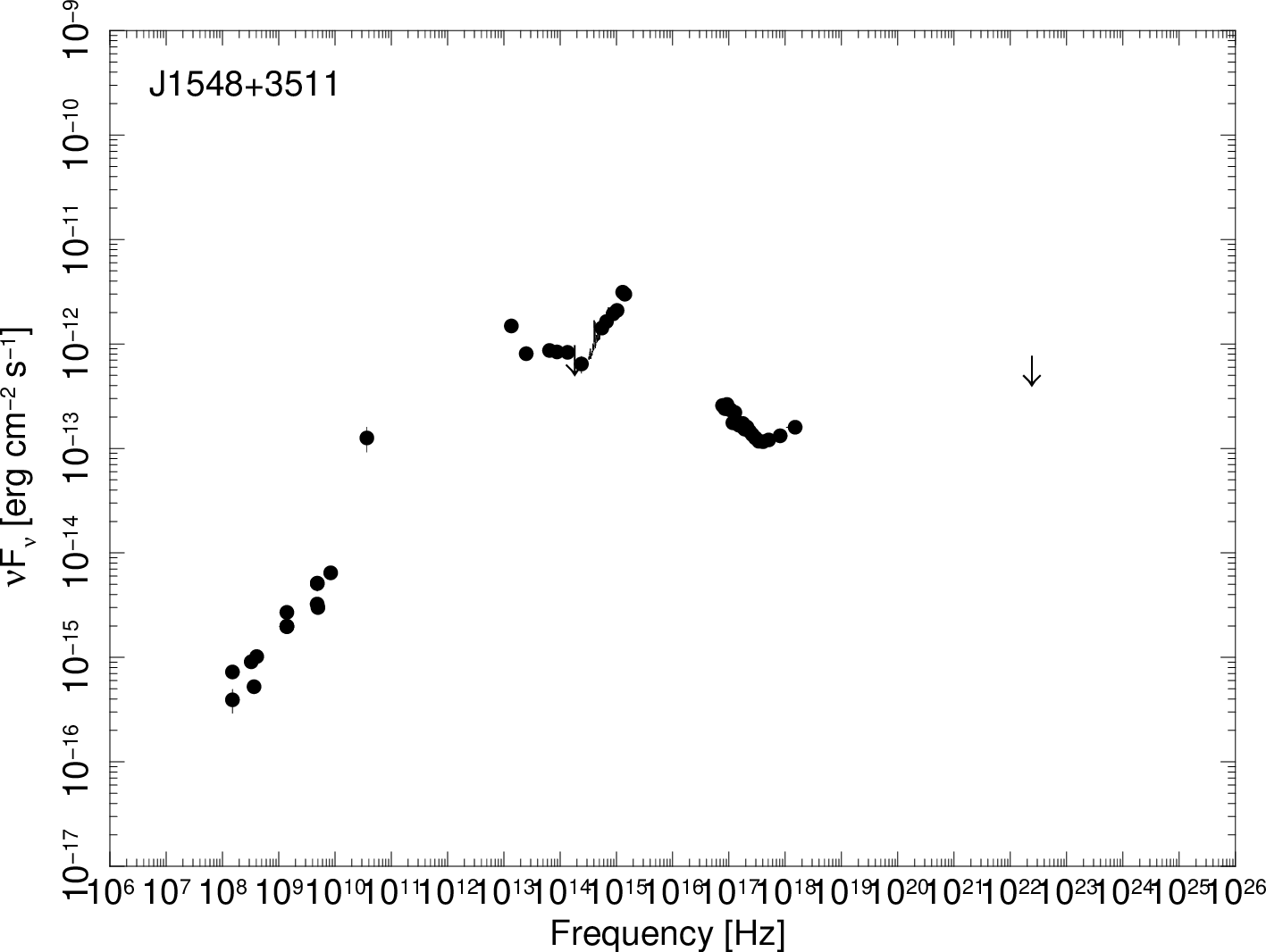}\\
\includegraphics[angle=270,scale=0.35]{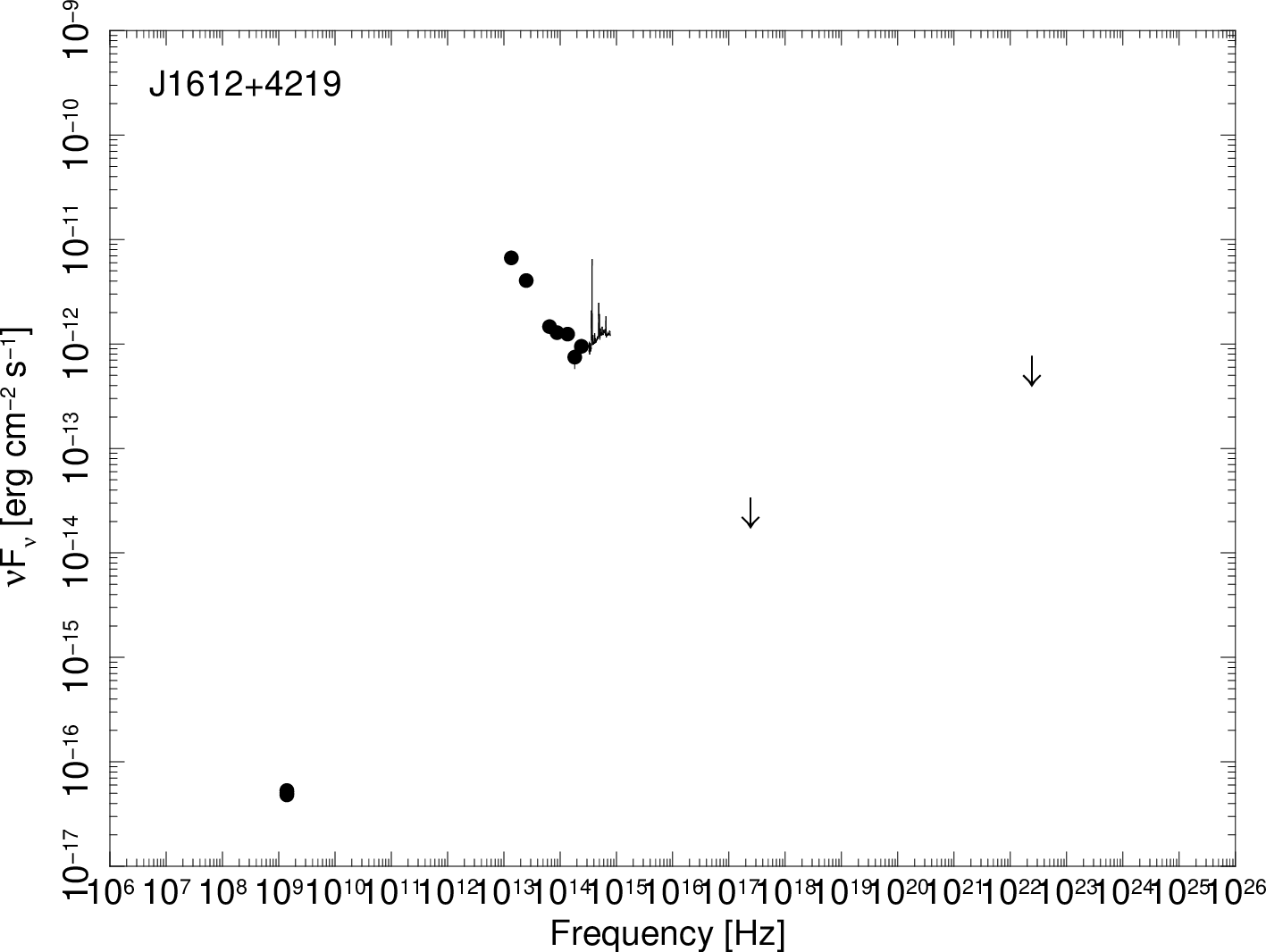}
\includegraphics[angle=270,scale=0.35]{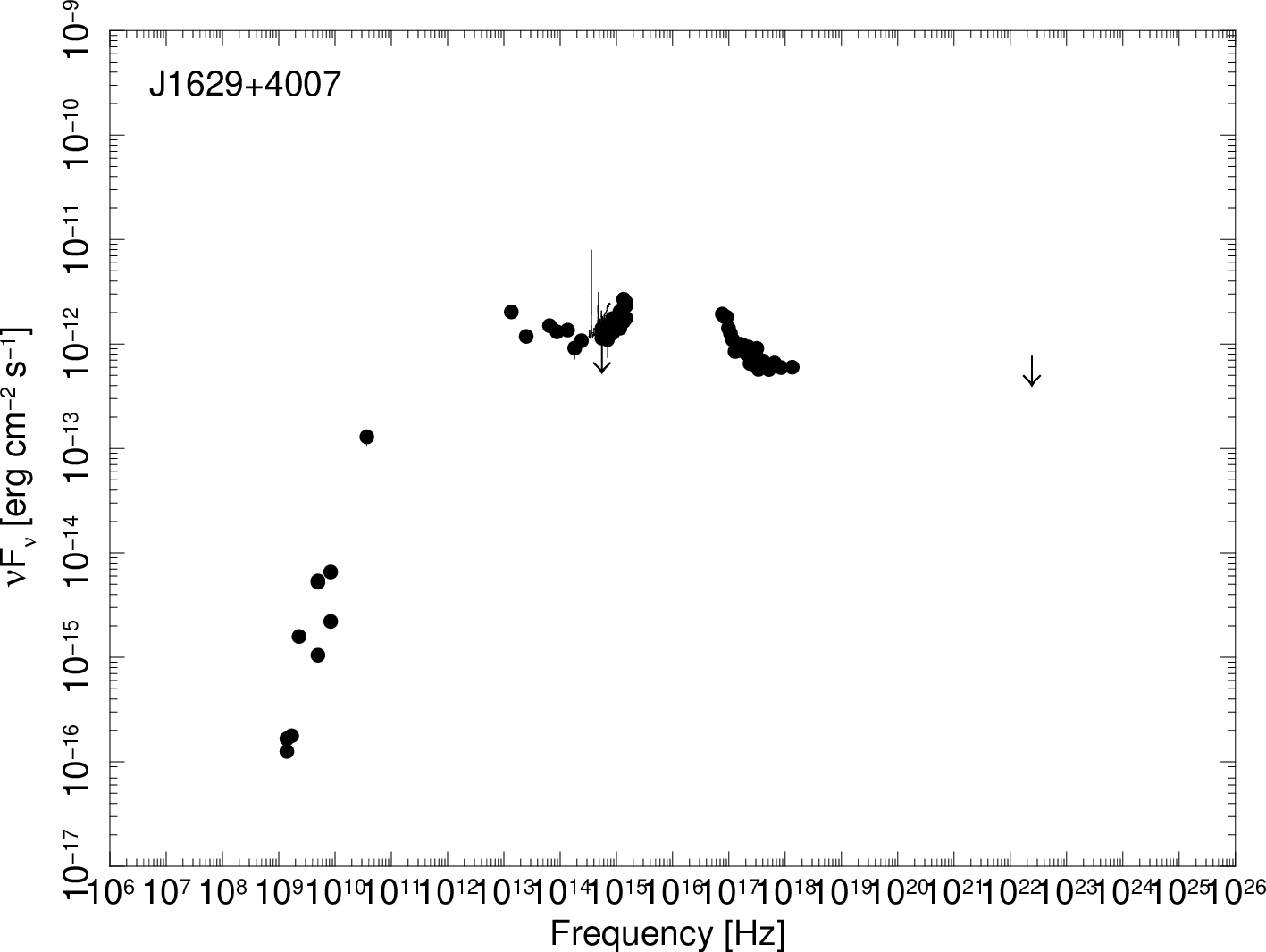}\\
\includegraphics[angle=270,scale=0.35]{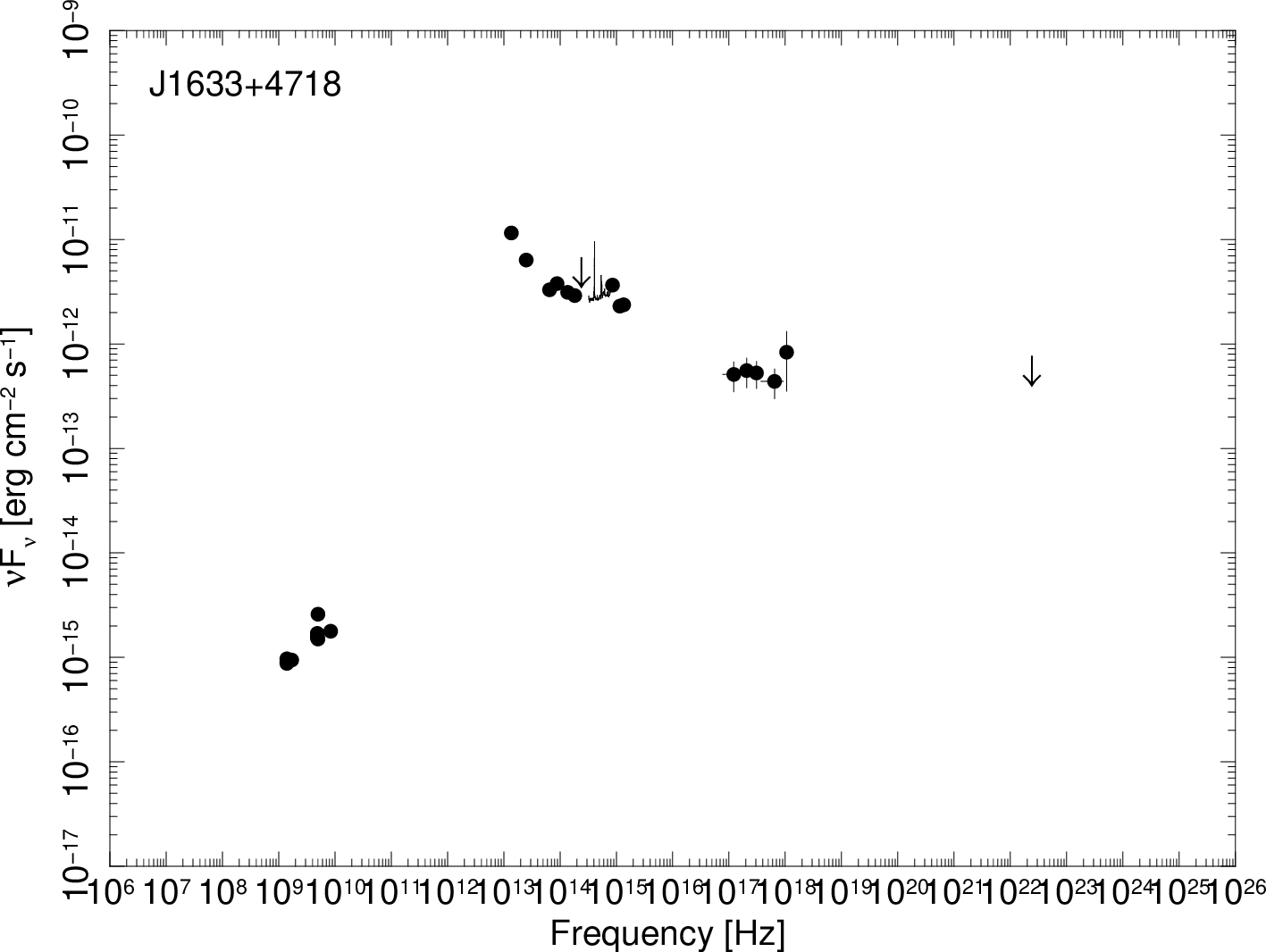}
\includegraphics[angle=270,scale=0.35]{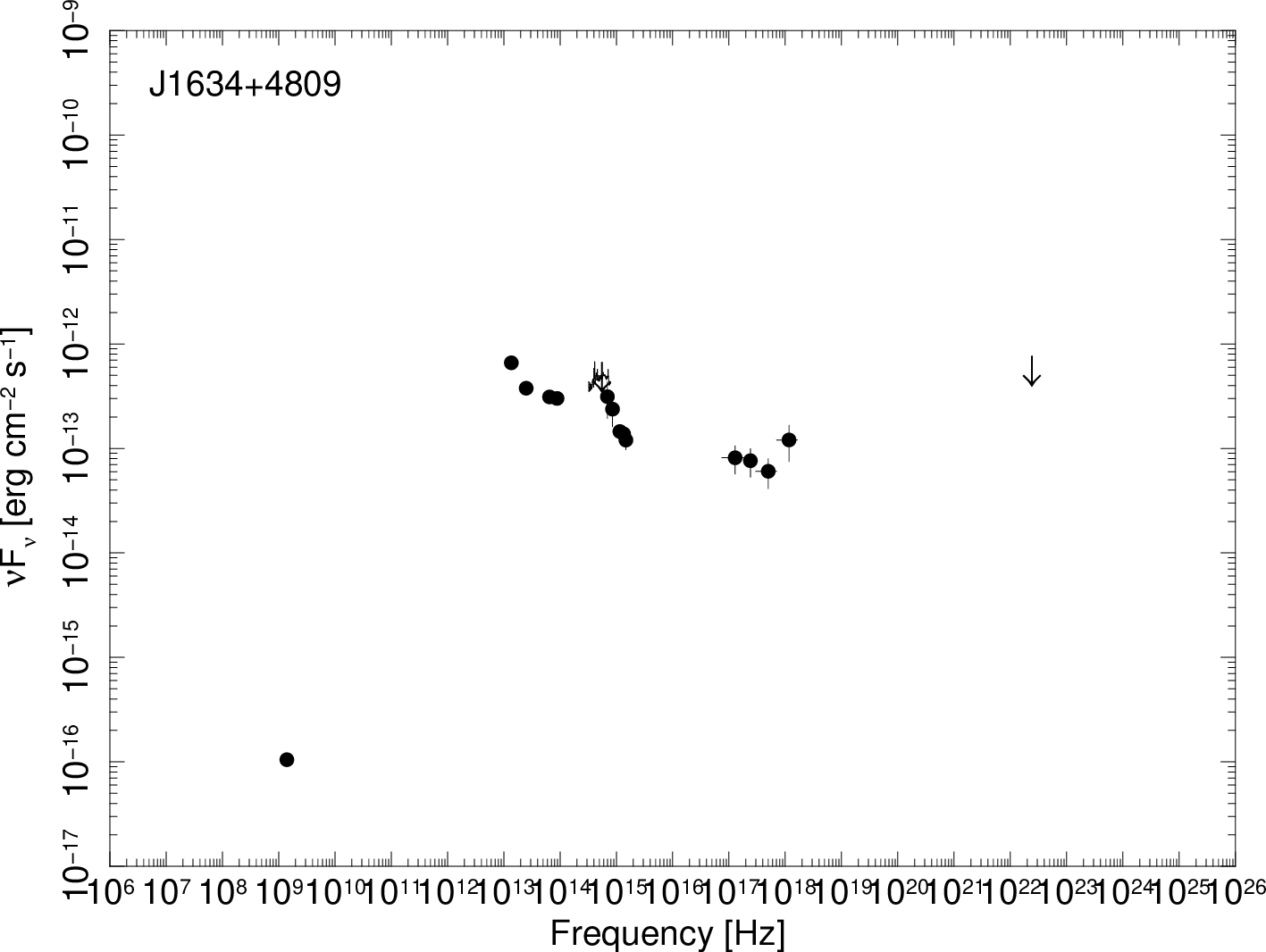}\\
\includegraphics[angle=270,scale=0.35]{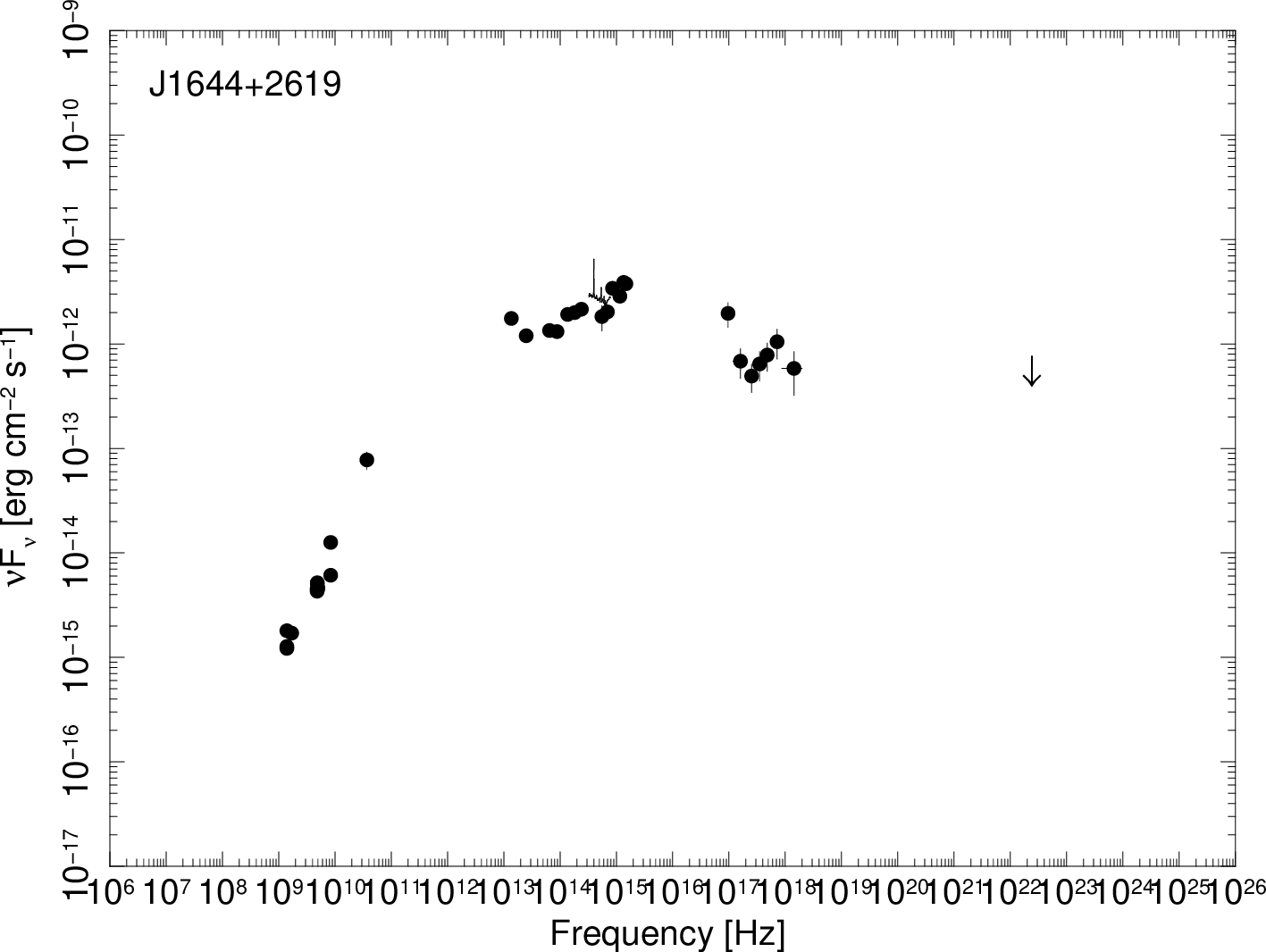}
\includegraphics[angle=270,scale=0.35]{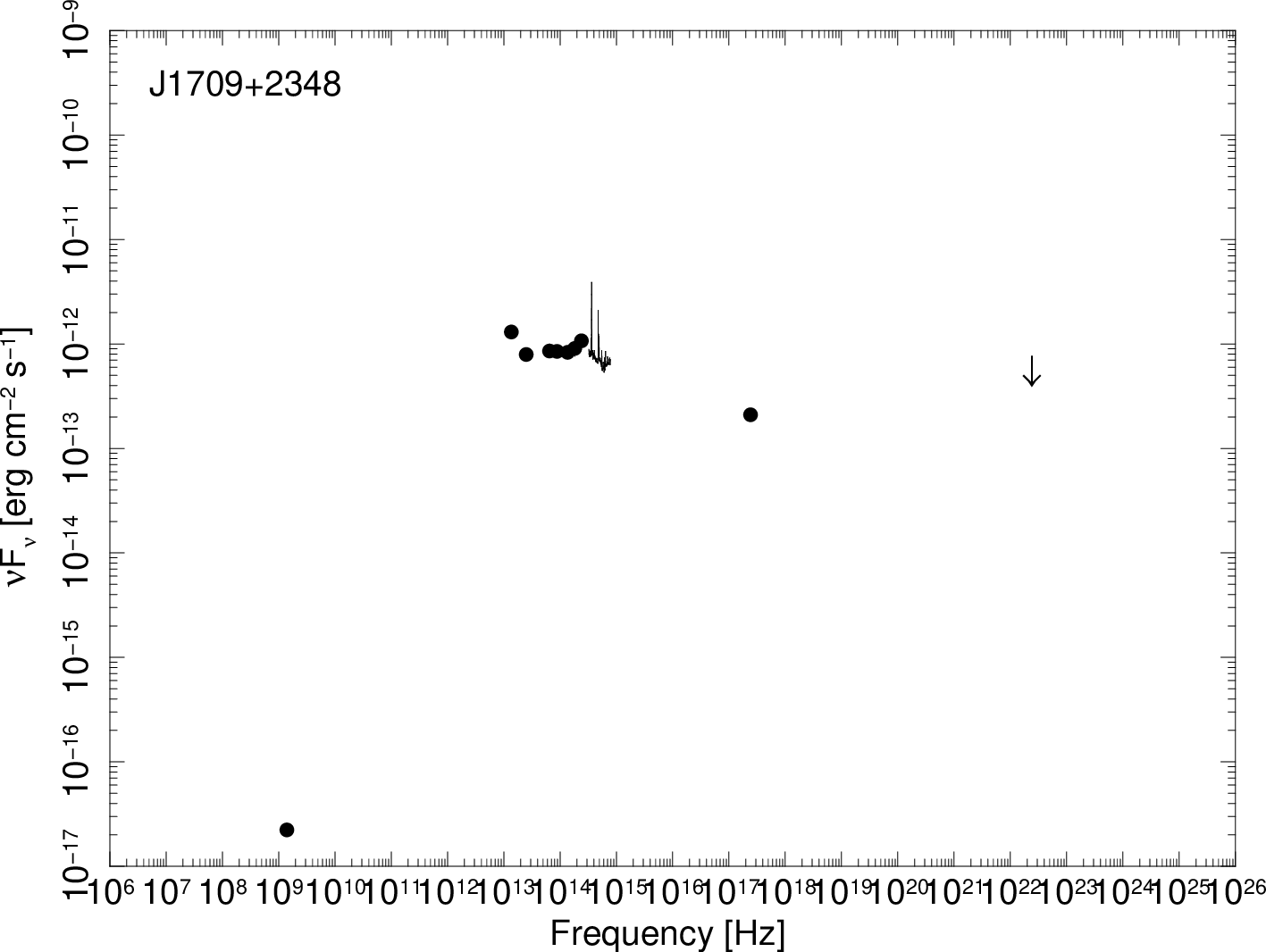}\\
\caption{Spectral Energy Distributions of the sources in the present sample. Data are corrected for the Galactic absorption. Points refer to detections; arrows are upper limits; the continuous lines are the optical spectra.} 
\label{fig:seds5}
\end{figure*}
}

\onlfig{
\begin{figure*}
\centering
\includegraphics[angle=270,scale=0.35]{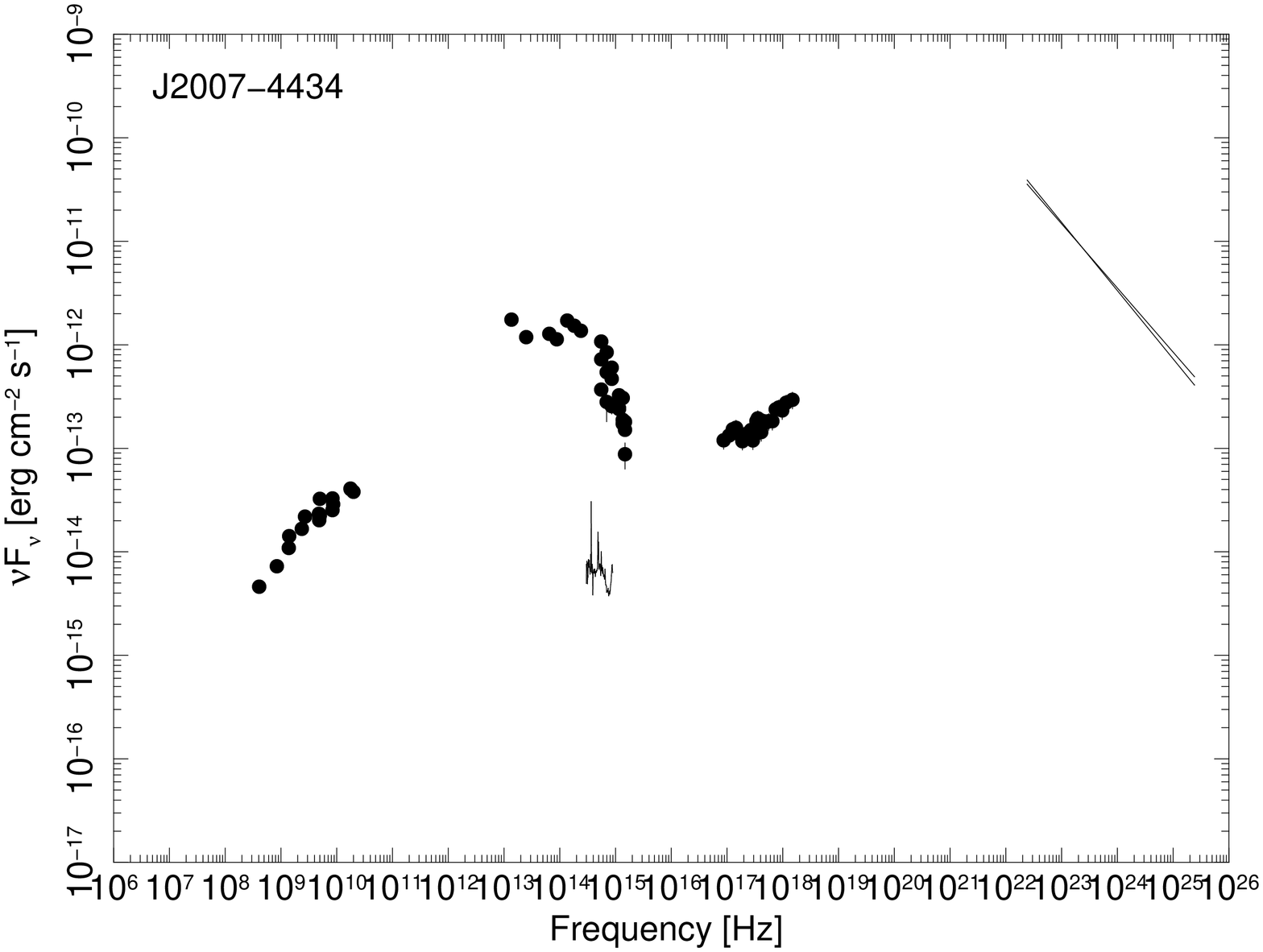}
\includegraphics[angle=270,scale=0.35]{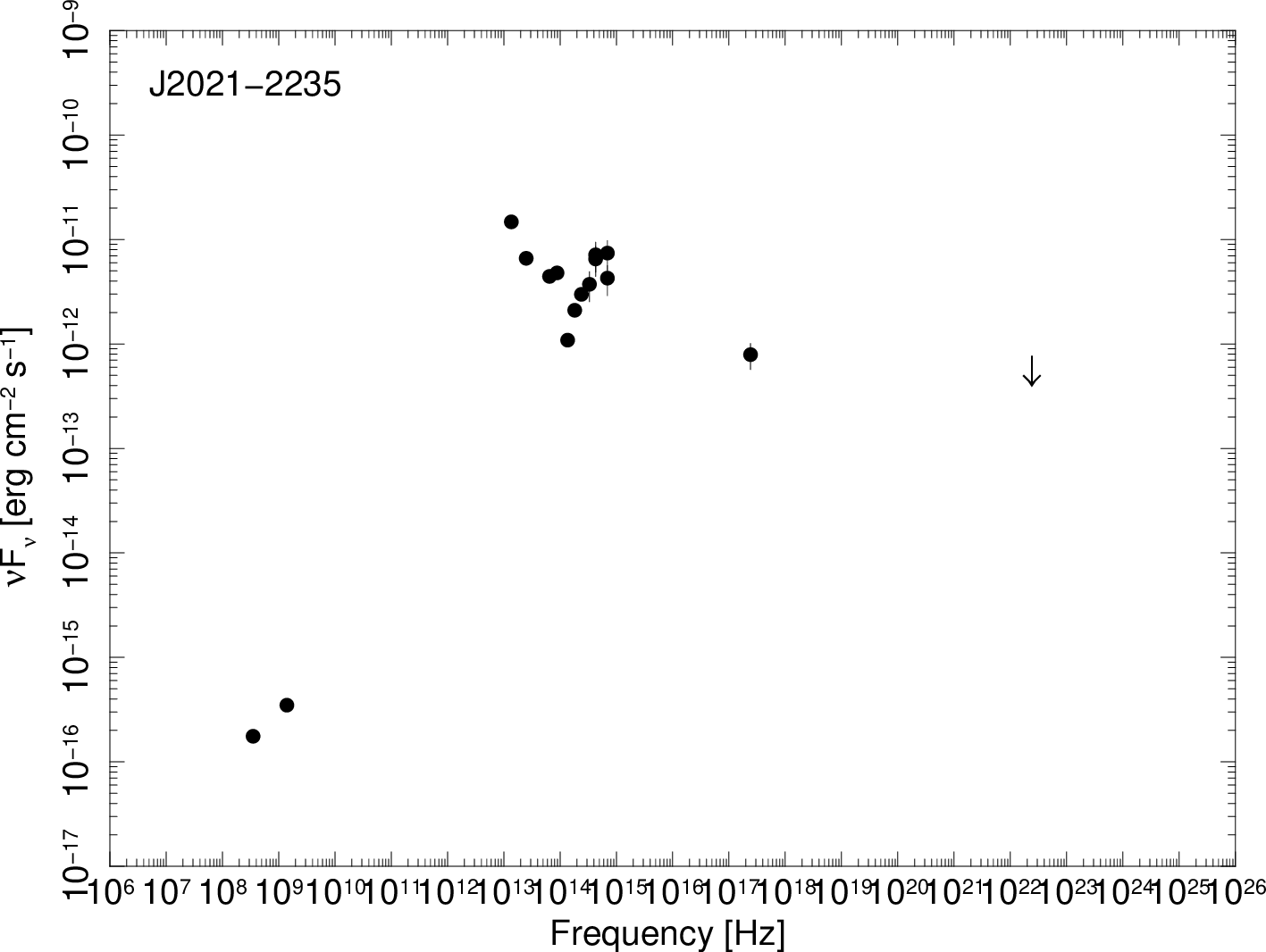}
\caption{Spectral Energy Distributions of the sources in the present sample. Data are corrected for the Galactic absorption. Points refer to detections; arrows are upper limits; the continuous lines are the optical spectra.} 
\label{fig:seds6}
\end{figure*}
}

\begin{acknowledgements}
We would like to thank the members of the Fermi LAT Collaboration -- David Thompson, Denis Bastieri, Jeremy Perkins, and Filippo D'Ammando -- for a critical review of the manuscript.

Part of the Swift observations have been supported by the contract ASI-INAF I/004/11/0.

The Mets\"ahovi team acknowledges the support from the Academy of Finland to our observing projects (numbers 212656, 210338, 121148, and others)

YYK and MML are partly supported by the Russian Foundation for Basic Research (project 13-02-12103). Y.Y.K.\ is also supported by the Dynasty Foundation.

BMP is supported by the NSF through grant AST-1008882.

This research has made use of data from the MOJAVE database that is maintained by the MOJAVE team (Lister et al. 2009, 2013). The MOJAVE program is supported under NASA Fermi grant NNX12AO87G. JLR acknowledges support from NASA through Fermi Guest Investigator grant NNX13AO79G.

This research has made use of data and/or software provided by the High Energy Astrophysics Science Archive Research Center (HEASARC), which is a service of the Astrophysics Science Division at NASA/GSFC and the High Energy Astrophysics Division of the Smithsonian Astrophysical Observatory. 

This research has made use of the NASA/IPAC Extragalactic Database (NED) which is operated by the Jet Propulsion Laboratory, California Institute of Technology, under contract with the National Aeronautics and Space Administration. 

This research has made use of the XRT Data Analysis Software (XRTDAS) developed under the responsibility of the ASI Science Data Center (ASDC), Italy.

Funding for SDSS-III has been provided by the Alfred P. Sloan Foundation, the Participating Institutions, the National Science Foundation, and the U.S. Department of Energy Office of Science. The SDSS-III web site is \url{http://www.sdss3.org/}. SDSS-III is managed by the Astrophysical Research Consortium for the Participating Institutions of the SDSS-III Collaboration including the University of Arizona, the Brazilian Participation Group, Brookhaven National Laboratory, Carnegie Mellon University, University of Florida, the French Participation Group, the German Participation Group, Harvard University, the Instituto de Astrofisica de Canarias, the Michigan State/Notre Dame/JINA Participation Group, Johns Hopkins University, Lawrence Berkeley National Laboratory, Max Planck Institute for Astrophysics, Max Planck Institute for Extraterrestrial Physics, New Mexico State University, New York University, Ohio State University, Pennsylvania State University, University of Portsmouth, Princeton University, the Spanish Participation Group, University of Tokyo, University of Utah, Vanderbilt University, University of Virginia, University of Washington, and Yale University. 

\end{acknowledgements}


\end{document}